\RequirePackage{lineno}
\documentclass[a4paper,10pt,oneside]{cpc-hepnp}
\usepackage{multicol}
\usepackage{CJK,upgreek,fancyhdr}
\usepackage{booktabs}
\usepackage{mathrsfs}
\usepackage{bbm}
\usepackage{amscd}
\usepackage[tbtags]{amsmath}
\usepackage{amssymb,bm,mathrsfs,bbm,amscd}
\usepackage{lastpage}
\usepackage{upgreek,fancyhdr}
\usepackage{color}
\usepackage{geometry}
\usepackage{xspace}
\usepackage{multirow}
\usepackage{hhline}
\usepackage{amssymb}
\usepackage{lineno}
\usepackage{overpic}
\usepackage{graphicx}
\usepackage{float}
\usepackage{dcolumn}
\usepackage{bm}
\usepackage{rotating}
\usepackage{epsfig,graphics,subfigure,psfrag}
\usepackage[colorlinks=true,linkcolor=blue,anchorcolor=black,citecolor=blue]{hyperref}
\newsavebox{\tablebox}
\usepackage{caption}
\usepackage{verbatim}
\usepackage{rotating}
\RequirePackage{rotating}
\DeclareCaptionLabelFormat{cont}{#1~#2\alph{ContinuedFloat}}
\let\oldequation\equation
\let\oldendequation\endequation

\renewenvironment{equation}
  {\linenomathNonumbers\oldequation}
  {\oldendequation\endlinenomath}

\begin{document}
\begin{CJK*}{UTF8}{gkai}

\fancyhead[co]{\footnotesize \boldmath Amplitude Analysis of the Decays $D^0\to\pi^+\pi^-\pi^+\pi^-$ and $\pi^+\pi^-\pi^0\pi^0$}

\footnotetext[0]{Received xxxx}

\title{\boldmath Amplitude Analysis of the Decays $D^0\to\pi^+\pi^-\pi^+\pi^-$ and $D^0\to\pi^+\pi^-\pi^0\pi^0$
\thanks{This work is supported in part by National Key R\&D Program of China under Contracts Nos. 2020YFA0406300, 2020YFA0406400; National Natural Science Foundation of China (NSFC) under Contracts Nos. 11625523, 11635010, 11735014, 11835012, 11935015, 11935016, 11935018, 11961141012, 12025502, 12035009, 12035013, 12061131003, 12105276, 12122509, 12192260, 12192261, 12192262, 12192263, 12192264, 12192265, 12221005, 12225509, 12235017; the Chinese Academy of Sciences (CAS) Large-Scale Scientific Facility Program; the CAS Center for Excellence in Particle Physics (CCEPP); Joint Large-Scale Scientific Facility Funds of the NSFC and CAS under Contract No. U1732263, U1832103, U1832207, U2032111; CAS Key Research Program of Frontier Sciences under Contracts Nos. QYZDJ-SSW-SLH003, QYZDJ-SSW-SLH040; 100 Talents Program of CAS; The Institute of Nuclear and Particle Physics (INPAC) and Shanghai Key Laboratory for Particle Physics and Cosmology; European Union's Horizon 2020 research and innovation programme under Marie Sklodowska-Curie grant agreement under Contract No. 894790; German Research Foundation DFG under Contracts Nos. 455635585, Collaborative Research Center CRC 1044, FOR5327, GRK 2149; Istituto Nazionale di Fisica Nucleare, Italy; Ministry of Development of Turkey under Contract No. DPT2006K-120470; National Research Foundation of Korea under Contract No. NRF-2022R1A2C1092335; National Science and Technology fund of Mongolia; National Science Research and Innovation Fund (NSRF) via the Program Management Unit for Human Resources \& Institutional Development, Research and Innovation of Thailand under Contract No. B16F640076; Polish National Science Centre under Contract No. 2019/35/O/ST2/02907; The Swedish Research Council; U. S. Department of Energy under Contract No. DE-FG02-05ER41374}
\vspace{-0.5in}
}

\maketitle

\begin{small}
\begin{center}
M.~Ablikim$^{1}$, M.~N.~Achasov$^{4,b}$, P.~Adlarson$^{75}$, O.~Afedulidis$^{3}$, X.~C.~Ai$^{80}$, R.~Aliberti$^{35}$, A.~Amoroso$^{74A,74C}$, Q.~An$^{71,58}$, Y.~Bai$^{57}$, O.~Bakina$^{36}$
% ...
\\
\vspace{0.2cm}
(BESIII Collaboration)\\
\vspace{0.2cm} {\it
$^{1}$ Institute of High Energy Physics, Beijing 100049, People's Republic of China\\
% ...
$^{3}$ Bochum  Ruhr-University, D-44780 Bochum, Germany\\
$^{4}$ Budker Institute of Nuclear Physics SB RAS (BINP), Novosibirsk 630090, Russia\\
% ...
$^{35}$ Johannes Gutenberg University of Mainz, Johann-Joachim-Becher-Weg 45, D-55099 Mainz, Germany\\
$^{36}$ Joint Institute for Nuclear Research, 141980 Dubna, Moscow region, Russia\\
% ...
$^{57}$ Southeast University, Nanjing 211100, People's Republic of China\\
$^{58}$ State Key Laboratory of Particle Detection and Electronics, Beijing 100049, Hefei 230026, People's Republic of China\\
% ...
$^{71}$ University of Science and Technology of China, Hefei 230026, People's Republic of China\\
% ...
$^{74}$ University of Turin and INFN, (A)University of Turin, I-10125, Turin, Italy; (B)University of Eastern Piedmont, I-15121, Alessandria, Italy; (C)INFN, I-10125, Turin, Italy\\
$^{75}$ Uppsala University, Box 516, SE-75120 Uppsala, Sweden\\
% ...
$^{80}$ Zhengzhou University, Zhengzhou 450001, People's Republic of China\\
\vspace{0.2cm}
% ...
$^{b}$ Also at the Novosibirsk State University, Novosibirsk, 630090, Russia\\
% ...
}

\end{center}
\vspace{-0.2cm}
\end{small}
\begin{abstract}
Using $e^+e^-$ annihilation data corresponding to an integrated luminosity of 2.93 $\rm fb^{-1}$ taken at the center-of-mass energy $\sqrt{s}=3.773$~GeV with the BESIII detector, a joint amplitude analysis is performed on the decays $D^0\to\pi^+\pi^-\pi^+\pi^-$ and $D^0\to\pi^+\pi^-\pi^0\pi^0$(non-$\eta$). The fit fractions of individual components are obtained, and large interferences among the dominant components of the decays $D^{0}\to a_{1}(1260)\pi$, $D^{0}\to\pi(1300)\pi$, $D^{0}\to\rho(770)\rho(770)$ and $D^{0}\to2(\pi\pi)_{S}$ are found in both channels. With the obtained amplitude model, the $CP$-even fractions of $D^0\to \pi^+\pi^-\pi^+\pi^-$ and $D^0\to\pi^+\pi^-\pi^0\pi^0$(non-$\eta$) are determined to be $(75.2~\pm~1.1_{\rm stat.}~\pm~1.5_{\rm syst.})\%$ and $(68.9~\pm~1.5_{\rm stat.}~\pm~2.4_{\rm syst.})\%$, respectively. The branching fractions of $D^0\to \pi^+\pi^-\pi^+\pi^-$ and $D^0\to\pi^+\pi^-\pi^0\pi^0$(non-$\eta$) are measured to be $(0.688~\pm~0.010_{\rm stat.}~\pm~0.010_{\rm syst.})\%$ and $(0.951~\pm~0.025_{\rm stat.}~\pm~0.021_{\rm syst.})\%$, respectively. The amplitude analysis provides an important model for binning strategy in the measurements of the strong phase parameters of $D^0 \to 4\pi$ when used to determine the CKM angle $\gamma (\phi_{3})$ via the $B^{-}\to D K^{-}$ decay.
\end{abstract}


\begin{multicols}{2}
\section{INTRODUCTION}
Precision measurements of the elements of the Cabibbo-Kobayashi-Maskawa (CKM) matrix and the test of the unitarity of the CKM triangle~\cite{Cabibbo:1963yz,Kobayashi:1973fv} are essential goals in the field of flavour physics. 
One of the three angles of the unitarity triangle, $\gamma(\phi_{3})\equiv \mathrm{arg}(-V_{ud}V_{ub}^{*}/V_{cd}V_{cb}^{*})$, can be measured with the tree-level decay $B^{\pm}\to DK^{\pm}$ through the interference between $B^{-}\to D^0 K^{-}$~$ (b\to c\bar{u}s)$ and  $B^{-}\to \bar{D}^0 K^{-} $~$(b\to u\bar{c}s)$ amplitudes, which is one of the most important measurements of LHCb and Belle II.
Several approaches have been proposed to measure the CKM angle $\gamma$ via the decay $B^{\pm}\to DK^{\pm}$~\cite{Gronau:1991dp, Atwood:1996ci, Giri:2003ty}. Here, the relative magnitude and phase between the $D^0$ and $\bar{D}^0$ mesons decay into the same final states, and the $D^0$ decay parameters ($e.g.$ $CP$-even fraction $F_+$) are the critical inputs.
With more data collected by the LHCb and Belle II experiments in the coming years, the decay parameters of $D^0$ ($\bar{D}^0$)  will become the dominant source of uncertainty in the $\gamma$ measurement. Therefore, precision measurements of the $D^0$ decay parameters are urgently required to improve the precision of the $\gamma$ measurement.

The decay $D^{0}\to4\pi$ is regarded as a sensitive mode~\cite{Harnew:2017tlp, LHCb:2019yan} to extract the CKM angle $\gamma$ via $B^{-}\to D K^{-}$.
Their $CP$-even fractions ($F_+$) and relative strong phase parameters in the different phase space (PHSP) bins ($c_i$/$s_i$) serve as the direct inputs in the GLW~\cite{Gronau:1991dp} and BPGGSZ~\cite{Giri:2003ty,Belle:2004bbr} methods, respectively. 
A reliable decay amplitude model of $D^{0}\to4\pi$ is critical to precisely extract  $F_+$ and a model-independent $c_i$/$s_i$~\cite{Harnew:2017tlp}, and to search for $CP$ violation in $D^0\to 4\pi$~\cite{LHCb:2016qbq}.
Moreover, the four-body $D^0$ hadronic decays provide an excellent platform to study the two-body decays $D^0\to VV$ and $D^0\to AP$, where $V$, $A$, and $P$ denote vector, axial-vector, and pseudo-scalar mesons, respectively. These decays thus enhance the understanding of the decay dynamics of the $D^0$ meson~\cite{Cheng:2003bn,Cheng:2010ry}. 

Currently, there are only a limited number of experimental studies of $D^0\to 4\pi$. The FOCUS experiment performed an amplitude analysis of $D^0\to\pi^+\pi^-\pi^+\pi^-$ based on $\sim6000$ candidate events with a background fraction of $\sim10\%$~\cite{FOCUS:2007ern}.
An amplitude analysis of  $D^0\to\pi^+\pi^-\pi^+\pi^-$ was also carried out with the CLEO-c data, which contain $\sim7000$ candidate events with a background fraction of $\sim20\%$~\cite{dArgent:2017gzv}. 
However, no amplitude analysis of $D^0\to\pi^+\pi^-\pi^0\pi^0$ has been performed yet. 
BESIII has collected 2.93~$\rm{fb}^{-1}$ of $e^+e^-$ collision data at the center-of-mass energy $\sqrt{s}=3.773$ GeV, where the $D\bar{D}$ pair is produced without any additional hadrons. This data sample provide an ideal environment for studying $D$ meson decays with the double tag (DT) technique~\cite{MARK-III:1985hbd,Li:2021iwf}. In this method, a single tag (ST) candidate requires that only one $D$ meson is reconstructed via ST mode. A DT candidate requires that both of $D$ and $\bar{D}$ are reconstructed via signal mode and ST mode, respectively. Based on this data sample and DT method with three ST modes $\bar{D}^{0}\to K^{+}\pi^-$, $\bar{D}^{0}\to K^{+}\pi^-\pi^0$ and $\bar{D}^{0}\to K^{+}\pi^-\pi^+\pi^-$, we report a joint amplitude analysis of $D^0\to\pi^+\pi^-\pi^+\pi^-$ and $D^0\to\pi^+\pi^-\pi^0\pi^0$(non-$\eta$). Furthermore, we determine the model-dependent $CP$-even fractions, the absolute branching fractions, and the fractions of individual components. Throughout this paper, the charge-conjugated processes are always implied.

\section{BESIII DETECTOR AND MONTE CARLO SIMULATION}
\label{MCsimulation}
The BESIII detector~\cite{BESIII:2009fln} records symmetric $e^+e^-$ collisions provided by the BEPCII storage ring~\cite{Yu:2016cof} in the center-of-mass energy range from 2.0 to 4.95~GeV, with a peak luminosity of $1 \times 10^{33}\;\text{cm}^{-2}\text{s}^{-1}$ achieved at $\sqrt{s} = 3.77\;\text{GeV}$. BESIII has collected large data samples in this energy region~\cite{BESIII:2020nme}. The cylindrical core of the $\rm{BESIII}$ detector covers 93\% of the full solid angle and consists of a helium-based multilayer drift chamber~(MDC), a plastic scintillator time-of-flight system~(TOF), and a CsI(Tl) electromagnetic calorimeter~(EMC), which are all enclosed in a superconducting solenoidal magnet providing a 1.0~T magnetic field. The solenoid is supported by an octagonal flux-return yoke with resistive plate counter muon identification modules interleaved with steel. The charged-particle momentum resolution at $1~{\rm GeV}/c$ is $0.5\%$, and the ${\rm d}E/{\rm d}x$ resolution is $6\%$ for electrons from Bhabha scattering. The EMC measures photon energies with a resolution of $2.5\%$ ($5\%$) at $1$~GeV in the barrel (end cap) region. The time resolution in the TOF barrel region is 68~ps, while
that in the end cap region is 110~ps. 

Simulated data samples produced with a {\sc geant4}-based~\cite{GEANT4:2002zbu} Monte Carlo (MC) package, which includes the geometric description~\cite{Huang:2022wuo} of the BESIII detector and the detector response, are used to determine detection efficiencies and to estimate backgrounds. The simulation models the beam
energy spread and initial state radiation (ISR) in the $e^+e^-$ annihilations with the generator {\sc kkmc}~\cite{Jadach:2000ir,Jadach:1999vf}. The inclusive MC sample including the production of $D\bar{D}$ pairs (including quantum coherence for the neutral $D$ channels), the non-$D\bar{D}$ decays of the $\psi(3770)$, the ISR production of the $J/\psi$ and $\psi(3686)$ states, and the continuum processes incorporated in {\sc kkmc} are generated to estimate the background and ST efficiencies. All particle decays are modeled with {\sc evtgen}~\cite{Lange:2001uf,Ping:2008zz} using branching fractions either taken from the Particle Data Group (PDG)~\cite{ParticleDataGroup:2022pth}, when available, or otherwise estimated with {\sc lundcharm}~\cite{Chen:2000tv,Yang:2014vra}. Final state radiation from charged final state particles is incorporated using the {\sc photos} package~\cite{Richter-Was:1992hxq}.
Signal MC samples of $D^0\to\pi^+\pi^-\pi^+\pi^-$ and $D^0\to\pi^+\pi^-\pi^0\pi^0$ generated uniformly in PHSP are used to normalize probability density functions (PDFs) in the amplitude analysis, while those generated according to the amplitude analysis results are used to estimate the DT efficiencies.

\section{EVENT SELECTION} \label{sec_sel}
Charged tracks detected in the MDC are required to be within a polar angle ($\theta$) range of $|\rm{cos\theta}|<0.93$, where $\theta$ is defined with respect to the $z$ axis, which is the symmetry axis of the MDC. The distance of closest approach of these charged tracks to the interaction point must be less than 10\,cm along the $z$ axis, and less than 1\,cm in the transverse plane.
Particle identification~(PID) for charged tracks combines measurements of the specific ionization energy loss in the MDC~(d$E$/d$x$) and the flight time in the TOF to form likelihoods $\mathcal{L}(h)~(h=K,\pi)$ for each hadron $h$ hypothesis. The charged kaons and pions are identified by comparing the likelihoods for the kaon and pion hypotheses, $\mathcal{L}(K)>\mathcal{L}(\pi)$ and $\mathcal{L}(\pi)>\mathcal{L}(K)$, respectively.

Photon candidates are identified using showers in the EMC.  The deposited energy of each shower must be $>25$~MeV in the barrel region ($|\cos \theta|< 0.80$) and $>50$~MeV in the end cap region ($0.86 <|\cos \theta|< 0.92$). To exclude showers that originate from charged tracks, the angle subtended by the EMC shower and the position of the closest charged track at the EMC must be $>10^{\circ}$ as measured from the interaction point. To suppress electronic noises and reject showers unrelated to the event, the difference between the EMC time and the event start time is required to be within [0, 700]\,ns. The $\pi^0$ candidates are reconstructed from pairs of photon candidates with invariant mass being in the interval (0.115, 0.150) GeV/$c^2$. To improve the momentum resolution, a kinematic fit constraining the two-photon invariant mass to the known $\pi^0$ mass~\cite{ParticleDataGroup:2022pth}  is performed, and the four-momenta updated by this kinematic fit are used in the following analysis.

The signal candidates of $D^0\to\pi^+\pi^-\pi^+\pi^-$ and $D^0\to\pi^+\pi^-\pi^0\pi^0$ are selected with the DT method~\cite{MARK-III:1985hbd,Li:2021iwf}. First, the ST $\bar{D}^{0}$ mesons are reconstructed with the three hadronic decay modes $\bar{D}^{0}\to K^{+}\pi^-$, $\bar{D}^{0}\to K^{+}\pi^-\pi^0$, and $\bar{D}^{0}\to K^{+}\pi^-\pi^+\pi^-$. Two kinematic variables, the energy difference with respect to the beam energy $\Delta E$ and the beam energy constrained mass $M_{\rm{bc}}$ are defined as 
\begin{eqnarray}
  &\Delta E=&E_{\bar{D}^{0}} - E_{\rm{beam}}, \\  \label{eq1}
  &M_{\rm{bc}} =&\sqrt{E_{\rm{beam}}^{2}-\vec{p}_{\bar{D}^{0}}^2}~,  \label{eq2}
\end{eqnarray}
where $E_{\rm{beam}}$ is the beam energy, and $E_{\bar{D}^{0}}$ and $\vec{p}_{\bar{D}^{0}}$ are the energy and momentum of the ST $\bar{D}^{0}$ candidate in the $e^+e^-$ center-of-mass frame. For multiple $\bar{D}^{0}$ candidates in each ST mode, only the one with the smallest $|\Delta E|$ is kept for further analysis. To reject the backgrounds from cosmic rays and Bhabha events in the ST mode $\bar{D}^{0}\to K^{+}\pi^-$, the requirements described in Ref.~\cite{BESIII:2014rtm} are applied. To reject the peaking background from the decay $\bar{D}^{0}\to K^{+}K_{S}^{0}\pi^-$ in the ST mode $\bar{D}^{0}\to K^{+}\pi^-\pi^+\pi^-$, the events with $|M(\pi^+\pi^-)-0.4976|<0.03$~GeV/$c^{2}$ are vetoed. To further reject combinatorial backgrounds, the $\bar{D}^{0}$ candidates are required to have $\Delta E$ within a given interval defined in Table~\ref{tab1}, about 3 times the resolution for each ST mode.
In the sub-sample containing ST candidates, the signal candidates of $D^0\to\pi^+\pi^-\pi^+\pi^-$ and $D^0\to\pi^+\pi^-\pi^0\pi^0$ are reconstructed with the $\pi^{\pm}$ and $\pi^0$ candidates which have not been used in the ST side (namely DT thereafter). Similar kinematic variables $\Delta E$ and $M_{\rm{bc}}$ are formed for signal $D^0$ candidates, and the corresponding $\Delta E$ requirements are listed in Table~\ref{tab1}. For multiple signal $D^0$ candidates, only the one giving the smallest $|\Delta E|$ is kept. 

\begin{table*}[htbp]
\caption{The $\Delta E$ requirements for different decay modes.}
\label{tab1}
\begin{center}
\begin{tabular}{l|c}
\hline
\hline
Decay mode		                                            	&$\Delta E$ (GeV) 	  \\
\hline
$\bar{D}^{0}\to K^{+}\pi^{-}$		           &	$(-0.027,0.026)$	\\
$\bar{D}^{0}\to K^{+}\pi^{-}\pi^{0}$	      &	$(-0.057,0.043)$	\\
$\bar{D}^{0}\to K^{+}\pi^{-}\pi^{+}\pi^{-}$&	$(-0.020,0.018)$    \\ 
\hline
$D^{0}\to \pi^{+}\pi^{-}\pi^{+}\pi^{-}$       &  $(-0.032,0.028)$     \\
$D^{0}\to \pi^{+}\pi^{-}\pi^{0}\pi^{0}$      &   $(-0.066,0.041)$    \\
\hline
\hline
\end{tabular}
\end{center}
\end{table*}

To improve the purity of signal candidates in the amplitude analysis, some further selection criteria are applied. 
The studies based on the inclusive MC sample indicate that most backgrounds from the continuum process $e^+e^- \to q\bar{q}$ include a $K_{S}^{0}$ in the final state. Therefore, common and secondary vertex fits are performed on all $\pi^{+}\pi^{-}$ pairs in the event to reconstruct the $K_{S}^{0}$ candidate. The candidate events are rejected if there exists any  $K_{S}^{0}$ candidate with $\pi^+\pi^-$ invariant mass within the interval $|M(\pi^{+}\pi^{-})-0.4976|<0.03~{\rm GeV}/c^{2}$ with the decay length greater than twice its resolution. 
The MC studies also show that most backgrounds from $e^+e^- \to D^+D^-$ contain the decay $D^{-}\to K^{+}\pi^{-}\pi^{-}$ due to its large branching fraction and a similar topology as the signal.
The candidate events are rejected if there exists any $K^{+}\pi^{-}\pi^{-}$ combinations with  $1.863<M_{\rm{bc}}(K^{+}\pi^{-}\pi^{-})<1.878~{\rm GeV}/c^{2}$ and $|\Delta E(K^{+}\pi^{-}\pi^{-})|<0.03 ~\rm{GeV}$.
There is also the background from the process $e^+e^- \to D^0\bar{D}^0$. Events with any $\pi^{+}\pi^{-}\pi^{0}$ combinations satisfying $1.859<M_{\rm{bc}}(\pi^{+}\pi^{-}\pi^{0})<1.873~{\rm GeV}/c^{2}$ and $-0.057<\Delta E(\pi^{+}\pi^{-}\pi^{0})<0.043 ~\rm{GeV}$ are rejected to eliminate the background of the decay $D^{0}\to\pi^{+}\pi^{-}\pi^{0}$ in the signal process $D^{0}\to\pi^{+}\pi^{-}\pi^{0}\pi^{0}$.
All of the above backgrounds do not form peaks in the $M_{\rm bc}$ distribution of the signal side.
There are also some backgrounds which have the same final states as those in the signal mode and produce peaks in the $M_{\rm bc}$ distributions of both the ST and signal sides.
To reject the peaking backgrounds from the decays $D^{0}\to K_{S}^{0}(\rightarrow\pi^{+}\pi^{-})\pi^{+}\pi^{-}$ and $D^{0}\to K^{-}(\rightarrow\pi^{+}\pi^{-}\pi^{-})\pi^{+}$ in the signal process $D^0\to\pi^+\pi^-\pi^+\pi^-$, events with $|M(\pi^{+}\pi^{-})-0.4976|<0.03~ {\rm GeV}/c^{2}$ or $M(\pi^{+}\pi^{-}\pi^{-})<0.51~{\rm GeV}/c^{2}$ are vetoed. To reject the peaking backgrounds from the decays $D^{0}\to K_{S}^{0}(\rightarrow\pi^{0}\pi^{0})\pi^{+}\pi^{-}$, $D^{0}\to K_{S}^{0}(\rightarrow\pi^{+}\pi^{-})\pi^{0}\pi^{0}$ and $D^{0}\to K^{-}(\rightarrow\pi^{-}\pi^{0})\pi^{+}\pi^{0}$ in the signal process $D^0\to \pi^{+}\pi^{-}\pi^0\pi^{0}$, events with $0.4376<M(\pi^{0}\pi^{0})<0.5276~{\rm GeV}/c^{2}$ or $|M(\pi^{+}\pi^{-})-0.4976|<0.03~{\rm GeV}/c^{2}$ or $0.4677<M(\pi^{-}\pi^{0})<0.5067~{\rm GeV}/c^{2}$ are vetoed. Since the interference between the decays $D^{0}\to \pi^{0}\eta(\rightarrow\pi^{+}\pi^{-}\pi^{0})$ and $D^{0}\to \pi^{+}\pi^{-}\pi^{0}\pi^{0}$ is negligible, events with $M(\pi^{+}\pi^{-}\pi^{0})<0.57~{\rm GeV}/c^{2}$ are also vetoed in the signal process $D^0\to \pi^{+}\pi^{-}\pi^0\pi^{0}$. 
Moreover, we reject events if the energy of the photon from $\pi^{0}$ decays is below 50 MeV, as this indicates an incorrect choice of a soft photon.

\section{BACKGROUND STUDY AND SIGNAL EXTRACTION}
\label{signalyield}
\begin{figure*}[htbp]
  \centering
\begin{overpic}[width=0.32\textwidth]{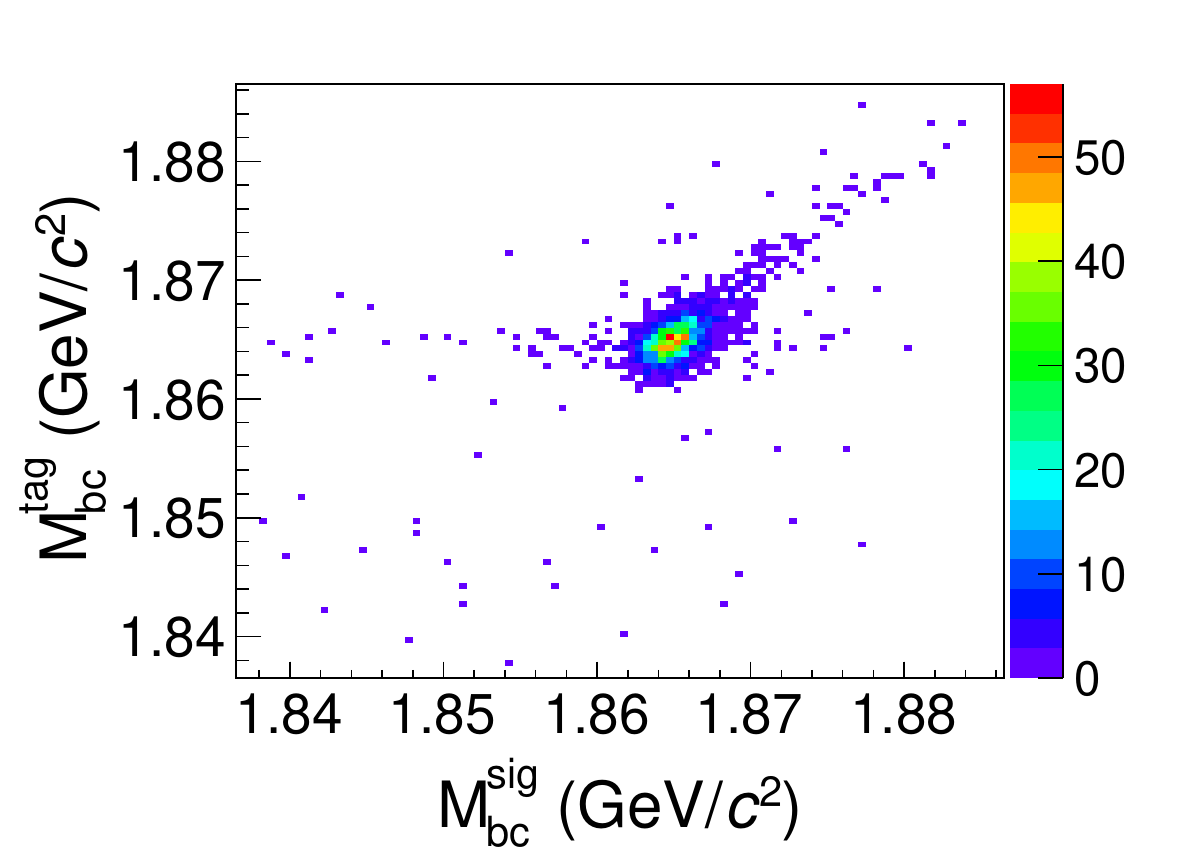}
   \end{overpic}
\begin{overpic}[width=0.32\textwidth]{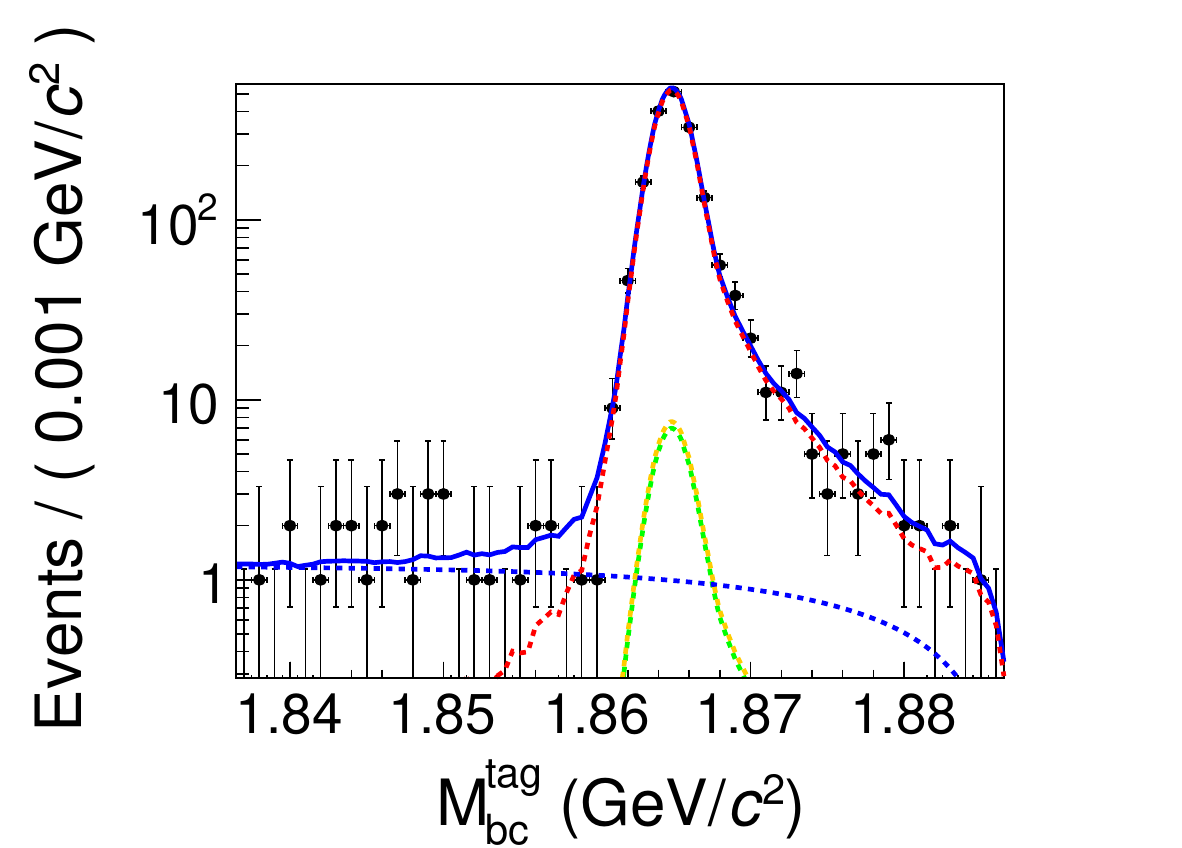}
   \end{overpic}
\begin{overpic}[width=0.32\textwidth]{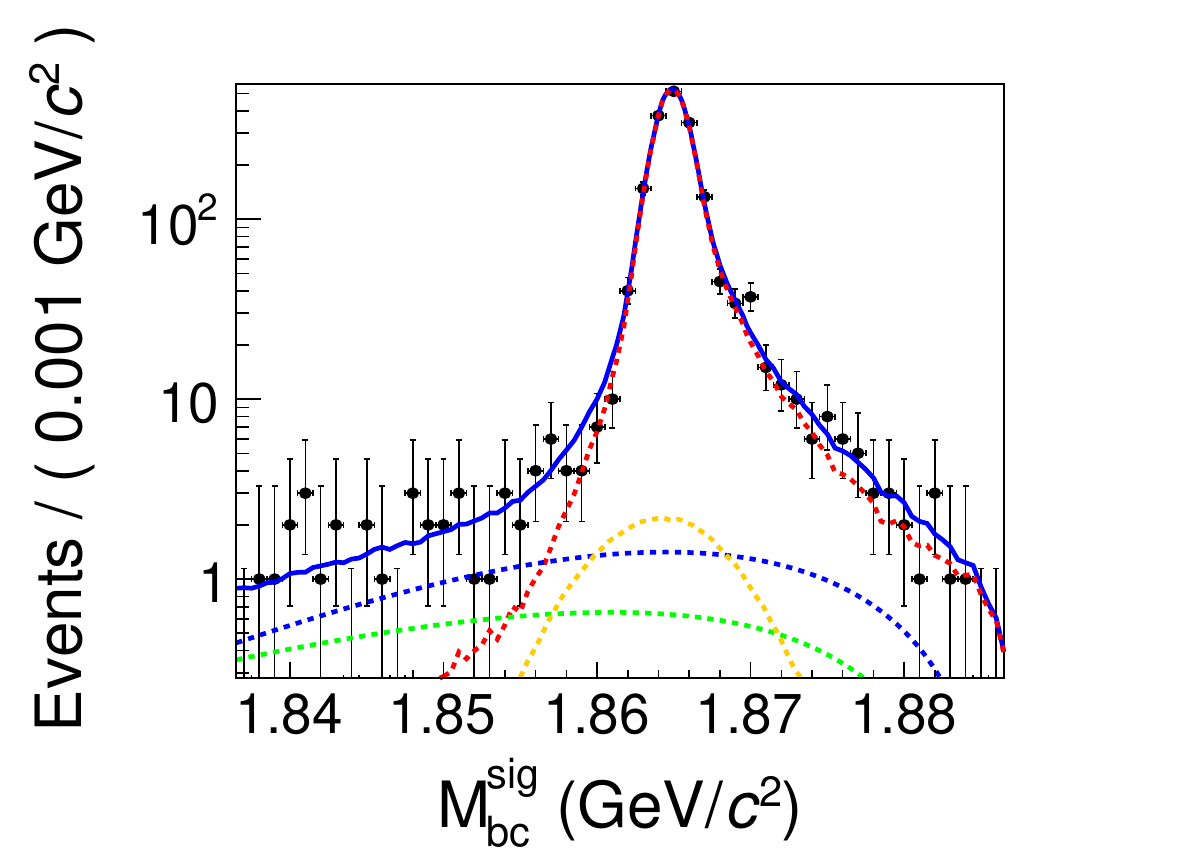}
   \end{overpic}

\begin{overpic}[width=0.32\textwidth]{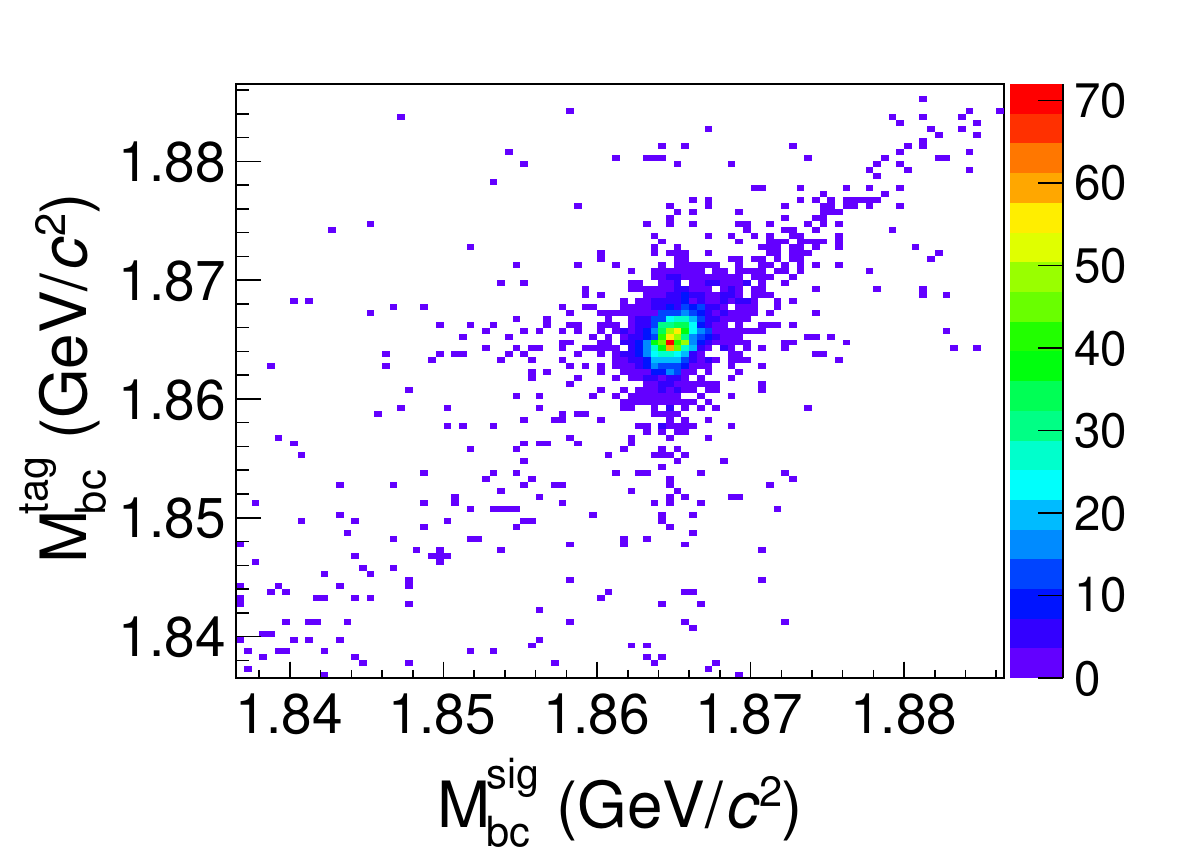}
   \end{overpic} 
 \begin{overpic}[width=0.32\textwidth]{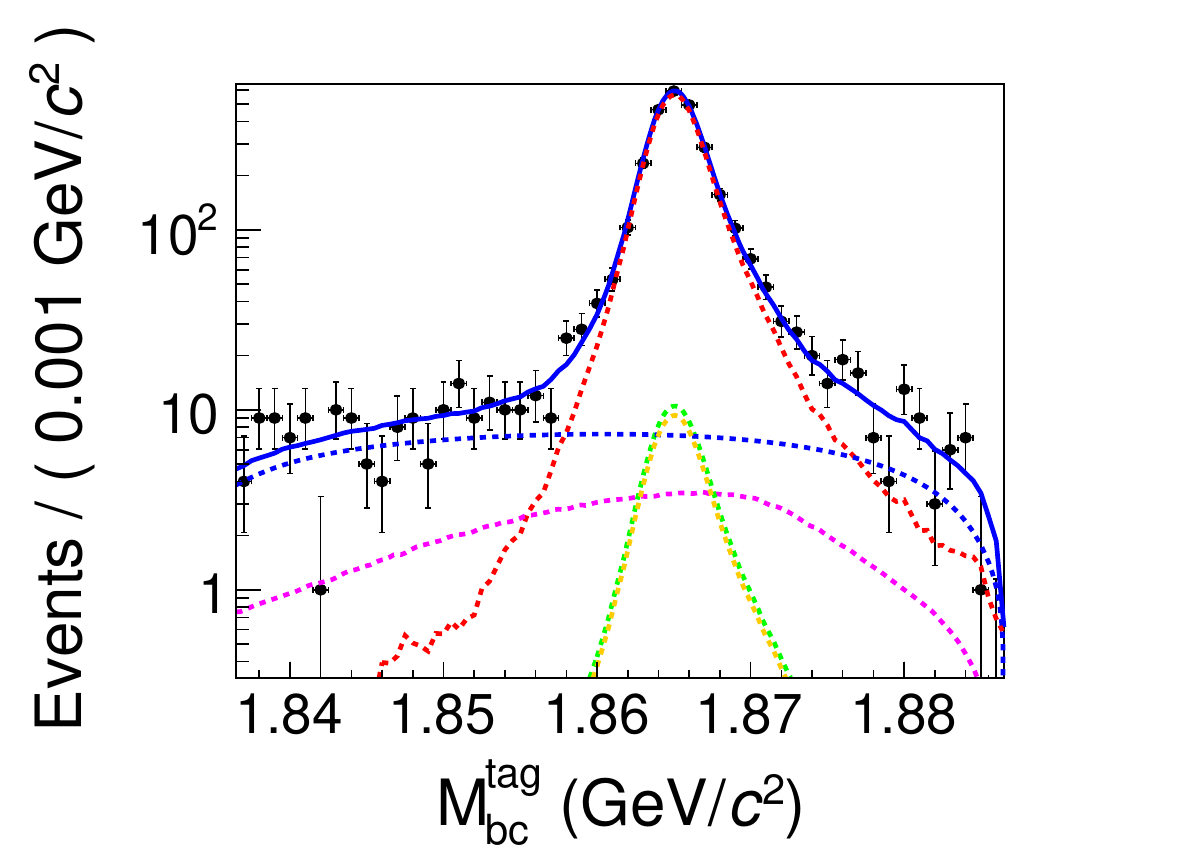}
   \end{overpic}
\begin{overpic}[width=0.32\textwidth]{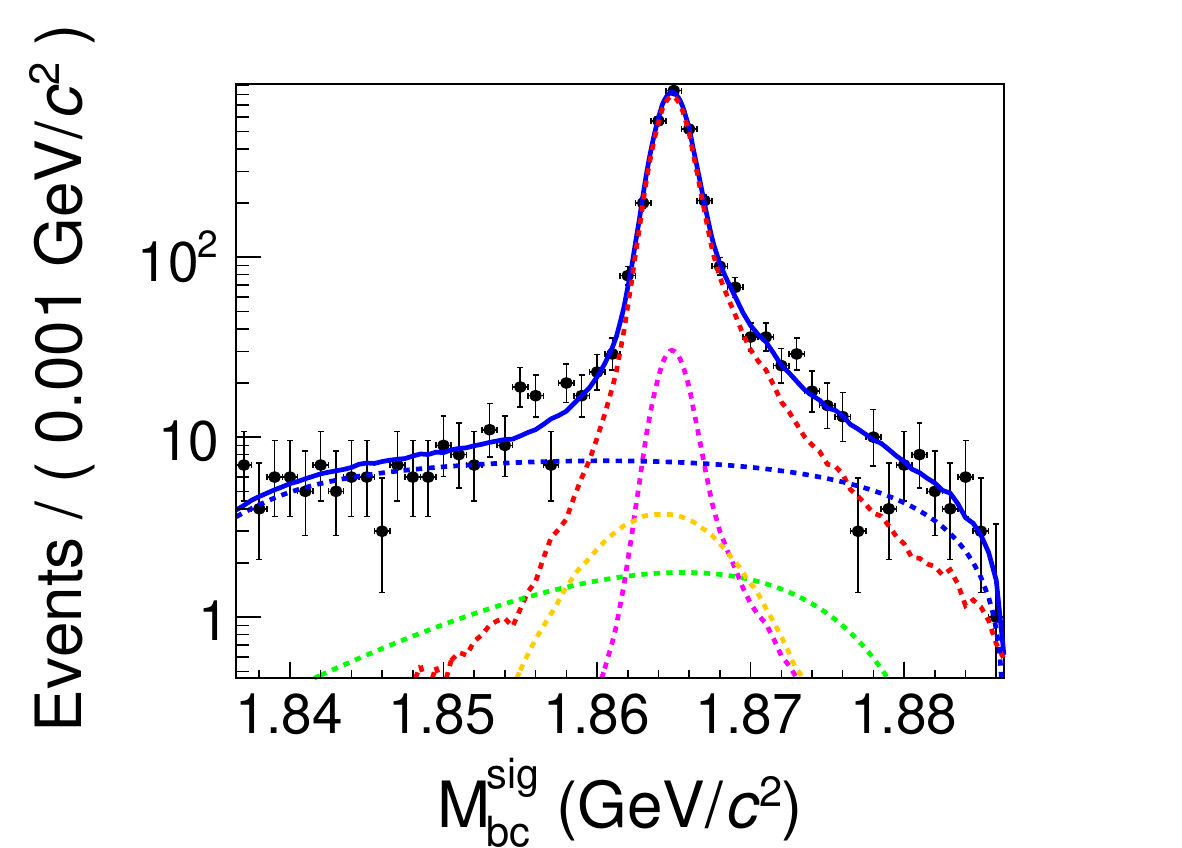}
   \end{overpic} 
  
   \begin{overpic}[width=0.32\textwidth]{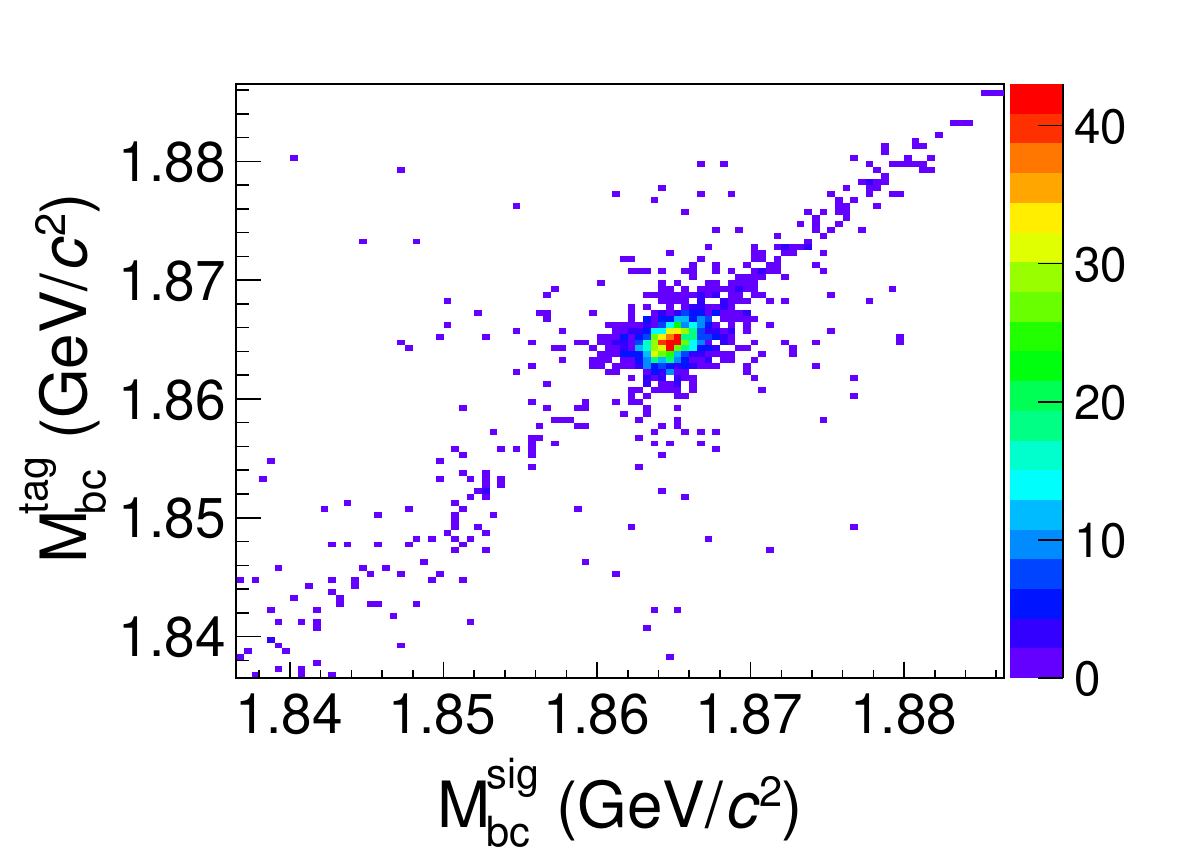}
   \end{overpic}   
   \begin{overpic}[width=0.32\textwidth]{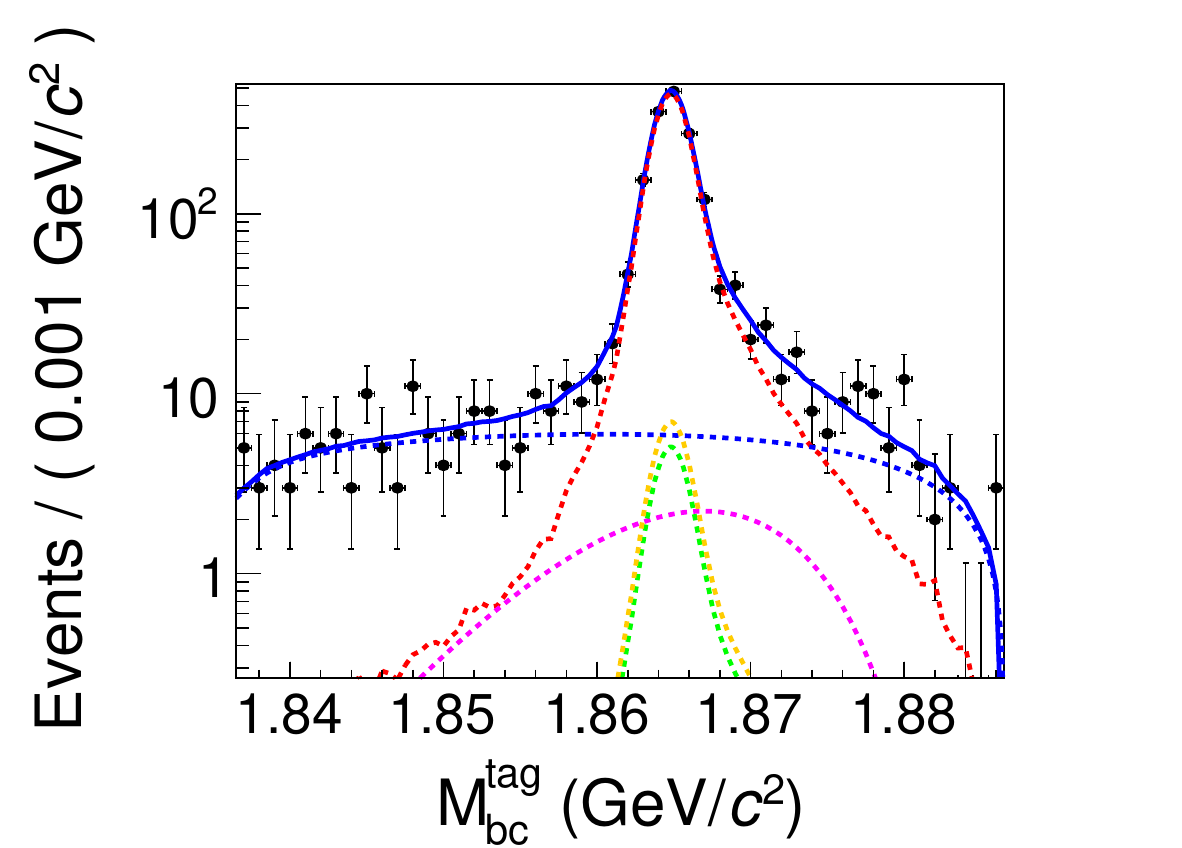}
   \end{overpic}  
   \begin{overpic}[width=0.32\textwidth]{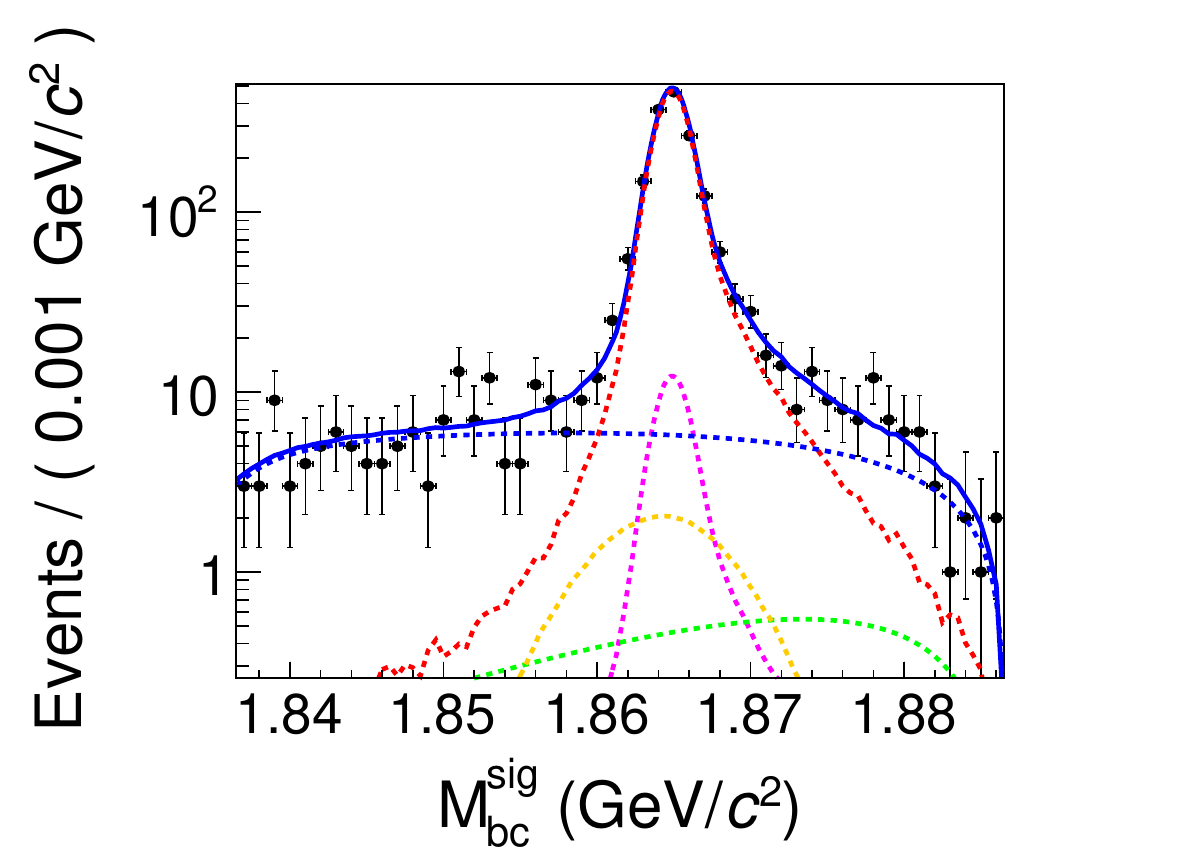}
   \end{overpic}
  \caption{The 2D distributions (left) of $M_{\rm bc}^{\rm tag}$ versus $M_{\rm bc}^{\rm sig}$ and the projections on $M_{\rm bc}^{\rm tag}$ (middle) and $M_{\rm bc}^{\rm sig}$ (right) of the 2D fits on the DT events tagged by $\bar{D}^{0}\to K^{+}\pi^-$ (first row), $\bar{D}^{0}\to K^{+}\pi^-\pi^0$ (second row), and $\bar{D}^{0}\to K^{+}\pi^-\pi^+\pi^-$ (third row) in the $D^0\to\pi^+\pi^-\pi^+\pi^-$ decay. 
The dots with error bars are data. The blue solid curves are the total fit results, and the red dashed curves show the signal. The green, blue, orange, and magenta dashed curves are BKGI,  BKGIII, BKGVI, and the combination of  BKGII and BKGV, respectively.}
  \label{fig:DT_C}
\end{figure*}

\begin{figure*}[htbp]
  \centering
  \begin{overpic}[width=0.32\textwidth]{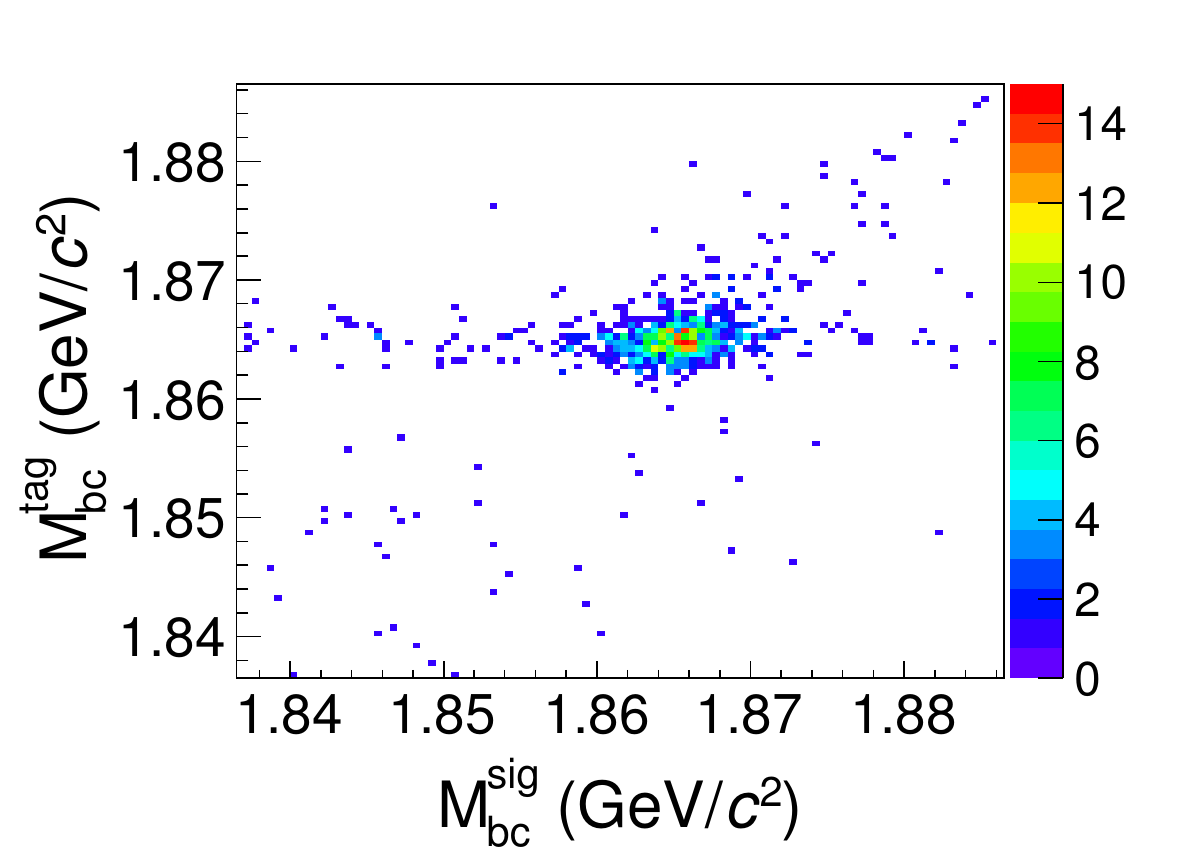}
   \end{overpic}
\begin{overpic}[width=0.32\textwidth]{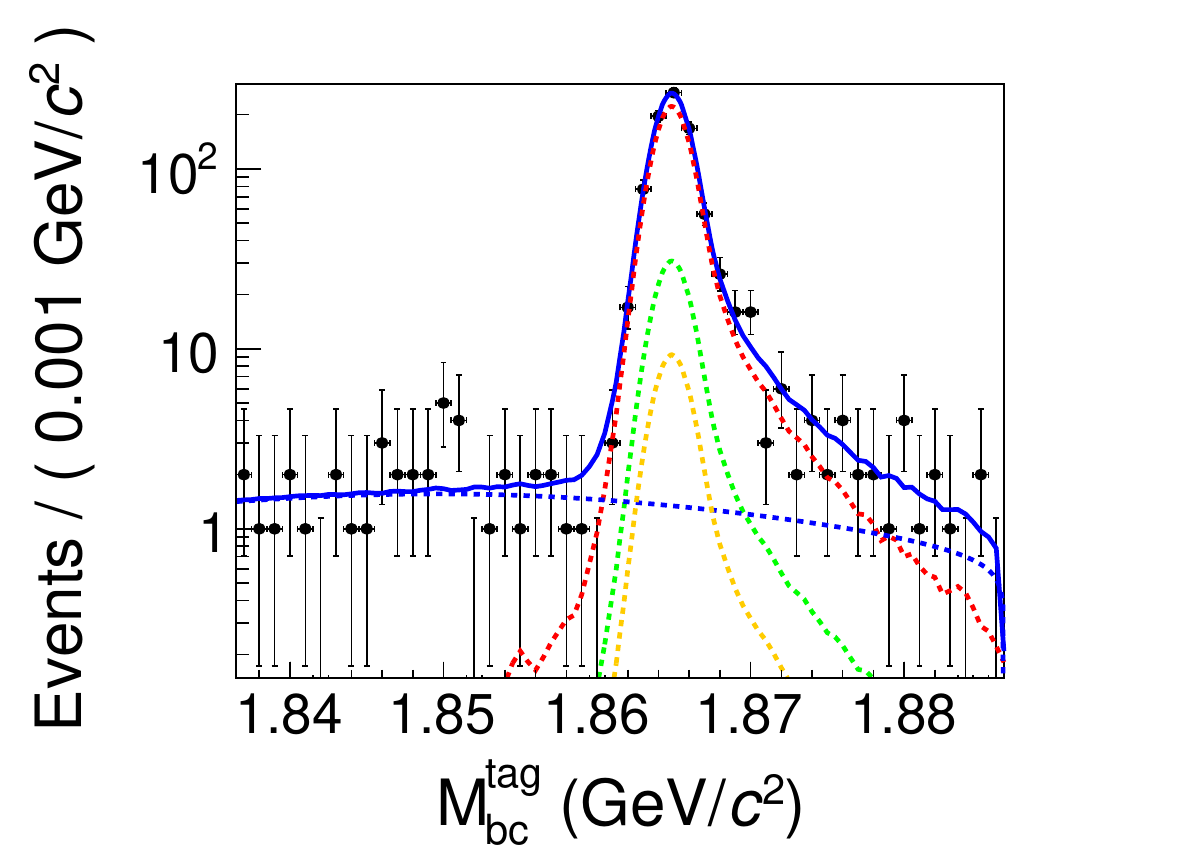}
   \end{overpic}
\begin{overpic}[width=0.32\textwidth]{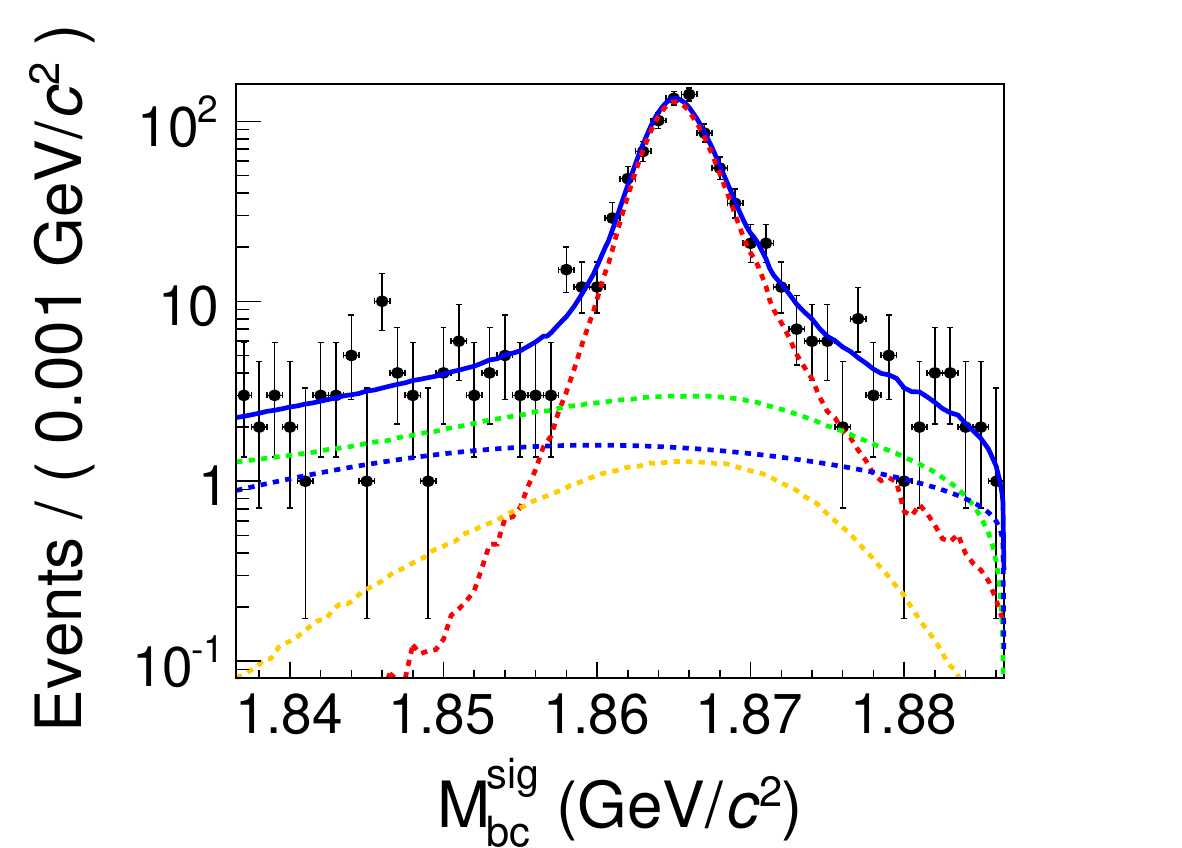}
   \end{overpic}

\begin{overpic}[width=0.32\textwidth]{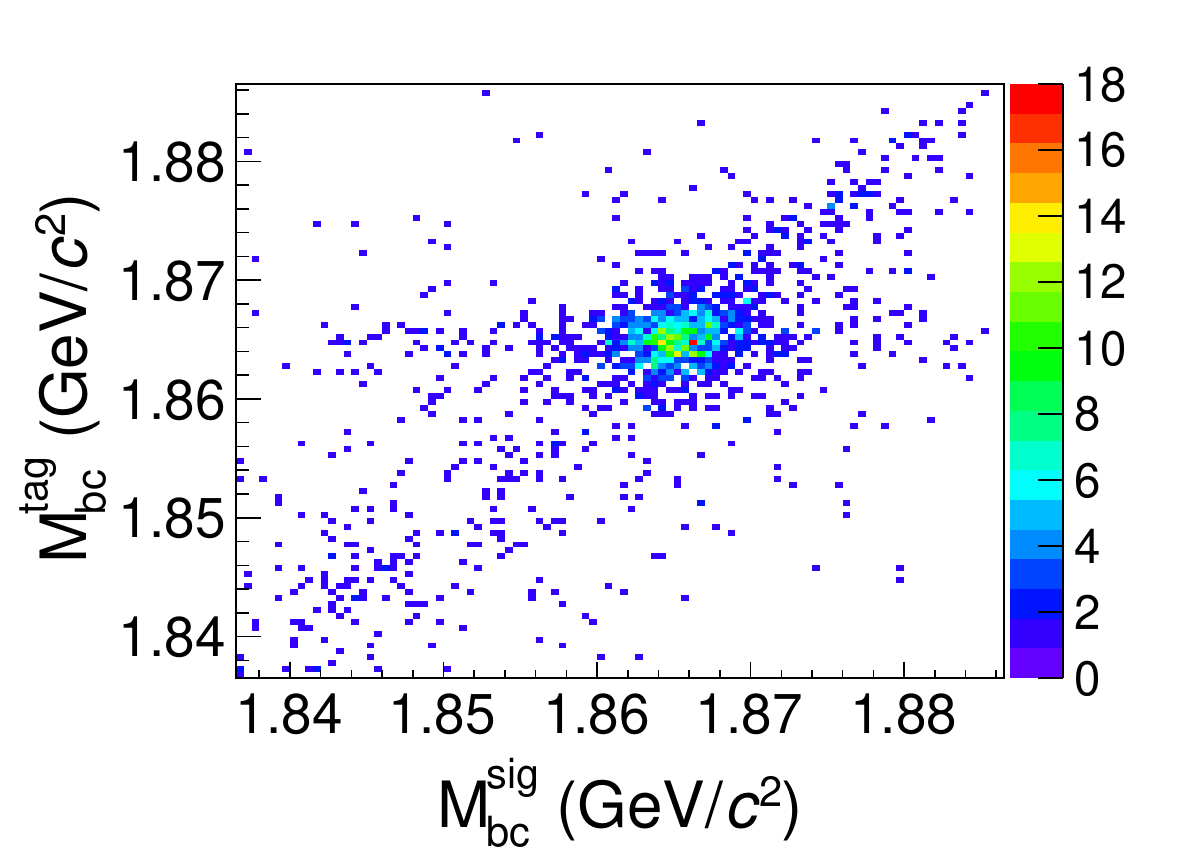}
   \end{overpic} 
 \begin{overpic}[width=0.32\textwidth]{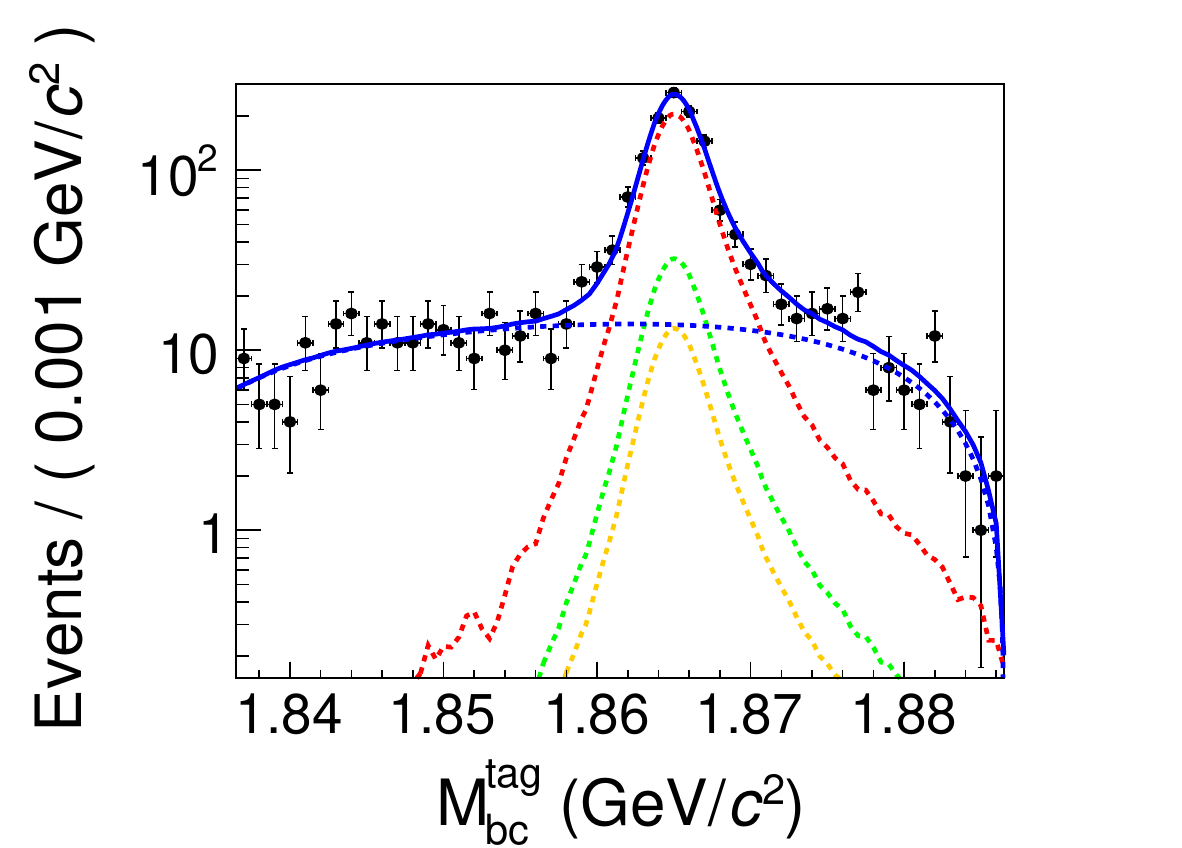}
   \end{overpic}
\begin{overpic}[width=0.32\textwidth]{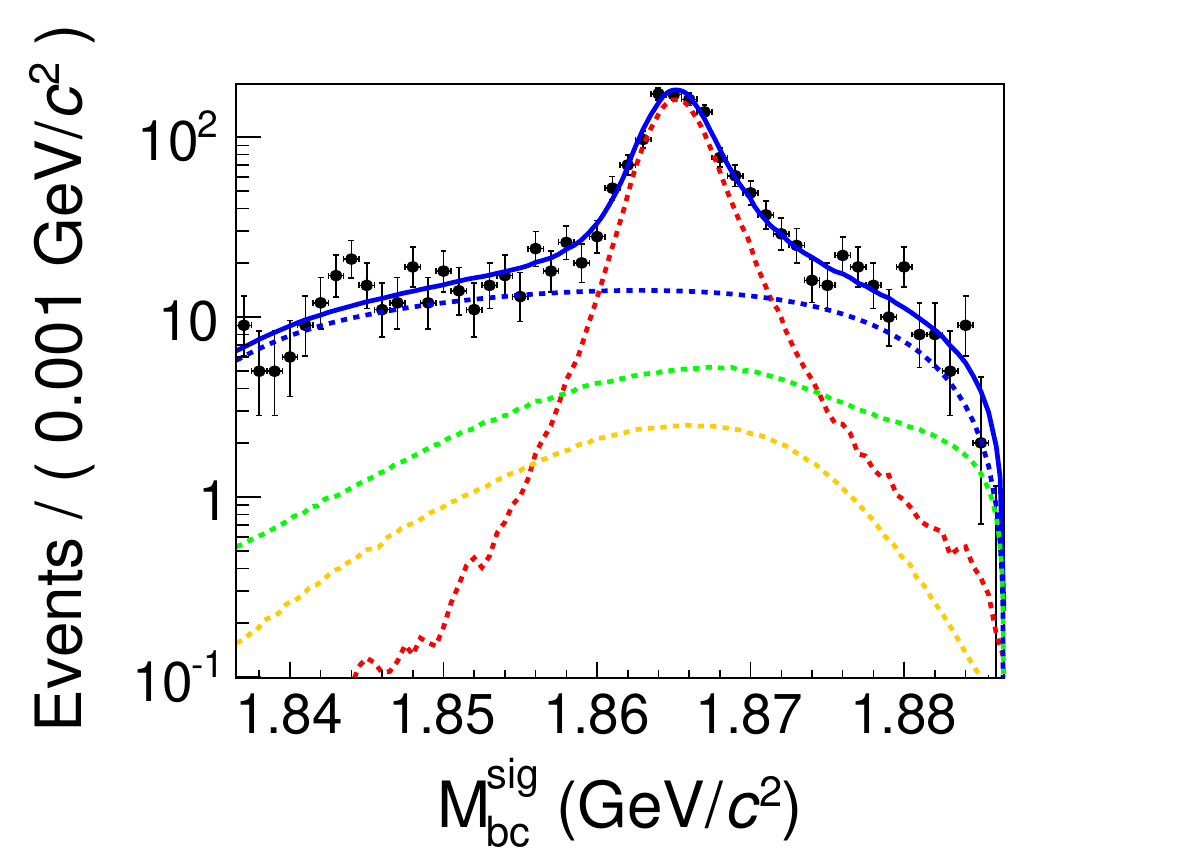}
   \end{overpic} 
  
   \begin{overpic}[width=0.32\textwidth]{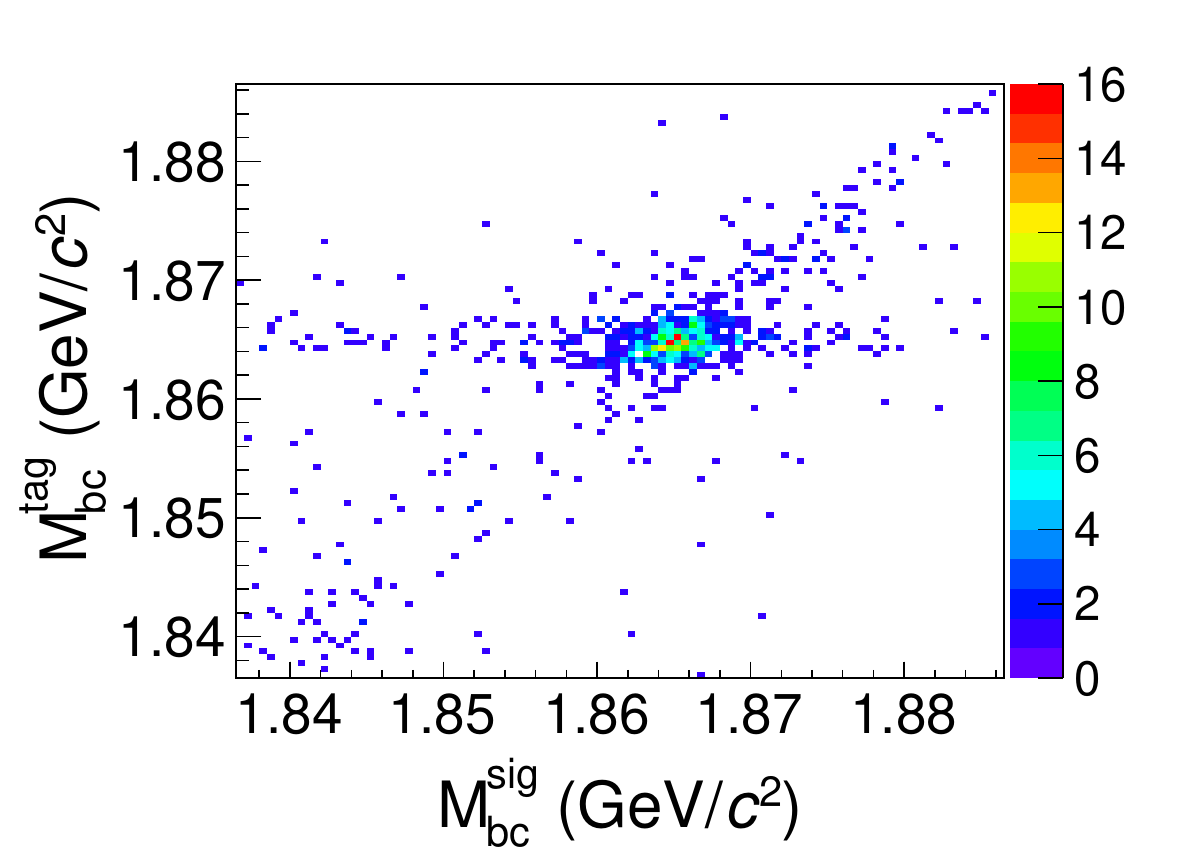}
   \end{overpic}   
   \begin{overpic}[width=0.32\textwidth]{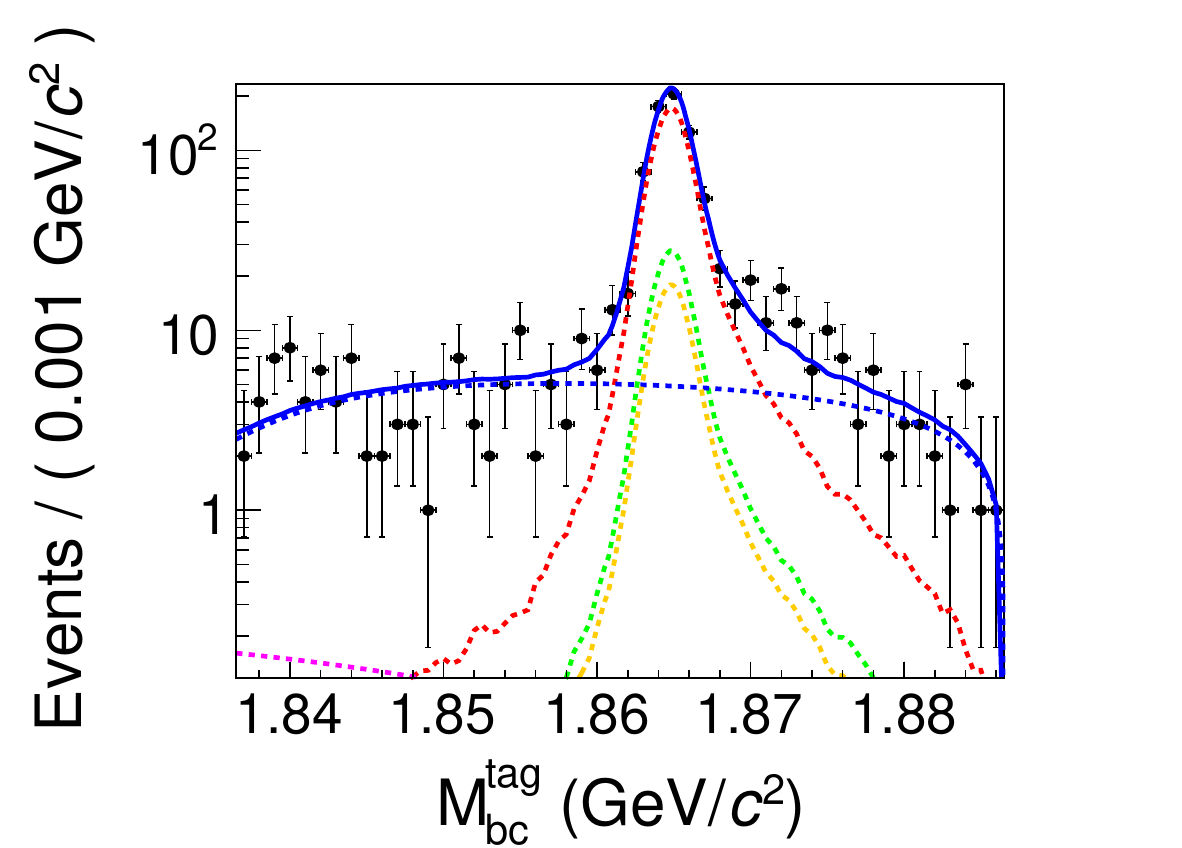}
   \end{overpic}  
   \begin{overpic}[width=0.32\textwidth]{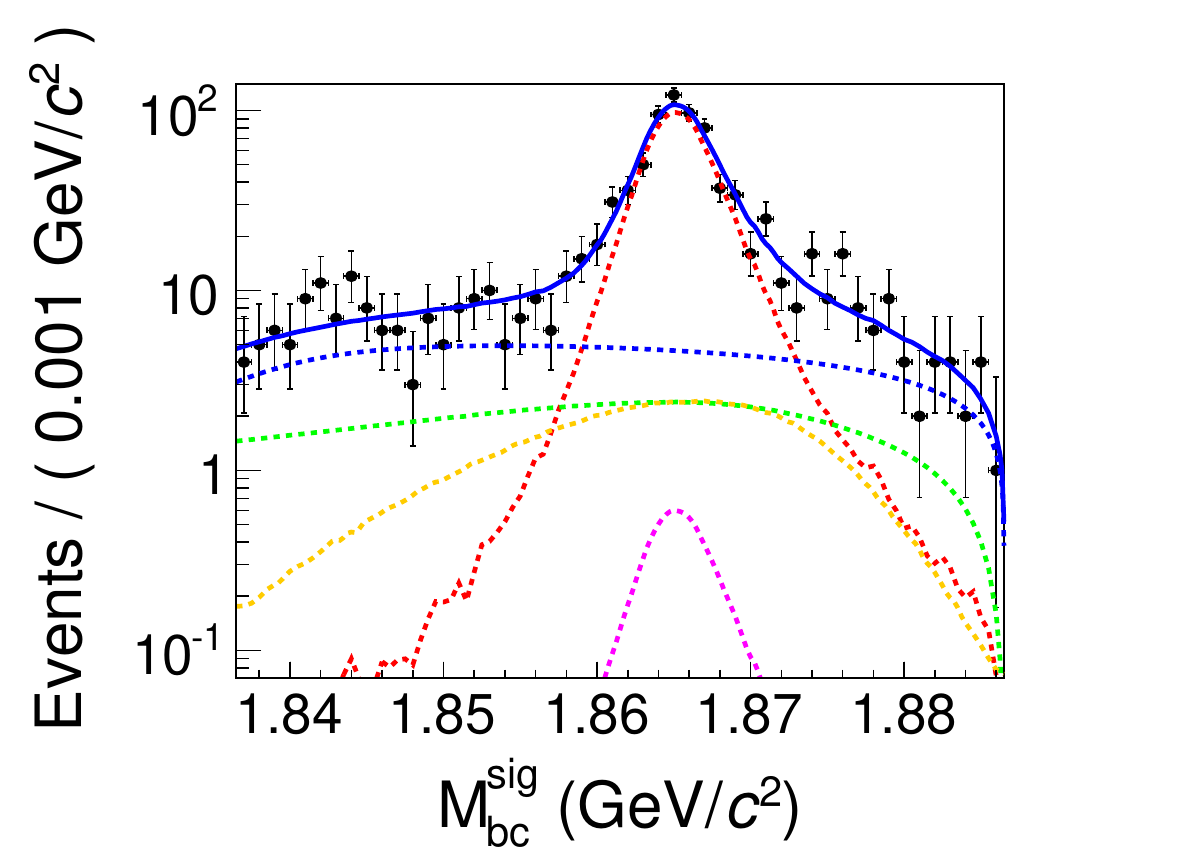}
   \end{overpic}
 \caption{The 2D distributions (left) of $M_{\rm bc}^{\rm tag}$ versus $M_{\rm bc}^{\rm sig}$ and the projections on $M_{\rm bc}^{\rm tag}$ (middle) and $M_{\rm bc}^{\rm sig}$ (right) of the 2D fits on the DT events tagged by $\bar{D}^{0}\to K^{+}\pi^-$ (first row), $\bar{D}^{0}\to K^{+}\pi^-\pi^0$ (second row) and $\bar{D}^{0}\to K^{+}\pi^-\pi^+\pi^-$ (third row) in the $D^0\to\pi^+\pi^-\pi^0\pi^0$ decay.
The dots with error bars are data. The blue solid curves are the total fit results, and the red dashed curves show the signal. The green dashed curves are the combination of BKGI and BKGIV, the magenta dashed curves are the combination of BKGII and BKGV,  the blue and orange dashed curves are BKGIII and BKGVI, respectively.}
  \label{fig:DT_N}
\end{figure*}

With all the above selection criteria, the two-dimensional (2D) distributions of $M_{\rm bc}^{\rm sig}$ versus  $M_{\rm bc}^{\rm tag}$ of the survived events are shown in the left columns of Figs.~\ref{fig:DT_C} and \ref{fig:DT_N}, where $M_{\rm bc}^{\rm sig}$ and $M_{\rm bc}^{\rm tag}$ are for the signal and tag sides, respectively.
Typically, the signal events are accumulated around the intersection of $M_{\rm bc}^{\rm sig}$ and $M_{\rm bc}^{\rm tag}$ at the $D^0$ nominal mass. The background, which arises from miscombinations both in the signal and tag sides (namely BKGI thereafter), is located at the diagonal band.
There are also the backgrounds from $e^+e^-\to D^0\bar{D}^0$ but with the wrong reconstruction of the signal side (namely BKGII) or tag side (namely BKGIII), distributed as the vertical and horizontal bands, respectively.
Detailed MC studies indicate that most of the backgrounds from  $e^+e^-\to D^0\bar{D}^0$ do not form individual peaks in the distribution of $M_{\rm bc}^{\rm sig}$ or $M_{\rm bc}^{\rm tag}$.
However, there are backgrounds from $e^+e^-\to D^0\bar{D}^0$ with $\bar{D}^0\to K^+\pi^-\pi^0$ (namely BKGIV) or $D^0\to \pi^+\pi^-\pi^0\pi^0$ (namely BKGV) with the wrong reconstruction of $\pi^0$, which are exactly the signal but with the wrong  reconstruction of $\pi^0$ and produce a relatively broad peak in the $M_{\rm bc}^{\rm tag}$ or $M_{\rm bc}^{\rm sig}$ distribution, respectively.
There are also the backgrounds from the decays $D^0\to K_S^0\pi^+\pi^-$, $D^0\to K_S^0\pi^0\pi^0$, $D^0\to K^-\pi^+\pi^0$ and $D^0\to K_S^0\eta^{\prime}$  (namely BKGVI), which have analogous final states as the signal with the specific decay modes of $K_S^0\to\pi^+\pi^-$, $K_S^0\to\pi^0\pi^0$, $K^-\to\pi^-\pi^0$ and $\eta^{\prime}\to\gamma\pi^+\pi^-$. These backgrounds are not directly distinguished from the signal in the $M_{\rm bc}^{\rm sig}$ and $M_{\rm bc}^{\rm tag}$ distributions and are estimated by MC simulation. The corresponding yields are summarized in Table~\ref{tab:Npeaks}.

\begin{table*}[htbp]
\caption{The estimated numbers of peaking background events. The uncertainties include the statistical uncertainties of the estimated background yields and the uncertainties of the quoted branching fractions of different background processes.}
\label{tab:Npeaks}
\begin{center}
\begin{tabular}{c|c|c|c|c|c|c}
\hline
\hline
Signal mode			&\multicolumn{3}{c|}{$\pi^{+}\pi^{-}\pi^{+}\pi^{-}$}  					&\multicolumn{3}{c}{$\pi^{+}\pi^{-}\pi^{0}\pi^{0}$}\\
\hline
Tag mode				& $K^{+}\pi^{-}$	&$K^{+}\pi^{-}\pi^{0}$	&$K^{+}\pi^{-}\pi^{+}\pi^{-}$			& $K^{+}\pi^{-}$	&$K^{+}\pi^{-}\pi^{0}$	&$K^{+}\pi^{-}\pi^{+}\pi^{-}$\\
\hline
$N(K_{S}^{0}\pi^{+}\pi^{-})$		&$24.7~\pm~2.3$	&$42.4~\pm~4.3$	&$22.9~\pm~2.6$			&$12.6~\pm~3.1$	&$27.7~\pm~6.2$	&$29.2~\pm~6.2$		\\
$N(K_{S}^{0}\pi^{0}\pi^{0})$		&	-		&	-		&	-				&$1.7~\pm~0.4$	&$4.3~\pm~0.9$	&$2.4~\pm~0.6$		\\
$N(K^{-}\pi^{+}\pi^{0})$		&	-		&	-		&	-				&$12.6~\pm~1.2$	&$23.1~\pm~2.2$	&$24.0~\pm~1.8$		\\
$N(K_{S}^{0}\eta')$			&	-		&	-		&	-				&$3.0~\pm~1.2$	&$3.9~\pm~1.7$	&$2.8~\pm~1.2$		\\
\hline
\hline
\end{tabular}
\end{center}
\end{table*}

\begin{figure*}[htbp]
  \centering
  \begin{overpic}[width=0.32\textwidth]{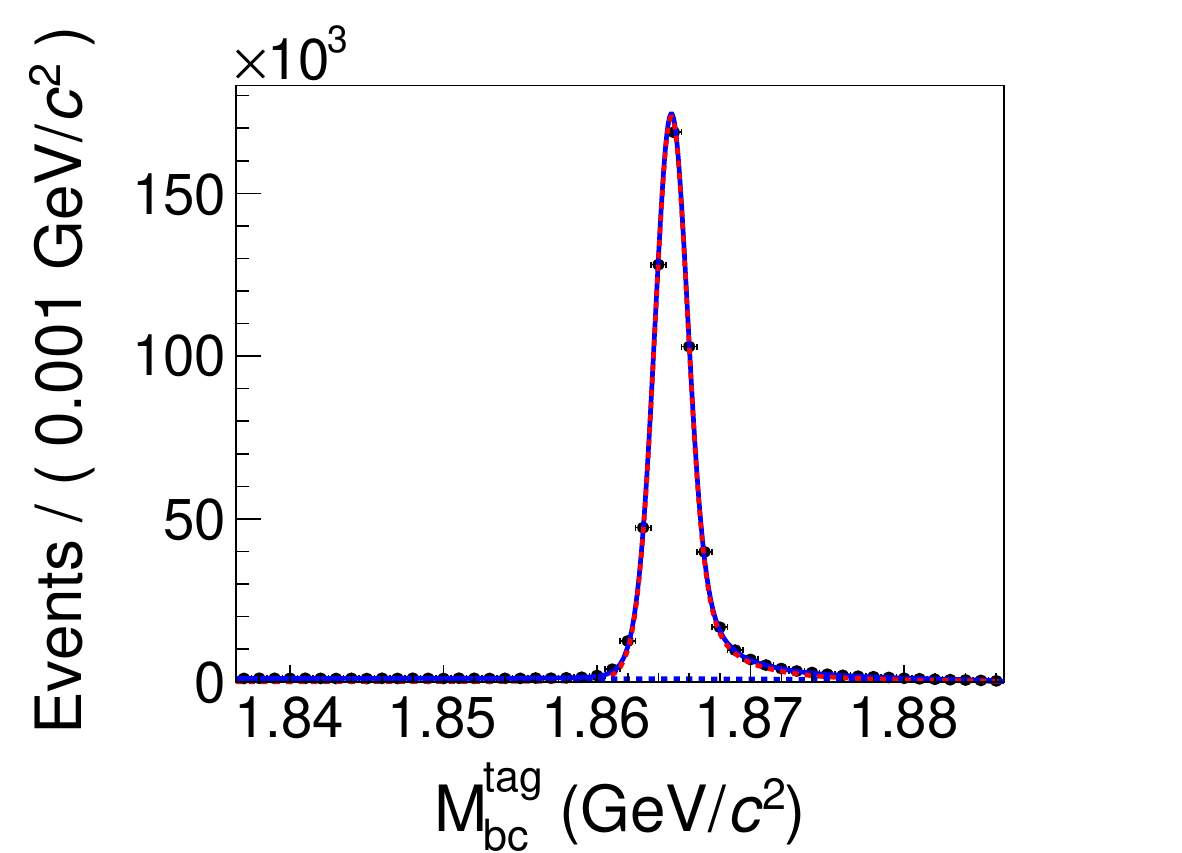}
  \end{overpic}
    \begin{overpic}[width=0.32\textwidth]{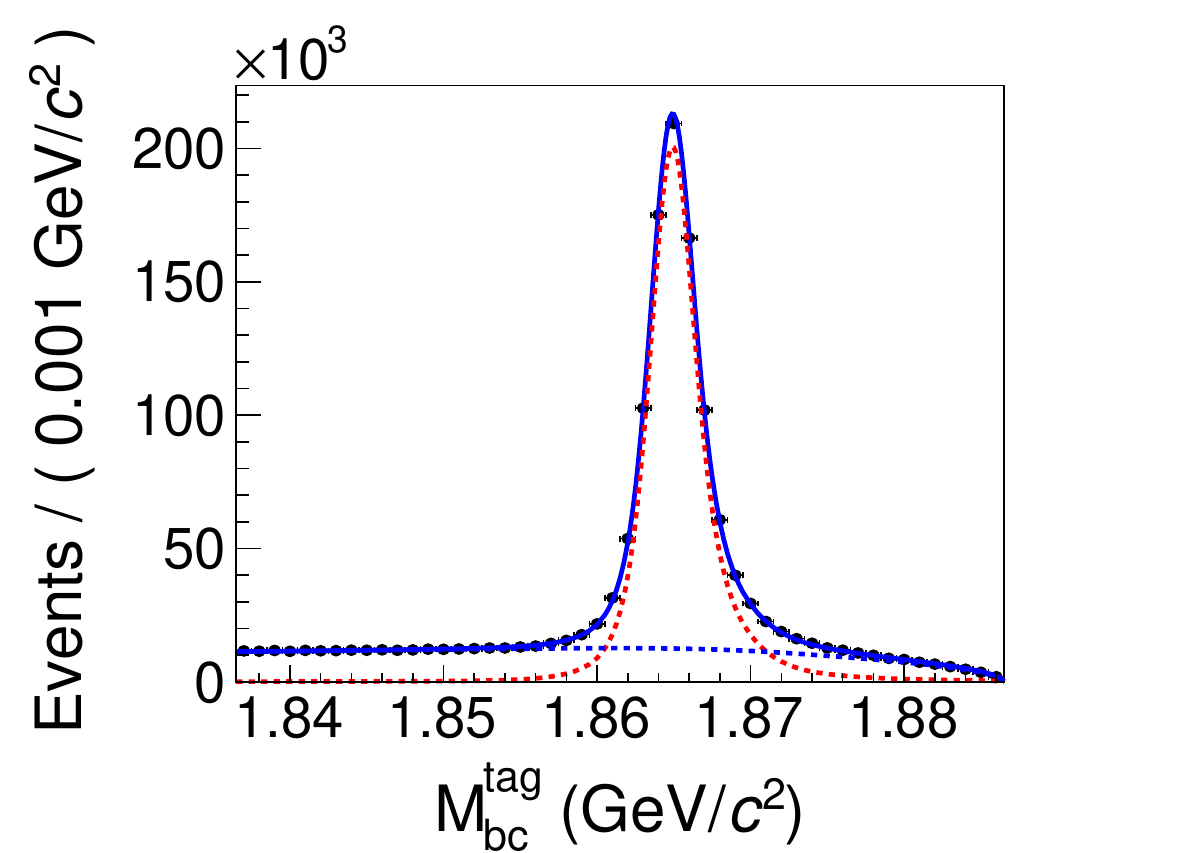}
  \end{overpic}
    \begin{overpic}[width=0.32\textwidth]{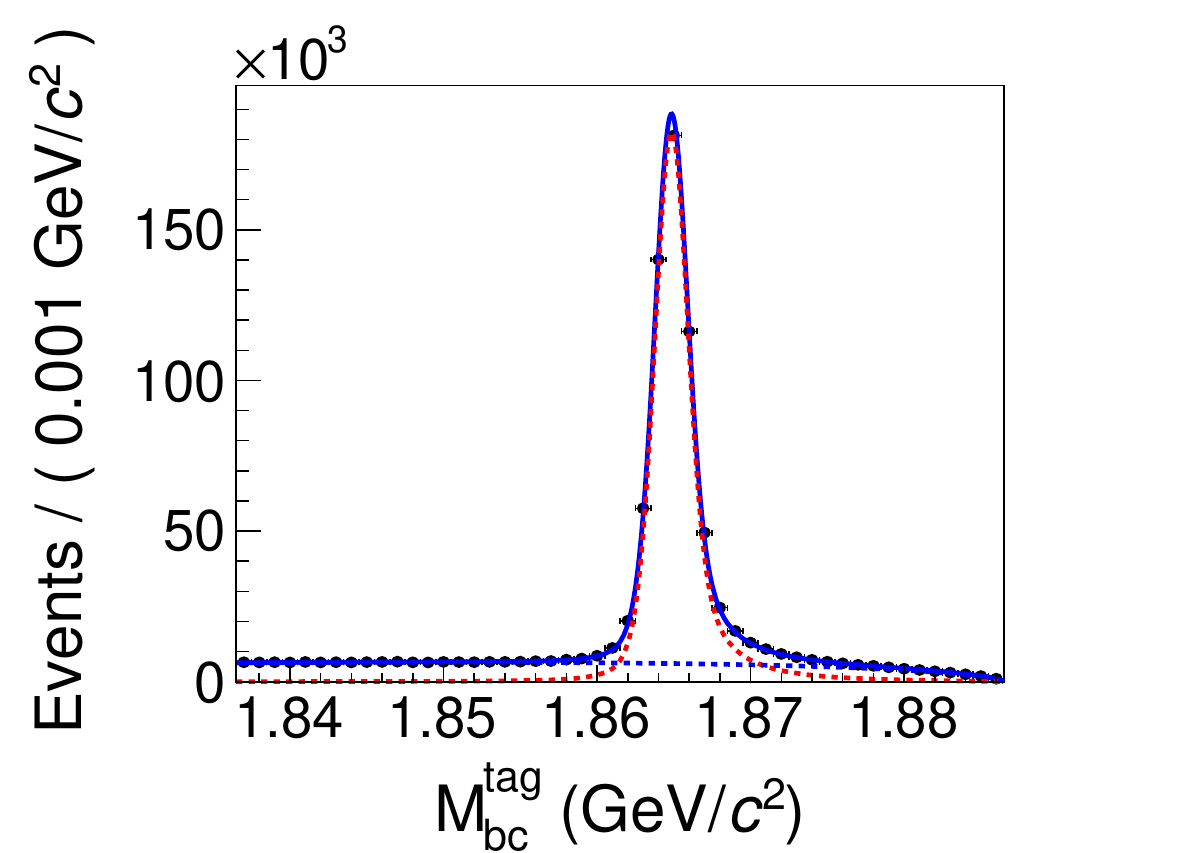}
  \end{overpic}
  \caption{The $M_{\rm bc}^{\rm tag}$ distributions of the ST candidates for $\bar{D}^{0}\to K^+\pi^-$ (left), $\bar{D}^{0}\to K^+\pi^-\pi^0$ (middle) and $\bar{D}^{0}\to K^+\pi^-\pi^+\pi^-$ (right). The dots with error bars are data, the blue solid curves are the total fit results, and the red and blue dashed curves are the signal and background, respectively.}
  \label{fig:ST}
\end{figure*}

To extract the ST and DT yields, unbinned maximum likelihood fits are performed on the $M_{\rm bc}^{\rm tag}$ distribution of the remaining ST candidates and the 2D distributions of $M_{\rm bc}^{\rm sig}$ versus  $M_{\rm bc}^{\rm tag}$ of the remaining DT events in individual tag modes.
In the fit to the $M_{\rm bc}^{\rm tag}$ distribution, the signal shape is described by the MC simulated shape of the truth-matched ST events, and the background shape is described by an ARGUS function with the cut-off parameter fixed at $1.8865$~GeV/$c^2$~\cite{ARGUS:1990hfq}. 
For the $\bar{D}^0\to K^+\pi^-\pi^0$ tag mode, an additional peaking background is considered which represents the wrong reconstruction of $\pi^0$ as discussed above. This peaking background shape is described by the MC-simulated shape of the truth-matched ST events convolved with a bifurcated Gaussian function with fixed parameters that are obtained from the fit to the corresponding simulated background events. The results of the fits to the $M_{\rm bc}^{\rm tag}$ distributions are shown in Fig.~\ref{fig:ST}. 

In the fit to the 2D distribution of $M_{\rm bc}^{\rm sig}$ (labelled as $y_1$) versus  $M_{\rm bc}^{\rm tag}$ (labelled as $y_2$), the signal is described by   
\begin{eqnarray}
S(y_1,y_2) \otimes G(y_1;\mu_{y_1},\sigma_{y_1}) \otimes G(y_2;\mu_{y_2},\sigma_{y_2}), \nonumber
 \end{eqnarray}
where $S(y_1,y_2)$ is the signal MC shape derived from the truth-matched events by using a smoothed 2D histogram, and $G(\mu,\sigma)$ is the Gaussian function describing the resolution difference between data and MC simulation by fixing the parameters $\mu_{y_{1(2)}}$ and $\sigma_{y_{1(2)}}$ to those from the fits on the one-dimensional $M_{\rm bc}$ distributions.

The BKGI is described by 
\begin{linenomath*}
\begin{align}
&T(y_1-y_2; \mu, \sigma(y_1+y_2), n) \times A(y_1; m_{y_1}, z'_{y_1}, \rho'_{y_1}) \nonumber \\ &\times A(y_2; m_{y_2}, z'_{y_2}, \rho'_{y_2}), \nonumber
 \end{align}
\end{linenomath*}
where $T$ is the student's function defined as $T(y; \mu, \sigma, n) = \frac{\Gamma(n/2+0.5)}{\sigma\sqrt{n\pi}\Gamma(n/2)}[1+\frac{1}{n}(\frac{y-\mu}{\sigma})^{2}]^{-\frac{n+1}{2}}$, $\sigma(y_1+y_2) = \sigma_0 + \sigma_{1}(y_1+y_2-m_{y_1}-m_{y_2})$ and  $A$ is the ARGUS function defined as $A(y;m,z,\rho) = y(1-\frac{y^2}{m^2})^{\rho}e^{z(1-\frac{y^2}{m^2})}$. In the fit, only the cut-off parameter $m_{y_{1(2)}}$ is fixed to $1.8865~{\rm GeV}/c^{2}$, the other parameters are free to vary. 

The BKGII and BKGIII are described by 
\begin{linenomath*}
 \begin{align}
 &[S(y_{2(1)}) \otimes G(y_{2(1)};\mu_{y_{2(1)}},\sigma_{y_{2(1)}})]  \nonumber\\ &\times A(y_{1(2)}; m_{y_{1(2)}}, z_{y_{1(2)}}, \rho_{y_{1(2)}}),\nonumber
 \end{align} 
 \end{linenomath*}
where $A$ is the ARGUS function as described above, $S(y_{2(1)})$ is the projection of $S(y_1,y_2)$ on $y_{2(1)}$ and $G$ is the same Gaussian function as described above. In the fit, the parameter $m_{y_{1(2)}}$ is fixed to $1.8865~{\rm GeV}/c^{2}$, $\rho_{y_{1(2)}}$ is fixed to the values obtained from the fits to the inclusive MC sample, and $z_{y_{1(2)}}$ is a free parameter.

The BKGIV is described by 
\begin{eqnarray} 
B_1(y_1,y_2) \otimes G(y_1;\mu_{y_1},\sigma_{y_1}) \otimes G(y_2;\mu_{y_2},\sigma_{y_2}), \nonumber
\end{eqnarray}
where $B_1(y_1,y_2)$ is the truth-matched signal MC shape $S(y_1,y_2)$ convolved with a bifurcated Gaussian function in $y_2$ and $G$ is the same Gaussian function as described above. In the fit, the parameters of the bifurcated Gaussian function are fixed to the values obtained from the fit to the $M_{\rm bc}^{\rm tag}$ distribution of the corresponding simulated background events.

The BKGV in $D^0\to\pi^+\pi^-\pi^0\pi^0$ is described by
\begin{eqnarray}  
B_2(y_1,y_2) \otimes G(y_1;\mu_{y_1},\sigma_{y_1}) \otimes G(y_2;\mu_{y_2},\sigma_{y_2}),  \nonumber
 \end{eqnarray}
where $B_2(y_1,y_2)$ is the truth-matched signal MC shape $S(y_1,y_2)$ convolved with a bifurcated Gaussian function in $y_1$ and $G$ is the same Gaussian function as described above. In the fit, the parameters of the bifurcated Gaussian function are fixed to the values obtained from the fit to the $M_{\rm bc}^{\rm sig}$ distribution of the corresponding simulated background events. 

The BKGVI is described by
\begin{eqnarray} 
B_3(y_1,y_2) \otimes G(y_1;\mu_{y_1},\sigma_{y_1}) \otimes G(y_2;\mu_{y_2},\sigma_{y_2}),  \nonumber
\end{eqnarray}
where $B_3(y_1,y_2)$ is the truth-matched MC shape $S(y_1,y_2)$ convolved with a bifurcated Gaussian function in $y_1$ and $G$ are the Gaussian functions. In the fit, the parameters of the bifurcated Gaussian functions are fixed to the value obtained from the fit to the $M_{\rm bc}^{\rm sig}$ distribution of the corresponding simulated background events from the inclusive MC sample, and the yields are fixed to those summarized in Table~\ref{tab:Npeaks}.

The projections of the 2D fits on $M_{\rm bc}^{\rm tag}$ and $M_{\rm bc}^{\rm sig}$ are shown in Figs.~\ref{fig:DT_C} and \ref{fig:DT_N}, respectively. The obtained ST and DT yields are summarized in Table~\ref{tab:Neff}.

\begin{table*}[htbp]
\caption{The ST and DT yields, efficiencies, and quantum correlated correction factors for three tag modes.}
\label{tab:Neff}
\begin{center}
\begin{tabular}{c|c|c|c}
\hline
\hline
Tag mode									&$\bar{D}^{0}\to K^+\pi^-$		&$\bar{D}^{0}\to K^+\pi^-\pi^0$		&$\bar{D}^{0}\to K^+\pi^-\pi^+\pi^-$  	\\
\hline
$N^{\rm ST}$									&	$549586~\pm~778$		&	$914531~\pm~1321$			&	$600316~\pm~856$			\\
$\epsilon^{\rm ST}$								&$0.6762~\pm~0.0004$		&	$0.2953~\pm~0.0001$		&	$0.3365~\pm~0.0002$			\\
\hline
$N^{\rm DT}_{\pi^+\pi^-\pi^+\pi^-}$					&	$1719~\pm~42$		&	$2560~\pm~56$			&	$1520~\pm~43$			\\
$\epsilon^{\rm DT}_{\pi^+\pi^-\pi^+\pi^-}$				&$0.2792~\pm~0.0006$		&	$0.1187~\pm~0.0002$		&	$0.1221~\pm~0.0003$			\\
\hline
$N^{\rm DT}_{\pi^+\pi^-\pi^0\pi^0}$ (non-$\eta$)			&	$721~\pm~32$			&	$917~\pm~39$				&	$562~\pm~29$			\\
$\epsilon^{\rm DT}_{\pi^+\pi^-\pi^0\pi^0}$ (non-$\eta$)	&$0.0818~\pm~0.0003$		&	$0.0319~\pm~0.0001$		&	$0.0334~\pm~0.0001$			\\
\hline	
$\frac{2rR\rm{cos}\delta}{1+r^2}$				&$-0.114~\pm~0.004$~\cite{HFLAV:2019otj}		&	$-0.067~\pm~0.005$~\cite{BESIII:2021eud}		&	$-0.046^{+0.013}_{-0.011}$~\cite{BESIII:2021eud}	\\					
\hline
\hline
\end{tabular}
\end{center}
\end{table*}

\section{AMPLITUDE FORMULA}
\label{amplitudeformula}
To improve the resolutions of kinematic variables, a one-constraint kinematic fit with the hypothesis of $D^0\to 4 \pi$ by constraining the $4\pi$ invariant mass to the known $D^0$ mass~\cite{ParticleDataGroup:2022pth} is performed on the candidate events, and the updated kinematic variables are used in the amplitude analysis. 
Using the GPUPWA framework~\cite{Berger:2010zza}, a joint amplitude analysis is performed on the candidate events of $D^0\to\pi^+\pi^-\pi^+\pi^-$ and $D^0\to\pi^+\pi^-\pi^0\pi^0$(non-$\eta$). The general amplitude of the $D^{0}\to f$ decay is given by
\begin{eqnarray}
   A_{f}(p) = \sum_{i} \Lambda_iU_i(p)~,\label{eq3}
\end{eqnarray}
where $U_i$ is the amplitude of the $i$-th intermediate process with a complex coupling factor $\Lambda_i$ and $p$ is a set of four-momenta of final states. For the $\bar{D}^{0}$ decay amplitude, assuming $CP$ conservation, the $CP$-conjugate PHSP $\bar{p}$ is defined by interchanging the charge of the final state and the reversal of three-momenta. Then, the amplitude of the $\bar{D}^{0}\to \bar{f}$ decay is given as
\begin{eqnarray}
   \bar{A}_{\bar{f}}(p)= \sum_{i}\Lambda_i\bar{U}_i(p)=\sum_{i}\Lambda_iU_i(\bar{p})~.\label{eq7}
\end{eqnarray}

Since the $D^{0}\bar{D}^0$ pair is produced in the $\psi(3770)$ decay, quantum correlation between $D^{0}$ and $\bar{D}^0$ needs to be considered. 
By ignoring the effects of  $CP$ violation and $D^{0}$-$\bar{D}^0$ mixing, the observed differential cross section $|M_f(p)|^{2}$ of $D^0\to f$ together with the tag mode $\bar{D}^0\to \bar{g}$ is 
 \begin{eqnarray}
\label{eq5}
|A_f(p)-r_{g}R_{g}e^{-i\delta_{g}}\bar{A}_f(p)|^2+r_{g}^2(1-R_{g}^2)|\bar{A}_f(p)|^2 ~,\label{eq19}
  \end{eqnarray}
where $r_{g}^2 = \frac{\int |\bar{A}_{g}|^2{\mathrm{d}\Phi_{g}}}{\int |{A}_{g}|^2{\mathrm{d}\Phi_{g}}}$ and $R_ge^{-i\delta_g} = \frac{ \int A_{g}^*\bar{A}_{g} {\mathrm{d}\Phi_{g}}}{\sqrt{\int |A_{g}|^2{\mathrm{d}\Phi_{g}}\int |\bar{A}_{g}|^2{\mathrm{d}\Phi_{g}}}}$. The values of $r$, $R$, $\delta$ for the three tag modes are summarized in Table~\ref{tab:rRdelta}. In practice, the second term in Eq.~\eqref{eq5} is ignored  due to the relatively small value of $r_{g}^2(1-R_{g}^2)$.

\begin{table*}[htbp]
\caption{The input values of $r$, $R$ and $\delta$ for the three tag modes.}
\label{tab:rRdelta}
\begin{center}
\begin{tabular}{l|c|c|c}
\hline
\hline
Mode		            &     	$K^-\pi^+$						           &  		$K^-\pi^+\pi^0$					&   $K^-\pi^+\pi^-\pi^+$ \\
\hline
$r(\%)$		       &$5.86~\pm~0.02$~\cite{HFLAV:2019otj}	              &  $4.41~\pm~0.11$~\cite{BESIII:2021eud} & $5.50~\pm~0.07$~\cite{BESIII:2021eud} \\
$R$		            & 1                                                                        &  $0.79~\pm~0.04$~\cite{BESIII:2021eud} & $0.44_{-0.10}^{+0.09}$~\cite{BESIII:2021eud}	  \\
$\delta(^{\circ})$  & $192.1^{+8.6}_{-10.2}$~\cite{HFLAV:2019otj} &  $196~\pm~11$~\cite{BESIII:2021eud}   & $161^{+28}_{-18}$~\cite{BESIII:2021eud}\\
\hline
\hline
\end{tabular}
\end{center}
\end{table*}

The amplitude $U_i$ is constructed with the spin factor,  Blatt-Weisskopf barrier factors~\cite{VonHippel:1972fg}, and propagators of resonances.
To construct $U_i$ of the four-body $D^0$ decay, the isobar model is applied in which the decay is factorized into subsequent two-body decay amplitudes~\cite{Mandelstam:1962ols,Herndon:1973yn,Brehm:1977yr}. The general amplitude of the $i$-th intermediate process in a four-body decay is given by
\begin{eqnarray}
   U_i(p)=S_i(p)B_{L_D}(p)P_{R_1}(p)B_{L_{R_1}}(p)P_{R_2}(p)B_{L_{R_2}}(p)~,\label{eq6}
\end{eqnarray}
where $S_i$ is the spin factor of the $i$-th decay amplitude, $B_{L_X}$ ($X=D$, $R_1$, $R_2$) are the Blatt-Weisskopf barrier factors for the $D$ meson and the resonances $R_1$ and $R_2$,  and $P_{R_1}$ and $P_{R_2}$ are the propagators of $R_{1}$ and $R_{2}$, respectively. The amplitude is constructed with the exchange symmetry for indistinguishable pions.

The spin factor is constructed with the covariant Zemach (Rarita-Schwinger) tensor formalism~\cite{Rarita:1941mf,Zemach:1965ycj,Chung:1997jn,Zou:2002ar} by combining pure-orbital-angular-momentum covariant tensors $\tilde{t}^{(L)}_{\mu_1\cdots \mu_l}$ and the momenta of parent particles together with Minkowski metric $g_{\mu\nu}$ and Levi-Civita symbol $\epsilon_{\mu\nu\lambda\sigma}$. For a process $a\to bc$, the covariant tensors $\tilde{t}^{(L)}_{\mu_1\cdots \mu_l}$ for the final states of pure orbital angular momentum $L$ are 
\begin{eqnarray}
   \tilde{t}^{(L)}_{\mu_1\cdots \mu_l} = (-1)^{L}P^{(L)}_{\mu_1\cdots \mu_L\mu'_1\cdots \mu'_L}(p_a)r^{\mu'_1}\cdots r^{\mu'_L}~,\label{eq8}
\end{eqnarray}
where $r=p_b-p_c$ and $P^{(L)}_{\mu_1\cdots \mu_L\mu'_1\cdots \mu'_L}(p_a)$ is the spin projection operator of the particle $a$,
 \begin{eqnarray}
 &P^{(0)}(p_a)~~~=&1~, \\ \label{eq9}
 &P^{(1)}_{\mu\mu'}(p_a)~~~=&-g_{\mu\mu'}+\frac{p_{a\mu}p_{a\mu'}}{p_a^2}~, \\ \label{eq10}
 &P^{(2)}_{\mu\nu\mu'\nu'}(p_a)=&\frac{1}{2}[P^{(1)}_{\mu\mu'}(p_a)P^{(1)}_{\nu\nu'}(p_a)+P^{(1)}_{\mu\nu'}(p_a)P^{(1)}_{\nu\mu'}(p_a)]  \nonumber\\
&&-\frac{1}{3}P^{(1)}_{\mu\nu}(p_a)P^{(1)}_{\mu'\nu'}(p_a)~.\label{eq11}
\end{eqnarray}
Following the isobar model, the spin factors of the four-body decay $D^{0}\to P_1P_2P_3P_4$ are summarized in Table~\ref{spinfactor}.

\begin{table*}[htbp]
\caption{Summary of spin factors in this analysis, where $S$, $P$, $V$, $A$, $T$ and $PT$ denote scalar, pseudo-scalar, vector, axial-vector, tensor, and pseudo-tensor mesons, respectively. $[S]$, $[P]$, and $[D]$ represent orbital angular momenta $L$ = 0, 1, and 2 in the decays, respectively.}
\label{spinfactor}
\begin{center}
\begin{tabular}{c|c}
\hline
\hline
Decay chain				&Spin factor	 \\
\hline
$D[S]\to P~P_1, ~P[S]\to S~P_2, ~S[S]\to P_3~P_4$     		&	1	\\
$D[S]\to P~P_1, ~P[P]\to V~P_2, ~V[P]\to P_3~P_4$     		&	$\tilde{t}_{\mu}(P)\tilde{t}^{\mu}(V)$	\\
$D[S]\to P~P_1, ~P[D]\to T~P_2, ~T[D]\to P_3~P_4$     		&	$\tilde{t}_{\mu\nu}(P)\tilde{t}^{\mu\nu}(T)$	\\
$D[P]\to A~P_1, ~A[S]\to V~P_2, ~V[P]\to P_3~P_4$     		&	$\tilde{t}_{\mu}(D)P^{\mu\nu}(A)\tilde{t}_{\nu}(V)$	\\
$D[P]\to A~P_1, ~A[D]\to V~P_2, ~V[P]\to P_3~P_4$		&	$\tilde{t}_{\mu}(D)\tilde{t}^{\mu\nu}(A)\tilde{t}_{\nu}(V)$	\\
$D[P]\to A~P_1, ~A[P]\to S~P_2, ~S[S]\to P_3~P_4$		&	$\tilde{t}_{\mu}(D)\tilde{t}^{\mu}(A)$	\\
$D[P]\to A~P_1,~ A[P]\to T~P_2, ~T[D]\to P_3~P_4$		&	$\tilde{t}_{\mu}(D)\tilde{t}_{\nu}(A)P^{\mu\nu\rho\sigma}(A)\tilde{t}_{\rho\sigma}(T)$	\\
$D[P]\to V_1~P_1, ~V_1[P]\to V_2~P_2, ~V_2[P]\to P_3~P_4$	&	$\tilde{t}_{\mu}(D)P^{\mu\nu}(V_1)\epsilon_{\nu\lambda\sigma\alpha}p_{V_1}^{\alpha}\tilde{t}^{\lambda}(V_1)\tilde{t}^{\sigma}(V_2)$	\\
$D[D]\to PT~P_1, ~PT[S]\to T~P_2, ~T[D]\to P_3~P_4$		&	$\tilde{t}_{\mu\nu}(D)P^{\mu\nu\alpha\beta}(PT)\tilde{t}_{\alpha\beta}(T)$	\\
$D[D]\to PT~P_1, ~PT[P]\to V~P_2, ~V[P]\to P_3~P_4$		&	$\tilde{t}_{\mu\nu}(D)P^{\mu\nu\alpha\beta}(PT)\tilde{t}_{\alpha}(PT)\tilde{t}_{\beta}(V)$	\\
$D[D]\to PT~P_1, ~PT[D]\to S~P_2, ~S[S]\to P_3~P_4$		&	$\tilde{t}_{\mu\nu}(D)\tilde{t}^{\mu\nu}(PT)$	\\
$D[D]\to T~P_1, ~T[D]\to V~P_2, ~V[P]\to P_3~P_4$		&	$\tilde{t}_{\mu\nu}(D)P^{\mu\nu\alpha\beta}(T)\epsilon_{\alpha\lambda\sigma\rho}p_{T}^{\rho}\tilde{t}_{\beta}^{\lambda}(T)P^{\sigma\gamma}(T)\tilde{t}_{\gamma}(V)$	\\
$D[D]\to T_1~P_1, ~T_1[P]\to T_2~P_2, ~T_2[D]\to P_3~P_4$			&	$\tilde{t}_{\mu\nu}(D)P^{\mu\nu\alpha\beta}(T_1)\epsilon_{\alpha\lambda\sigma\rho}p_{T_1}^{\rho}\tilde{t}^{\lambda}(T_1)\tilde{t}^{\sigma}_{\beta}(T_2)$	\\
$D[S]\to S_1~S_2, ~S_1[S]\to P_1~P_2, ~S_2[S]\to P_3~P_4$     		&	1	\\
$D[P]\to V~S, ~V[P]\to P_1~P_2, ~S[S]\to P_3~P_4$     				&	$\tilde{t}_{\mu}(D)\tilde{t}^{\mu}(V)$	\\
$D[S]\to V_1~V_2, ~V_1[P]\to P_1~P_2, ~V_2[P]\to P_3~P_4$     		&	$\tilde{t}_{\mu}(V_1)\tilde{t}^{\mu}(V_2)$	\\
$D[P]\to V_1~V_2, ~V_1[P]\to P_1~P_2, ~V_2[P]\to P_3~P_4$     		&	$\epsilon_{\mu\nu\alpha\beta}p_{D}^{\beta}\tilde{t}^{\mu}(D)\tilde{t}^{\nu}(V_1)\tilde{t}^{\alpha}(V_2)$	\\
$D[D]\to V_1~V_2, ~V_1[P]\to P_1~P_2, ~V_2[P]\to P_3~P_4$     		&	$\tilde{t}_{\mu\nu}(D)\tilde{t}^{\mu}(V_1)\tilde{t}^{\nu}(V_2)$	\\
$D[D]\to T~S, ~T[D]\to P_1~P_2, ~S[S]\to P_3~P_4$     				&	$\tilde{t}_{\mu\nu}(D)\tilde{t}^{\mu\nu}(T)$	\\
$D[P]\to T~V, ~T[D]\to P_1~P_2, ~V[P]\to P_3~P_4$     				&	$\tilde{t}_{\mu}(D)\tilde{t}^{\mu\nu}(T)\tilde{t}_{\nu}(V)$	\\
$D[D]\to T~V, ~T[D]\to P_1~P_2, ~V[P]\to P_3~P_4$     				&	$\epsilon_{\mu\nu\alpha\beta}p_{D}^{\beta}\tilde{t}^{\mu\rho}(D)\tilde{t}^{\nu}_{\rho}(T)\tilde{t}^{\alpha}(V)$	\\
$D[S]\to T_1~T_2, ~T_1[D]\to P_1~P_2, ~T_2[D]\to P_3~P_4$     		&	$\tilde{t}_{\mu\nu}(T_1)\tilde{t}^{\mu\nu}(T_2)$	\\
$D[P]\to T_1~T_2, ~T_1[D]\to P_1~P_2, ~T_2[D]\to P_3~P_4$     		&	$\epsilon_{\mu\nu\alpha\beta}p_{D}^{\beta}\tilde{t}^{\mu}(D)\tilde{t}^{\nu\rho}(T_1)\tilde{t}_{\rho}^{\alpha}(T_2)$	\\
$D[D]\to T_1~T_2, ~T_1[D]\to P_1~P_2, ~T_2[D]\to P_3~P_4$     		&	$\tilde{t}_{\mu\nu}(D)\tilde{t}^{\mu\rho}(T_1)\tilde{t}^{\nu}_{\rho}(T_2)$	\\
\hline
\hline
\end{tabular}
\end{center}
\end{table*}

The Blatt-Weisskopf barrier factors $B_L(q)$ are derived by assuming a square well interaction potential as
\begin{eqnarray}
 & B_{L=0}(q)=&1~, \\\label{eq12}
 & B_{L=1}(q)=&\sqrt{\frac{2}{q^2+q_R^2}}~,\\ \label{eq13}
 & B_{L=2}(q)=&\sqrt{\frac{13}{q^4+3q^2q_R^2+9q_R^4}}~, \label{eq14}
\end{eqnarray}
where $q$ is the momentum of daughter particle in the rest frame of the mother particle, $L$ is the orbital angular momentum, and $q_R=1/R$ is a hadron ``scale'' parameter ($R$ denotes the radius of the centrifugal barrier). In this analysis, the radius $R$ is taken to be $5.0~{\rm{GeV}^{-1}}c$ for $D^{0}$ mesons, and $3.0~{\rm{GeV}^{-1}}c$ for other intermediate resonances.

Generally, the propagators of resonances are described by a relativistic Breit-Wigner function
\begin{eqnarray}
   P(s)=\frac{1}{m_0^2-s-i\sqrt{s}\Gamma(s)}~,\label{eq15}
\end{eqnarray}
where $m_0$ is the nominal mass of the resonance. For a resonance decaying into two scalar particles $a\to bc$, $\Gamma_{a\to bc}(s)$ is given by
\begin{eqnarray}
   \Gamma_{a\to bc}(s)=\Gamma_0^{a\to bc}\left(\frac{q}{q_0}\right)^{2L+1}\left(\frac{m_0}{\sqrt{s}}\right)^{2}\left(\frac{B_L(q)}{B_L(q_0)}\right)^2~,\label{eq16}
\end{eqnarray}
where $\Gamma_0^{a\to bc}$ is the nominal width when $s=m_0^2$ and  $q_0$ is the corresponding momentum of the daughter particle in the  rest frame of the mother particle. 

In this analysis, the propagators of $\rho(770)$ and $\rho(1450)$ are described by the Gounaris-Sakurai parametrization~\cite{Gounaris:1968mw}.  The propagator of $f_0(980)$ in the decay  $a_1(1420)\to f_0(980)\pi$ is described by a Flatt$\rm{\acute{e}}$ parametrization of $\pi\pi$ and $KK$ coupled channels with parameters from Ref.~\cite{BES:2004twe}.

The K-matrix parametrization \cite{Anisovich:2002ij,BaBar:2008inr} instead of the Breit-Wigner formula is adopted for $\pi\pi$ S-wave. The P-vector parametrization of the K-matrix for the $\pi\pi$ S-wave with five coupled channels $\pi\pi$, $KK$, $\pi\pi\pi\pi$, $\eta\eta$, $\eta\eta'$ and five poles is written as
  \begin{eqnarray}
  F_{\mu}(s) =   [I-iK(s)\rho(s)]^{-1}_{\mu\nu}P_{\nu}(s)~,\label{eq_F}
  \end{eqnarray}
where $I$ is the identity matrix, $K$ is the K-matrix describing the scattering process, $\rho$ is the PHSP matrix and $P$ is the initial production vector (P-vector). The indices $\mu$ and $\nu$ denote the coupled channels ($\pi\pi$, $KK$, $\pi\pi\pi\pi$, $\eta\eta$, and $\eta\eta'$), and only $F_{\mu}(s)$ of the $\pi\pi$ component is used as the propagator of the $\pi\pi$ S-wave. The K-matrix is given by
  \begin{eqnarray}
  K_{\mu\nu}(s) =  \left(\sum_{\alpha}\frac{g^{\alpha}_{\mu}g^{\alpha}_{\nu}}{m_{\alpha}^2-s}+f_{\mu\nu}^{\rm scatt}\frac{1-s_{0}^{\rm scatt}}{s-s_{0}^{\rm scatt}}\right)f_{A0}(s)~, \label{eq_K}
  \end{eqnarray}
which contains five poles ($\alpha = 1$ to $5$). The parameters of the K-matrix are fixed to those in Ref.~\cite{BaBar:2008inr}. The P-vector is written as
  \begin{eqnarray}
  P_{\nu}(s) =  \left(\sum_{\alpha}\frac{\beta_{\alpha}g^{\alpha}_{\nu}}{m_{\alpha}^2-s}+f_{\nu}^{\rm prod}\frac{1-s_{0}^{\rm prod}}{s-s_{0}^{\rm prod}}\right)~, \label{eq_P}
  \end{eqnarray}
  where the poles are the same as for the K-matrix. The parameters $\beta_{\alpha}$ and $f_{1\nu}^{\rm prod}$ are free in the fit. For the parameter $s_0^{\rm prod}$ in the P-vector, we assume that they take the same value in all $\pi\pi$ S-waves and fix them to $-5~{\rm GeV}^{2}/c^{4}$ since they are insensitive to any choice if $s_0^{\rm prod}\leq-5~{\rm GeV}^{2}/c^{4}$ in our fit.

In practice, there are two $\pi\pi$ S-waves in the decay $D^0\to S_1S_2$,
and the corresponding propagator is 
    \begin{eqnarray}
  F'_{\mu\nu}(s_1,s_2)&=&    [I-iK(s_1)\rho(s_1)]^{-1}_{\mu\rho}[I-iK(s_2)\rho(s_2)]^{-1}_{\nu\sigma} \nonumber\\ 
				&&	\times P_{\rho\sigma}(s_1,s_2)~, \label{eq_F'}
  \end{eqnarray}
  where $s_1$ and $s_2$ are the invariant masses squared of $\pi\pi$, $P_{\rho\sigma}(s_1,s_2)$ is the expansion of the product of two P-vectors $P_{\rho}(s_1)$ and $P_{\sigma}(s_2)$, and the corresponding coefficients of each term are taken to be independent parameters in the fit. Benefiting from the exchange symmetry of two $\pi\pi$ S-waves in the amplitude, $P_{\rho\sigma}(s_1,s_2)$ can be written as
\begin{eqnarray}
P_{\rho\sigma}(s_1,s_2)&=&\sum_{\alpha,\beta}^{\alpha\le\beta}a_{\alpha,\beta}\bigg[\frac{g^{\alpha}_{\rho}g^{\beta}_{\sigma}}{(m_{\alpha}^2-s_1)(m_{\beta}^2-s_2)} \nonumber \\
&&+ \frac{g^{\beta}_{\rho}g^{\alpha}_{\sigma}}{(m_{\beta}^2-s_1)(m_{\alpha}^2-s_2)}\bigg]    \nonumber \\
&&+\sum_{\alpha}b_{\alpha,\rho}\frac{g^{\alpha}_{\sigma}(1-s_{0}^{\rm prod})}{(s_1-s_{0}^{\rm prod})(m_{\alpha}^2-s_2)}  \nonumber \\
&&+ \sum_{\alpha}b_{\alpha,\sigma}\frac{g^{\alpha}_{\rho}(1-s_{0}^{\rm prod})}{(s_2-s_{0}^{\rm prod})(m_{\alpha}^2-s_1)} \nonumber \\  
&&+c_{[\rho,\sigma]}\frac{(1-s_{0}^{\rm prod})^2}{(s_1-s_{0}^{\rm prod})(s_2-s_{0}^{\rm prod})}, \label{eq_P2}             
  \end{eqnarray}
where $c_{[\rho,\sigma]}=c_{\rho,\sigma} + c_{\sigma,\rho}$, and  $c_{[\rho,\sigma]}$ with $\rho>\sigma$ is fixed to zero in the fit. Similar to one $\pi\pi$ S-wave, only ($\pi\pi,\pi\pi$) component of $F'_{\mu\nu}$ is used. 

\begin{figure*}[htbp]
  \centering
  \begin{overpic}[width=0.245\textwidth]{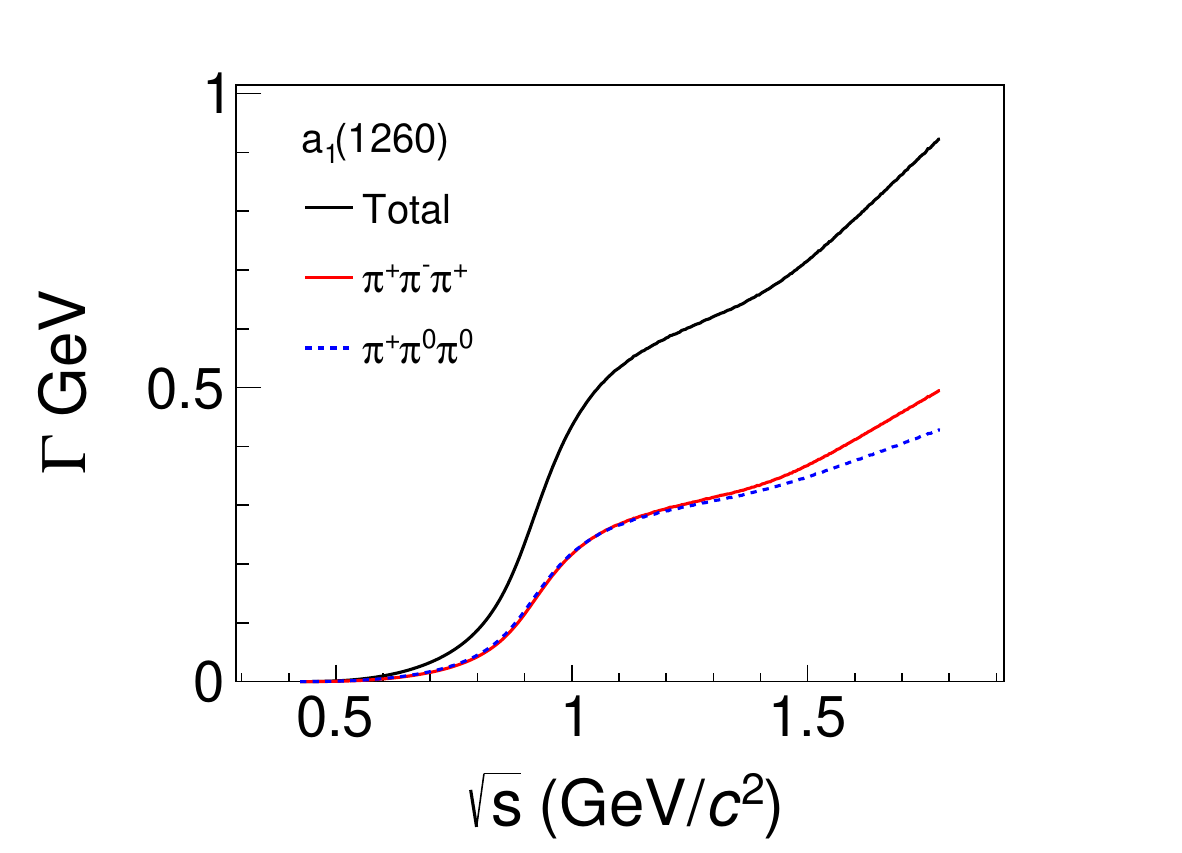}
  \end{overpic}
    \begin{overpic}[width=0.245\textwidth]{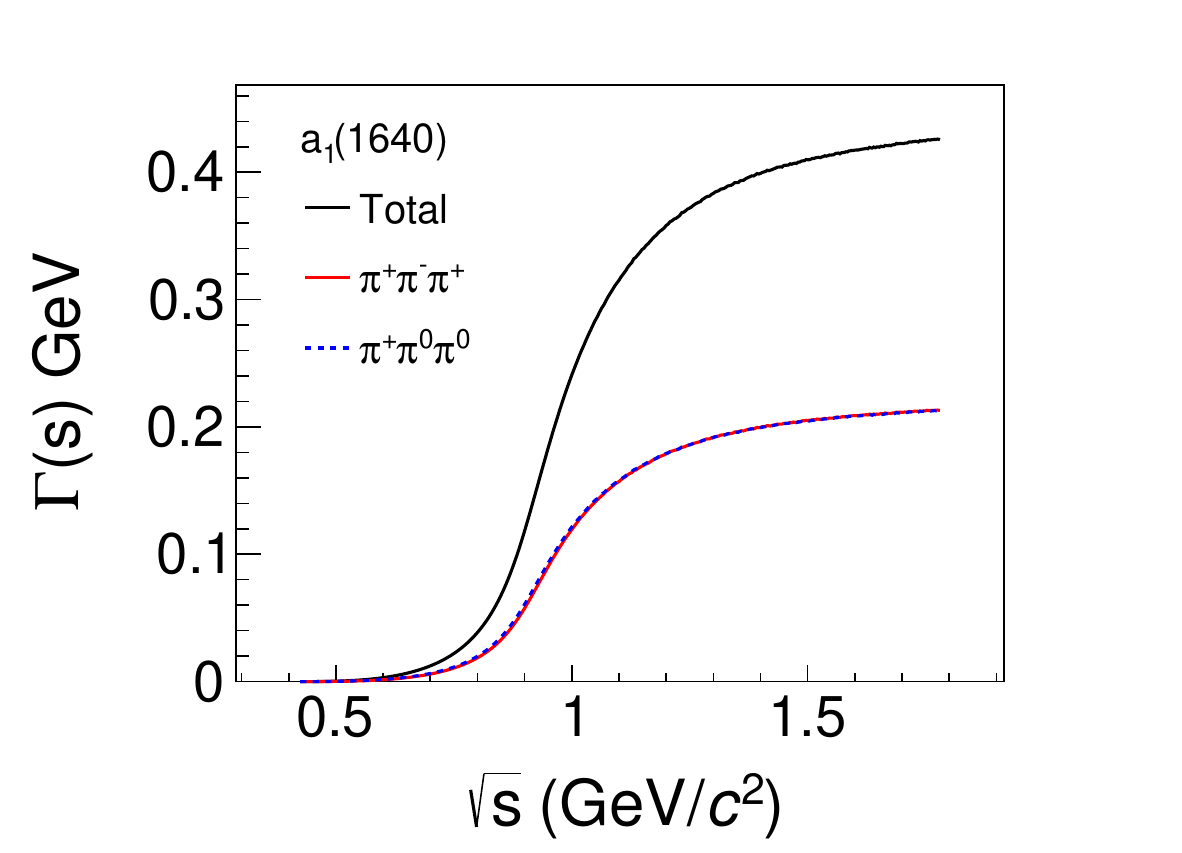}
  \end{overpic}
    \begin{overpic}[width=0.245\textwidth]{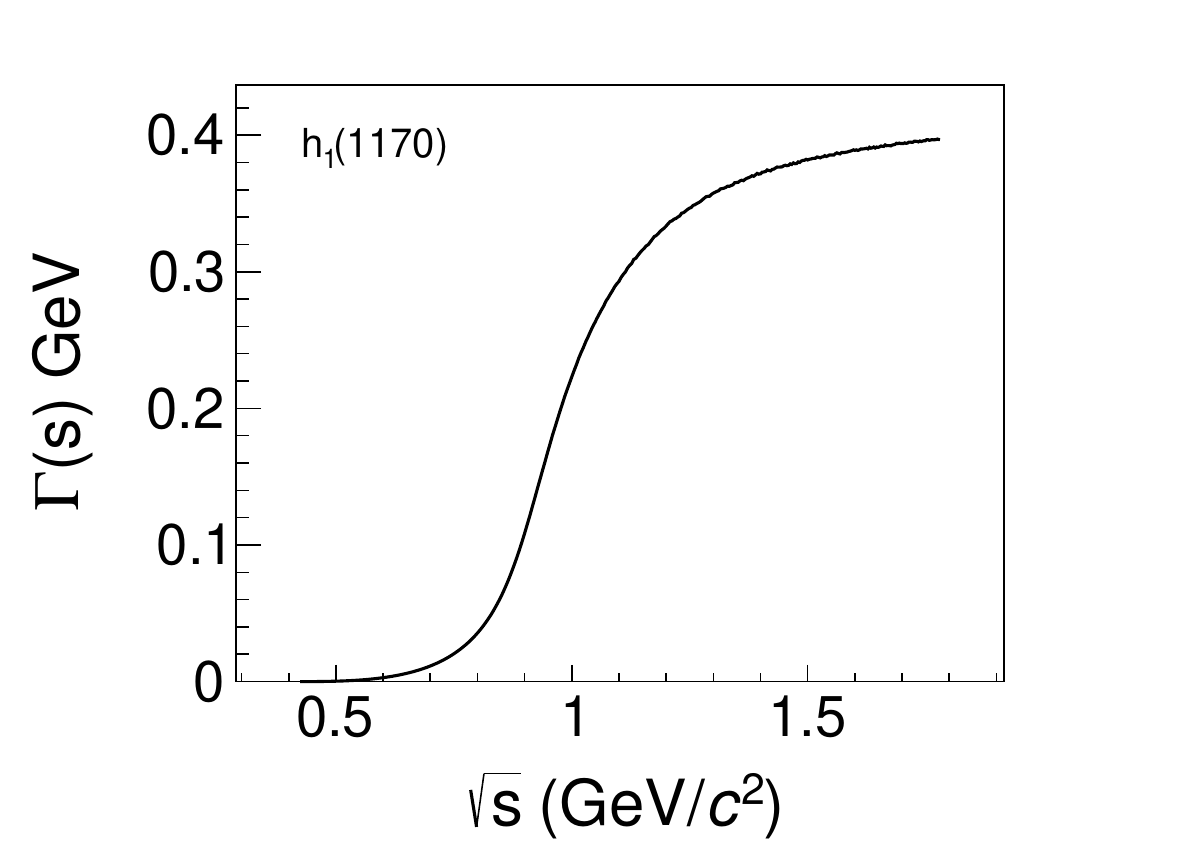}
  \end{overpic}
   \begin{overpic}[width=0.245\textwidth]{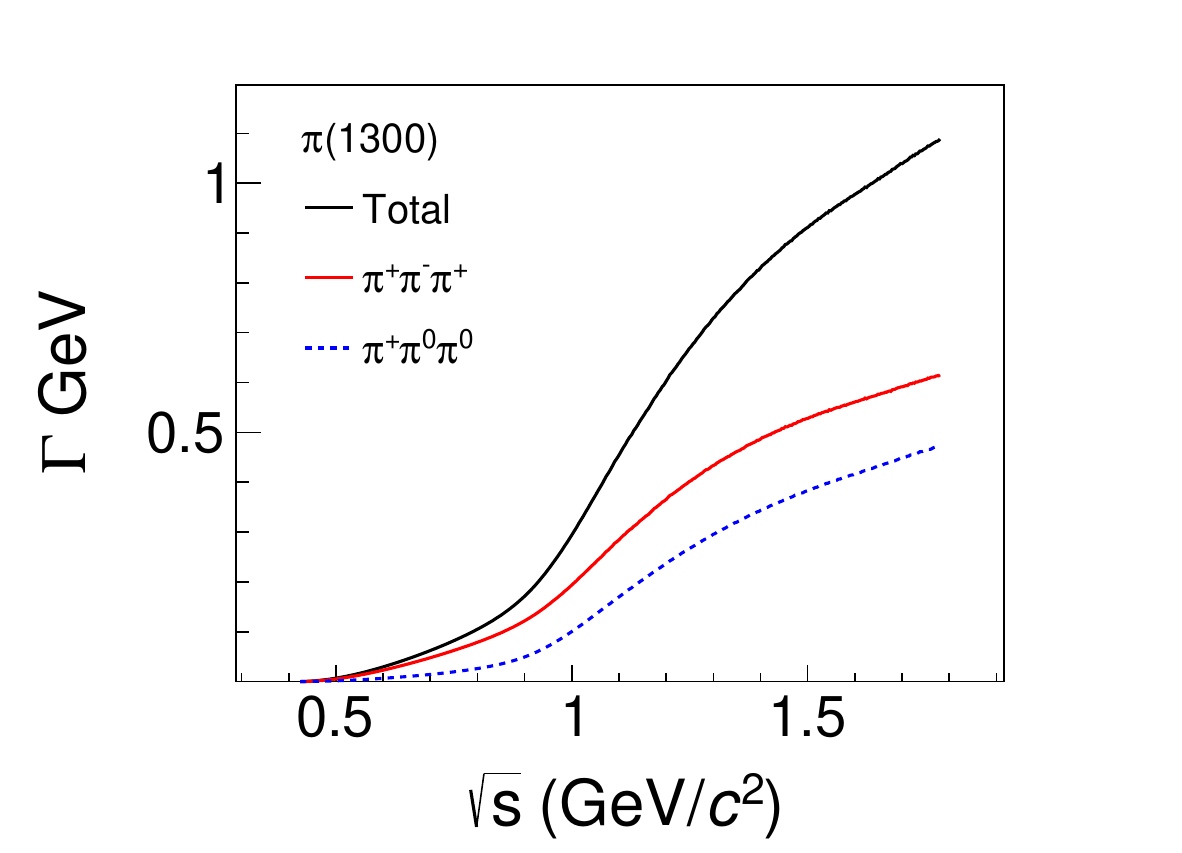}
  \end{overpic}
  
  \caption{The $\Gamma(s)$ of $a_1(1260)$, $a_{1}(1640)$, $h_1(1170)$ and $\pi(1300)$ resonances obtained in the amplitude analysis.}
  \label{fig:Gamma}
  \end{figure*}

The propagators of the resonances $a_1(1260)$, $a_{1}(1640)$, $h_1(1170)$ and $\pi(1300)$ decaying into $3\pi$ are described by relativistic Breit-Wigner functions with the coupled channel $3\pi$, where $\Gamma(s)$ are obtained by integrating the amplitude squared over PHSP
\begin{eqnarray}
   \Gamma_{a\to bcd}(s) \propto \frac{m_0}{\sqrt{s}}\int \sum_{\rm spin}|A_{a \to bcd}|^2 \mathrm{d}\Phi_3~\label{eq17}~,
\end{eqnarray}
based on the amplitudes $A_{R\to3\pi}$ obtained in this analysis. Figure \ref{fig:Gamma} shows the $\Gamma(s)$ obtained in the amplitude analysis. For other resonances decaying into $3\pi$, a relativistic Breit-Wigner function with a constant $\Gamma(s)$ is used. The parameters of $a_1(1260)$ and $\pi(1300)$ are determined in the fit and others are fixed to their respective PDG values~\cite{ParticleDataGroup:2022pth}.
Due to its small contribution, the $a_1(1420)$ is described as a relativistic Breit-Wigner function with a constant $\Gamma(s)$, even though it has been regarded as the effect of the Triangle Singularity~\cite{COMPASS:2020yhb}. Only the decay of $a_1(1420)\to f_0(980)\pi$ is taken into account, and the corresponding resonant parameters are taken from Ref.~\cite{COMPASS:2018uzl}.

To take into account the resolution effect for the narrow resonances $\phi(1020)$ and $\omega(782)$,  the shapes of corresponding relativistic Breit-Wigner functions in the $M(\pi^+\pi^-\pi^0)$ distribution are convolved with a Gaussian function. The parameters of this Gaussian function are obtained by the fit to the $\eta$ peak in the $M(\pi^+\pi^-\pi^0)$ distribution with the Breit-Wigner function of $\eta$ convolved with the same Gaussian function.

In this analysis, two decay topologies, $D\to R_1 R_2$ and $D\to R_1\pi \to R_2\pi\pi$, are considered. Meanwhile, isospin symmetry is considered by applying the Clebsch-Gordan (CG) coefficients between the isospin processes. The isospin states for $u$ and $d$ quarks are
 \begin{eqnarray}
|u\rangle:|\frac{1}{2}, \frac{1}{2}\rangle,~|\bar{u}\rangle:-|\frac{1}{2},-\frac{1}{2}\rangle,~|d\rangle:|\frac{1}{2}, -\frac{1}{2}\rangle,~|\bar{d}\rangle:|\frac{1}{2},\frac{1}{2}\rangle. \label{eq18}
  \end{eqnarray}
Based on the quark components of hadrons, the isospin states of the hadrons are shown in Table~\ref{tab:isospin}.
The isospin amplitudes for typical processes are written as
\end{multicols}
\vspace{-0.8cm}
 \begin{eqnarray}
A(D^{0}\to a_1^+\pi^{-})&=& A(\rho^0_{\pi^{+}\pi^{-}}\pi^{+},\pi^{-}) + A(\rho^+_{\pi^{+}\pi^{0}}\pi^{0},\pi^{-}) - c[2A(f_{\pi^{+}\pi^{-}}\pi^{+},\pi^{-}) + A(f_{\pi^{0}\pi^{0}}\pi^{+},\pi^{-})], \\
A(D^{0}\to a_1^0\pi^{0})~&=& -A(\rho^+_{\pi^{+}\pi^{0}}\pi^{-},\pi^{0}) + A(\rho^-_{\pi^{0}\pi^{-}}\pi^{+},\pi^{0}) - c[2A(f_{\pi^{+}\pi^{-}}\pi^{0},\pi^{0}) + A(f_{\pi^{0}\pi^{0}}\pi^{0},\pi^{0})], \\
A(D^{0}\to a_1^-\pi^{+})&=& -A(\rho^0_{\pi^{+}\pi^{-}}\pi^{-},\pi^{+}) - A(\rho^-_{\pi^{0}\pi^{-}}\pi^{0},\pi^{+}) - c[2A(f_{\pi^{+}\pi^{-}}\pi^{-},\pi^{+}) + A(f_{\pi^{0}\pi^{0}}\pi^{-},\pi^{+})], \\
A(D^{0}\to h_1^0\pi^{0})~&=& -A(\rho^+_{\pi^{+}\pi^{0}}\pi^{-},\pi^{0}) - A(\rho^-_{\pi^{0}\pi^{-}}\pi^{+},\pi^{0}) + A(\rho^0_{\pi^{+}\pi^{-}}\pi^{0},\pi^{0}),\\
A(D^{0}\to f f')~~~&=& 4A(f_{\pi^{+}\pi^{-}},f'_{\pi^{+}\pi^{-}}) + 2[A(f_{\pi^{+}\pi^{-}},f'_{\pi^{0}\pi^{0}})+ A(f_{\pi^{0}\pi^{0}},f'_{\pi^{+}\pi^{-}})] + A(f_{\pi^{0}\pi^{0}},f'_{\pi^{0}\pi^{0}}),\\
A(D^{0}\to \rho^0 f)~~~&=& 2A(\rho^0_{\pi^{+}\pi^{-}},f_{\pi^{+}\pi^{-}}) + A(\rho^0_{\pi^{+}\pi^{-}},f_{\pi^{0}\pi^{0}}).
  \end{eqnarray}
\begin{multicols}{2}
\noindent Here, $a_1$ and $\rho$ are isospin vectors, whereas $h_1$ and $f^{(')}$ are isospin scalars. The normalization factors are dropped here and $c$ is the relative difference (including magnitude and phase) between two isospin processes which is determined in the amplitude analysis.

\begin{table*}[htbp]
\caption{The quark components and isospin states for different hadrons used in the analysis.}
\label{tab:isospin}
\begin{center}
\begin{tabular}{c|c|c|c}
\hline
\hline
		&$\pi^+,b^+,\rho^+,a^+$				&$\pi^0,b^0,\rho^0,a^0$								&$\pi^-,b^-,\rho^-,a^-$  \\
\hline
$I=1$		&$|u\bar{d}\rangle:|1,1\rangle$			&$|\frac{u\bar{u}-d\bar{d}}{\sqrt{2}}\rangle:-|1,0\rangle$		&$|d\bar{u}\rangle:-|1,-1\rangle$\\
\hline
		&\multicolumn{3}{c}{$h,\omega,\phi,f$}\\
\hline
$I=0$		&\multicolumn{3}{c}{$|(u\bar{u}+d\bar{d})+g(s\bar{s})\rangle:-|0,0\rangle$}\\
\hline
\hline
\end{tabular}
\end{center}
\end{table*}

\section{AMPLITUDE FIT PROCESS AND RESULT}\label{sec_fit}
In the amplitude fit, the sPlot technique~\cite{Pivk:2004ty} is applied to deal with the effect of background. 
In the sPlot method, the four-momenta of final states are chosen as control variables $x$, $M_{\rm bc}^{\rm sig}$ and $M_{\rm bc}^{\rm tag}$ are chosen as discriminating variables $y$, and the weight for each candidate event is obtained by an unbinned maximum likelihood fit on the 2D distribution of $M_{\rm bc}^{\rm sig}$ versus $M_{\rm bc}^{\rm tag}$, as described in Sec.~\ref{signalyield}. 

According to the sPlot technique and based on the above DT fit, the weight $W_s(y)$ of each candidate event is calculated as 
\begin{eqnarray}
   W_s(y) = \frac{\sum_{i}V_{si}F_{i}(y)}{\sum_{i}N_jF_j(y)+\sum_{k}N'_kF'_k(y)}, \label{eq_sw}
\end{eqnarray}
where the index $s$ denotes the signal, $F_{i(j)}(y)$ are the PDFs of the signal ($i(j)=s$) and backgrounds ($i(j)\neq s$) from BKGI to BGKV with floating yields, $F'_{k}(y)$ are the PDFs of BKGVI with fixed yields, and $N_{i(j)}$ and $N_k^{'}$ are the corresponding yields for each component. 
The inverse of covariance matrix $V_{ij}$ is calculated with
 \begin{eqnarray}
   V_{ij}^{-1}=\sum_{n\in {\rm Data}}\frac{F_i(y_n)F_j(y_n)}{(\sum_{k}N_kF_k(y_n)+\sum_{l}N'_lF'_l(y_n))^{2}}, \label{eq_vij}
  \end{eqnarray}
where the summations for indices $k$ and $l$ run over all the components of $F_k(y)$ and $F'_l(y)$ as described above, respectively. The summation for index $n$ runs over all the events in the DT fit. 

Based on the sPlot technique, $W_s(y)$ defined in Eq.~\eqref{eq_sw} includes the effects of the backgrounds with the fixed number of events. To take into account these effects, the coefficients $c_{i}$ are calculated as
\begin{eqnarray}
   c_{i} = \sum_{j}V_{sj}\nu_{ij}, \label{eq_cn}
\end{eqnarray}
where the summation runs over the signal and backgrounds from BKGI to BKGV with floating yields in the fit,  $i$ represents the different components in the BKGVI, and $\nu_{ij}$ is
 \begin{eqnarray}
   \nu_{ij}=\sum_{n\in {\rm Data}}\frac{F'_i(y_n)F_j(y_n)}{(\sum_{k}N_kF_k(y_n)+\sum_{l}N'_lF'_l(y_n))^{2}}. \label{eq_nuij}
 \end{eqnarray}

With the obtained $W_s(y)$ and $c_i$, the $x$ distribution of all candidate events with the weight $W_s(y)$ is the real signal together with the contributions of the peaking background in BKGVI, 
\begin{eqnarray} 
N_{s}P_s(x)  +  \sum_{j} c_{j}N'_{j}P'_{j}(x),\label{eq_splot}
\end{eqnarray}
where $P_s(x)$ and $P'_{j}(x)$ are the $x$ distributions of the signal and the peaking background $j$ in the BKGVI and $N_{s}$  and $N'_{j}$  are the corresponding yields obtained from the above 2D fit and MC simulation, respectively. 

In practice, the PDF of observing a signal event with the given final kinematic $p$ is written as 
 \begin{eqnarray}
   P_{s}(p) = \frac{\epsilon_{s}(p)|M_f(p)|^{2}\phi_4(p)}{\int\epsilon_{s}(p)|M_f(p)|^{2}\mathrm{d}\Phi_{4}}~,\label{eq34}
  \end{eqnarray}
where $\epsilon_s(p)$ is the signal efficiency, $|M_f(p)|^2$ is the differential cross section as discussed in Sec.~\ref{amplitudeformula}, and $\phi_4(p)$ is the PHSP density. 
The normalization factor is calculated by MC integration with the PHSP signal MC sample after event selection,
 \begin{eqnarray}
   \int \epsilon_{s}(p)|M_f(p)|^{2}\mathrm{d}\Phi_{4} \propto \frac{1}{N_{\rm MC}}\sum_{i=1}^{N_{\rm MC}} |M_f(p_i)|^{2}.\label{eq35}
 \end{eqnarray}
 where the $N_{\rm MC}$ is the number of events of the PHSP signal MC.
According to Eq.~\eqref{eq_splot}, the weighted likelihood $\mathrm{ln}L$ is given by
\begin{eqnarray}
    f\left[\sum_{i \in {\rm Data}} W_s(y_i)\mathrm{ln}P_{s}(p_i) - \sum_{j}\sum_{i\in {\rm BMC}_j} \omega_{j} \mathrm{ln}P_{s}(p_i)\right], \label{eq_lh}
\end{eqnarray}
where the second term in the bracket is the contribution from the peaking background, which is estimated with the corresponding simulated  background MC(BMC). The normalization factor $\omega_j$ is given by
 \begin{eqnarray}
 \omega_j = c_{j}N_{{\rm BKG}_j}^{\rm Data}/N_{{\rm BKG}_j}^{\rm MC}.
\end{eqnarray}
Here, $c_j$ is obtained from Eq.~\eqref{eq_cn}, and $N_{{\rm BKG}_j}^{\rm Data}$ and $N_{{\rm BKG}_j}^{\rm MC}$ are the background yields in data and simulated background events, respectively.
The factor $f$ in Eq.~\eqref{eq_lh},
\begin{eqnarray}
   f = \frac{\sum_{i\in {\rm Data}}W_s(y_i) + \sum_{j}\omega_jN^{\rm MC}_{{\rm BKG}_j}}{\sum_{i\in {\rm Data}}W^2_s(y_i) + \sum_{j}\omega_j^2N^{\rm MC}_{{\rm BKG}_j}}~,
  \end{eqnarray}
is the global factor to correct the statistical bias in the weighted maximum likelihood fit.

The total likelihood function in this analysis is summed over the two signal channels and three tag modes, 
\begin{eqnarray}
   \mathrm{ln}L_{\rm total} = \sum_{i\in {\rm tag}} \left(\mathrm{ln}L_{i}^{\pi^+\pi^-\pi^+\pi^-} + \mathrm{ln}L_{i}^{\pi^+\pi^-\pi^0\pi^0}\right),
  \end{eqnarray}  
and the free parameters are optimized via a maximum likelihood fit using the MINUIT~\cite{James:1975dr} package.

Generally, all the possible intermediate processes including the resonances listed in Table~\ref{tab:reslist} and based on $J^{PC}$ conservation are considered in the fit. The only exception are $D^0$ decays, where parity conservation is not required. Only processes with a significance greater than $5\sigma$ are kept during the fit unless otherwise noted.
Here, the significance of a specific amplitude is calculated according to the Wilks's Theorem by comparing the change of log-likelihood (2$\Delta$ln$L$) to the expected values from the chi-square distribution ($\chi^{2}_{\Delta N_{\rm para}}$) with the number of degrees of freedom (NDF) equal to the change of the numbers of fit parameters ($\Delta N_{\rm para}$). 
If the significance of the amplitudes containing the isospin vector in the $3\pi$ invariant mass spectrum is greater than 5$\sigma$, the corresponding isospin partners with significance greater than 3$\sigma$ are also kept. 
For the P-vector of the $\pi\pi$ S-wave, only the parameters before the 1st, 2nd poles and $\pi\pi$, $KK$ non-resonant terms are considered, and the others are fixed to zero in the nominal fit. Meanwhile, only those with significance greater than $3\sigma$ are kept.

\begin{table*}[htbp]
\caption{Resonances considered in this analysis.}
\label{tab:reslist}
\begin{center}
\begin{tabular}{l|cccccc}
\hline
\hline
						&$J^{P}=0^+$		&$J^{P}=0^-$	&$J^{P}=1^+$		&$J^{P}=1^-$		&$J^{P}=2^+$	&$J^{P}=2^{-}$		\\
\hline
\multirow{2}{*}{$\pi\pi$}		&$(\pi\pi)_{S}$		&			&				&$\rho(770)$		&$f_{2}(1270)$	&		\\		
						&				&			&				&$\rho(1450)$		&			&\\
\hline
\multirow{4}{*}{$\pi\pi\pi$}	&				&$\pi(1300)$	&$a_{1}(1260)$		&$\omega(782)$		&$a_{2}(1320)$	&$\pi_{2}(1670)$	\\
						&				&			&$a_{1}(1420)$		&$\phi(1020)$		&			&		\\
						&				&			&$a_{1}(1640)$		&$\pi_{1}(1400)$	&			&		\\
						&				&			&$h_{1}(1170)$		&$\pi_{1}(1600)$	&			&		\\
\hline
\hline
\end{tabular}
\end{center}
\end{table*}

To find the optimal solution, the baseline model which contains the processes from Refs.~\cite{FOCUS:2007ern,dArgent:2017gzv} is built up first. Next, starting from the baseline model, the significance of each possible process is tested and the most significant one among those satisfying the significance requirement is added to the current model. This step is repeated until no additional processes can be added. After this, the significances of individual processes in the existing model are tested again and those that do not satisfy the significance requirement are removed. 
The above steps are repeated until all the processes in the model satisfy the significance requirement and no further ones can be added. With this strategy, the nominal amplitude model is obtained and no multiple solutions are found.

\begin{figure*}[htbp]
  \centering
  \begin{overpic}[width=0.245\textwidth]{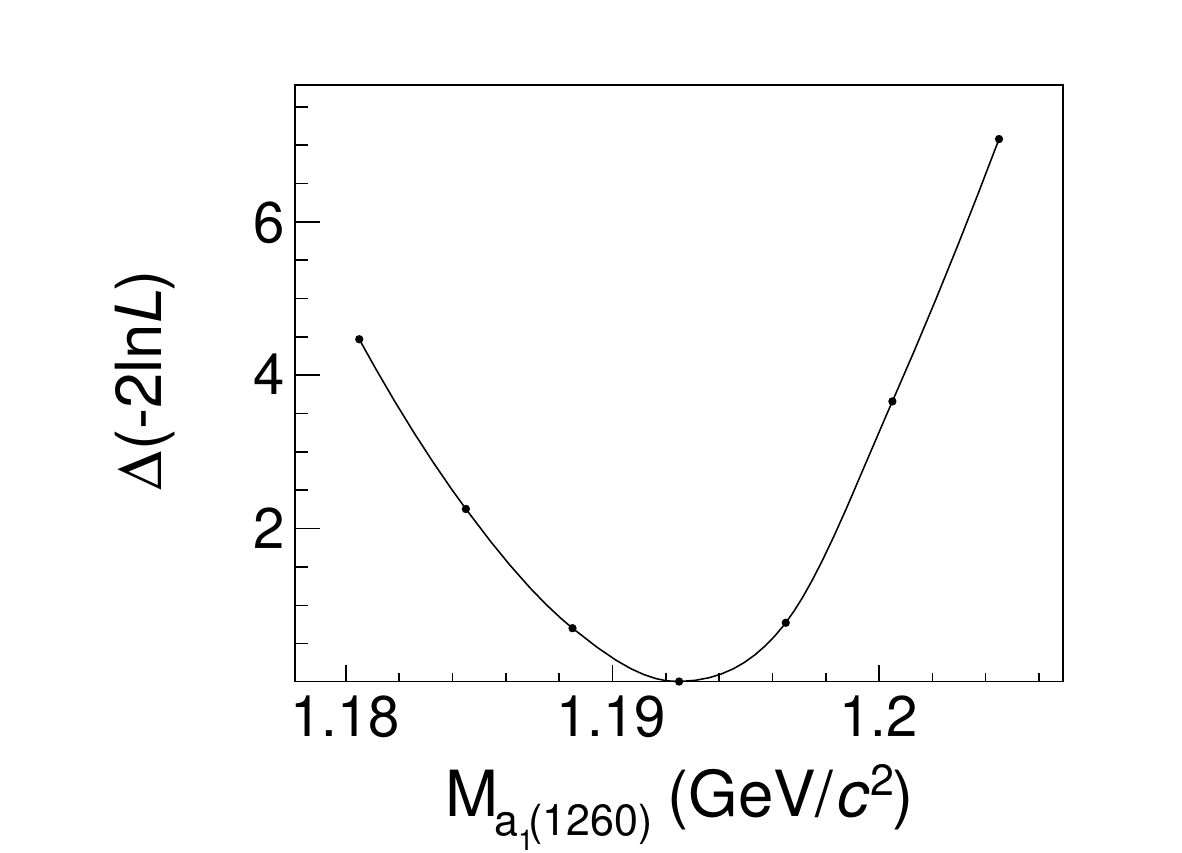}
  \end{overpic}
     \begin{overpic}[width=0.245\textwidth]{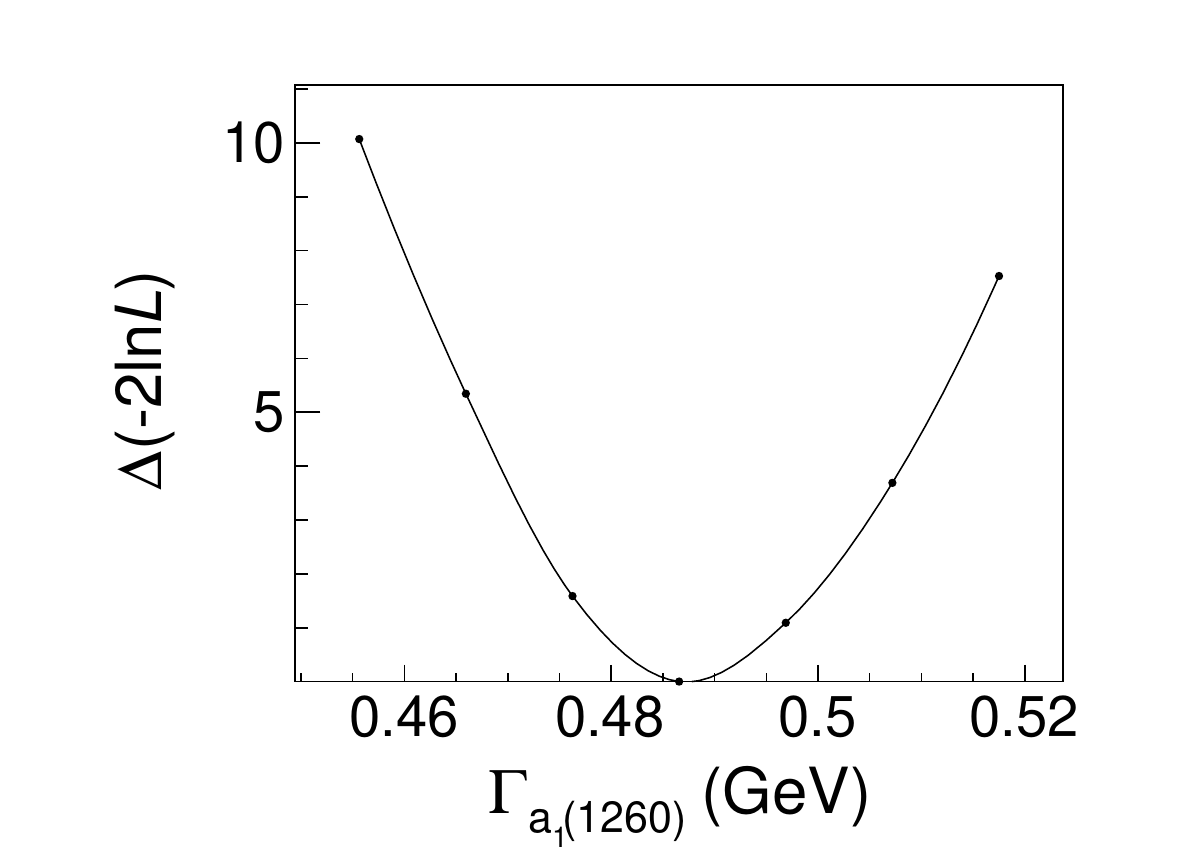}
  \end{overpic}
    \begin{overpic}[width=0.245\textwidth]{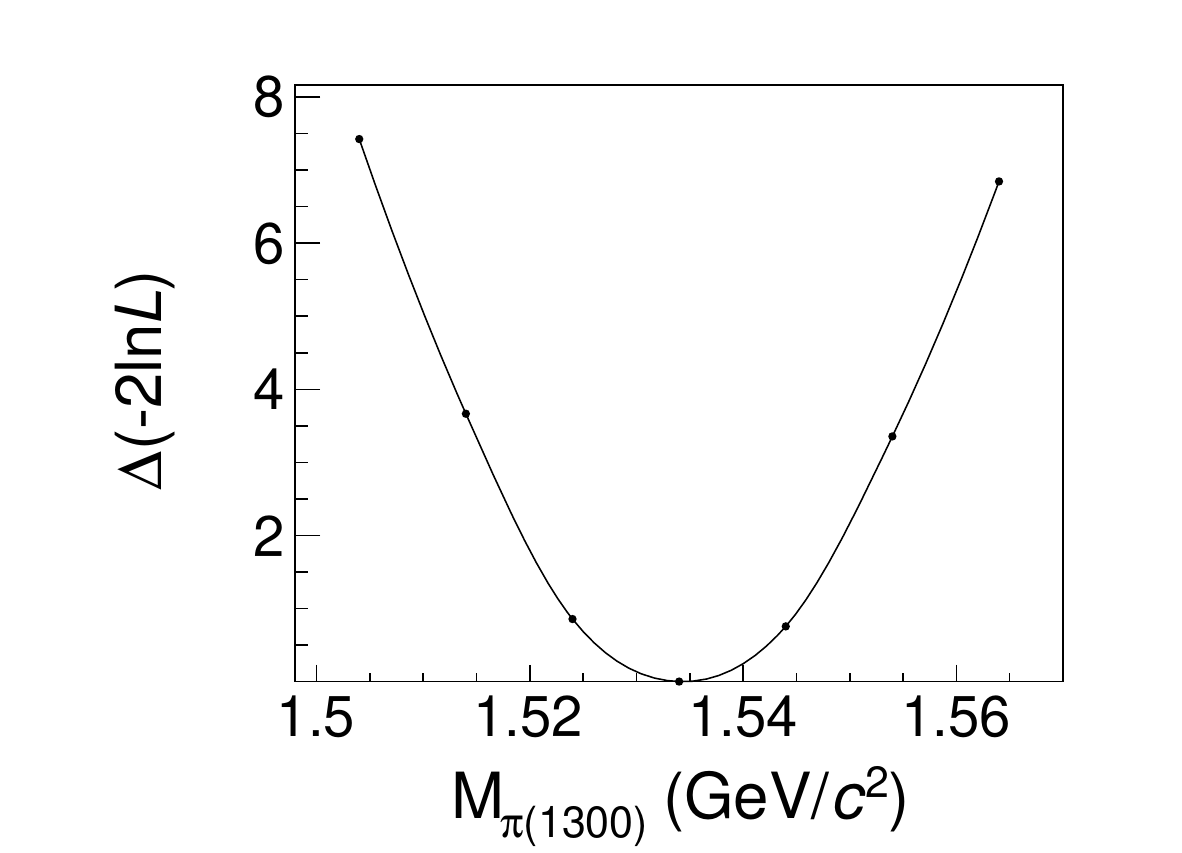}
  \end{overpic}
   \begin{overpic}[width=0.245\textwidth]{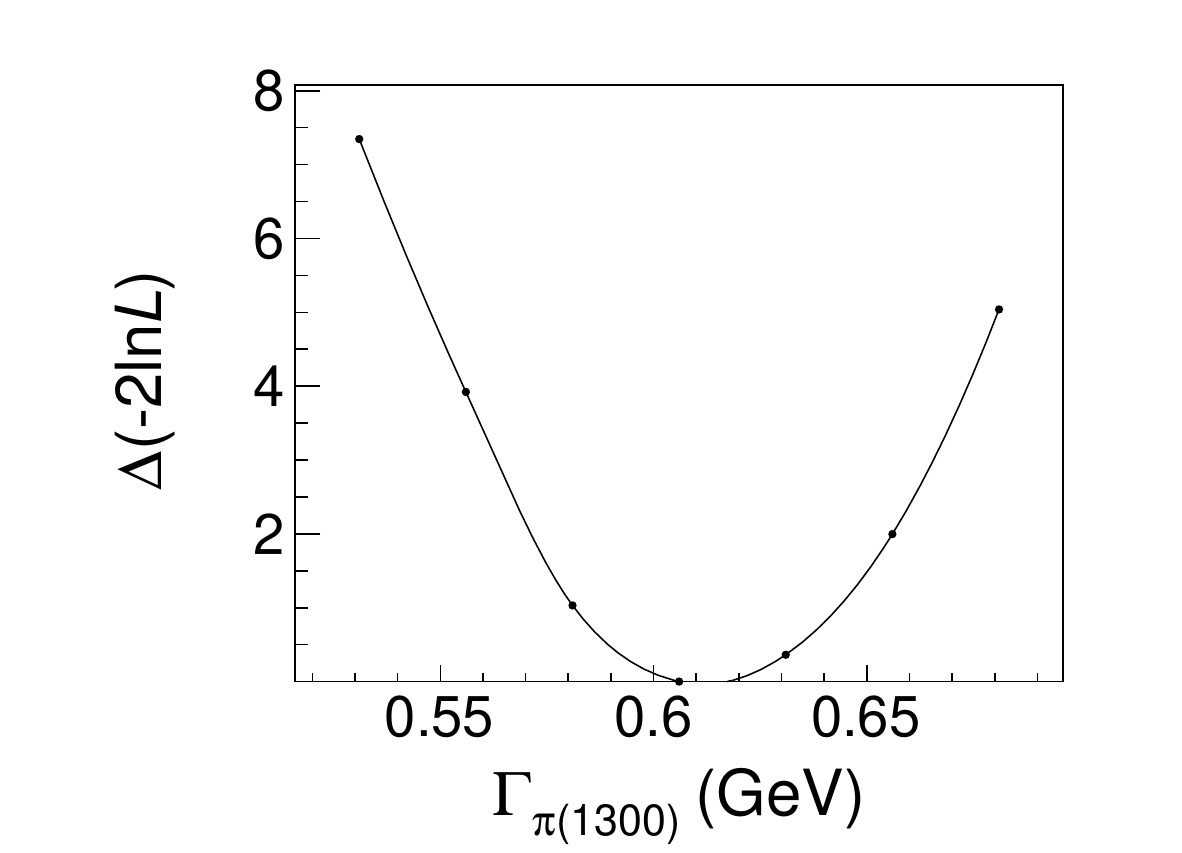}
  \end{overpic}
  \caption{Likelihood scans for masses and widths of $a_{1}(1260)$ and $\pi(1300)$.}
  \label{fig:respara}
\end{figure*}

\begin{figure*}[htbp]
  \centering
  \begin{overpic}[width=0.32\textwidth]{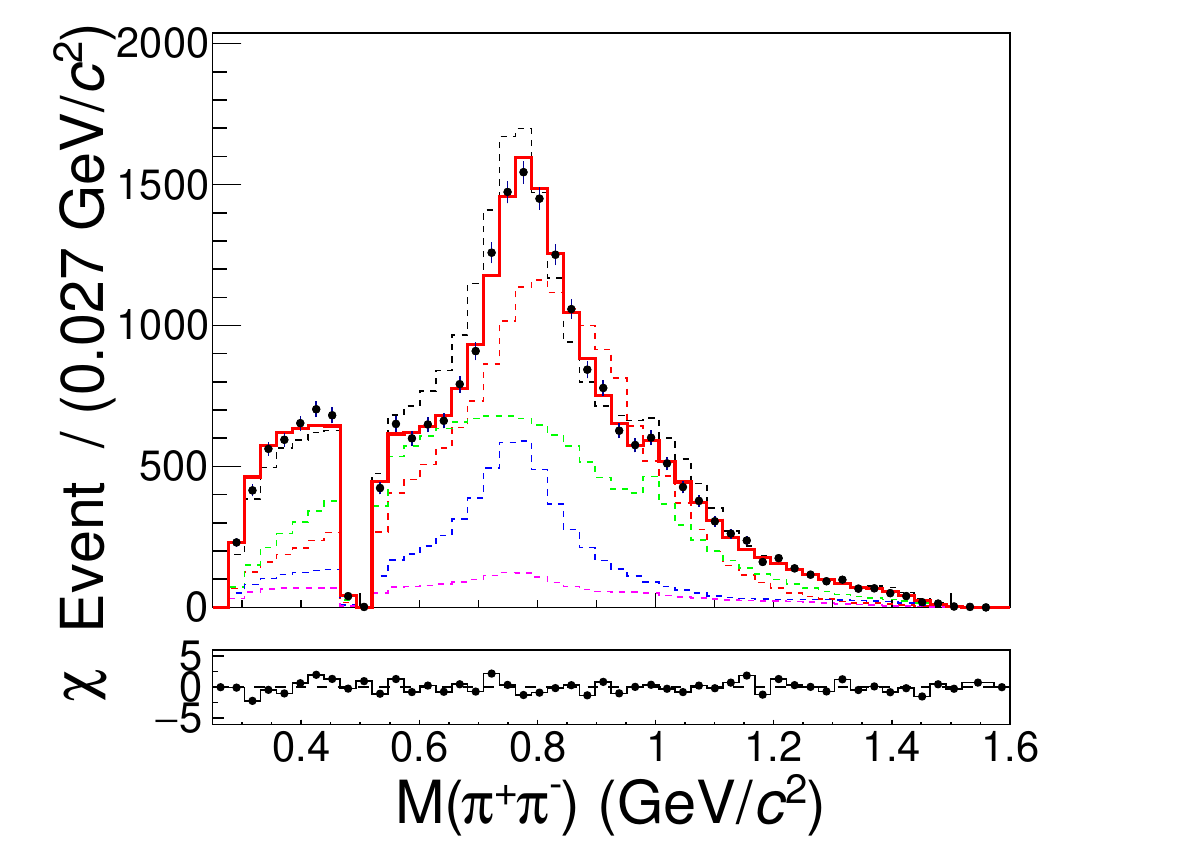}
  \end{overpic}
  \begin{overpic}[width=0.32\textwidth]{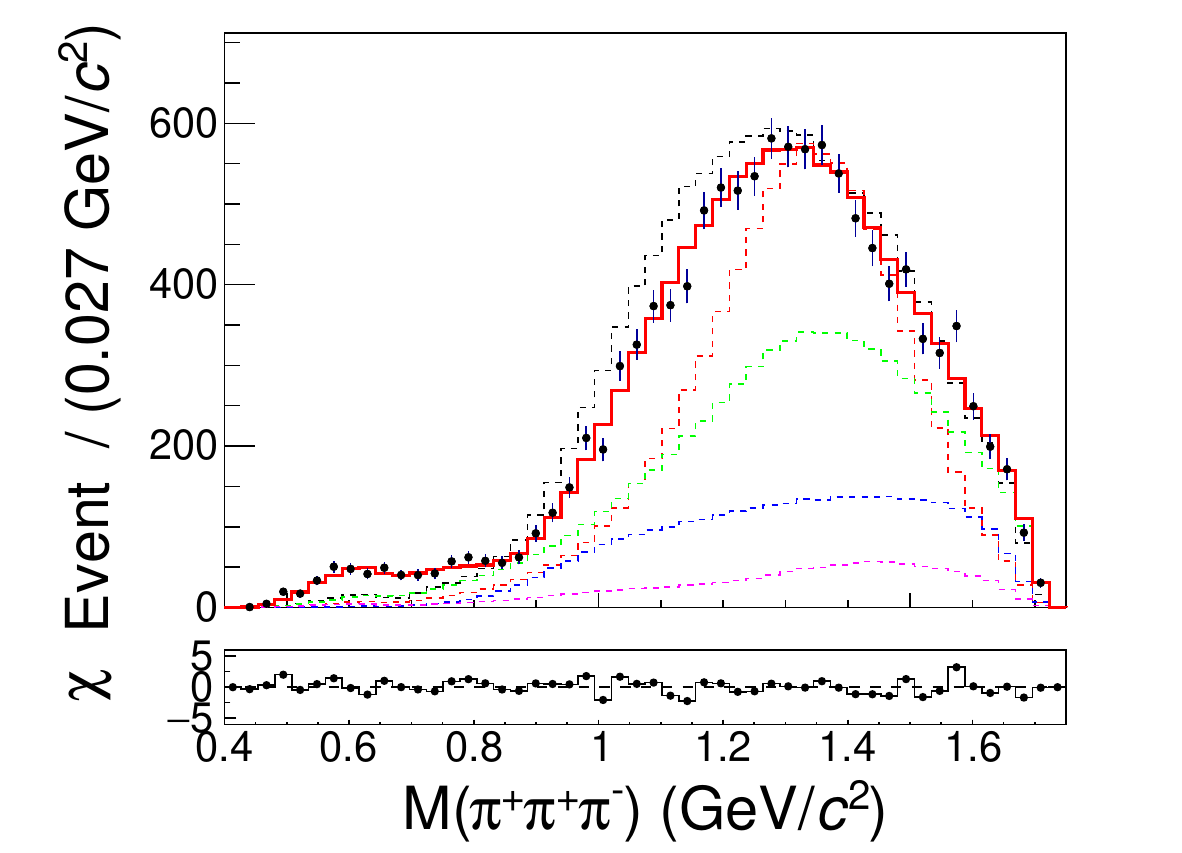}
  \end{overpic}
  \begin{overpic}[width=0.32\textwidth]{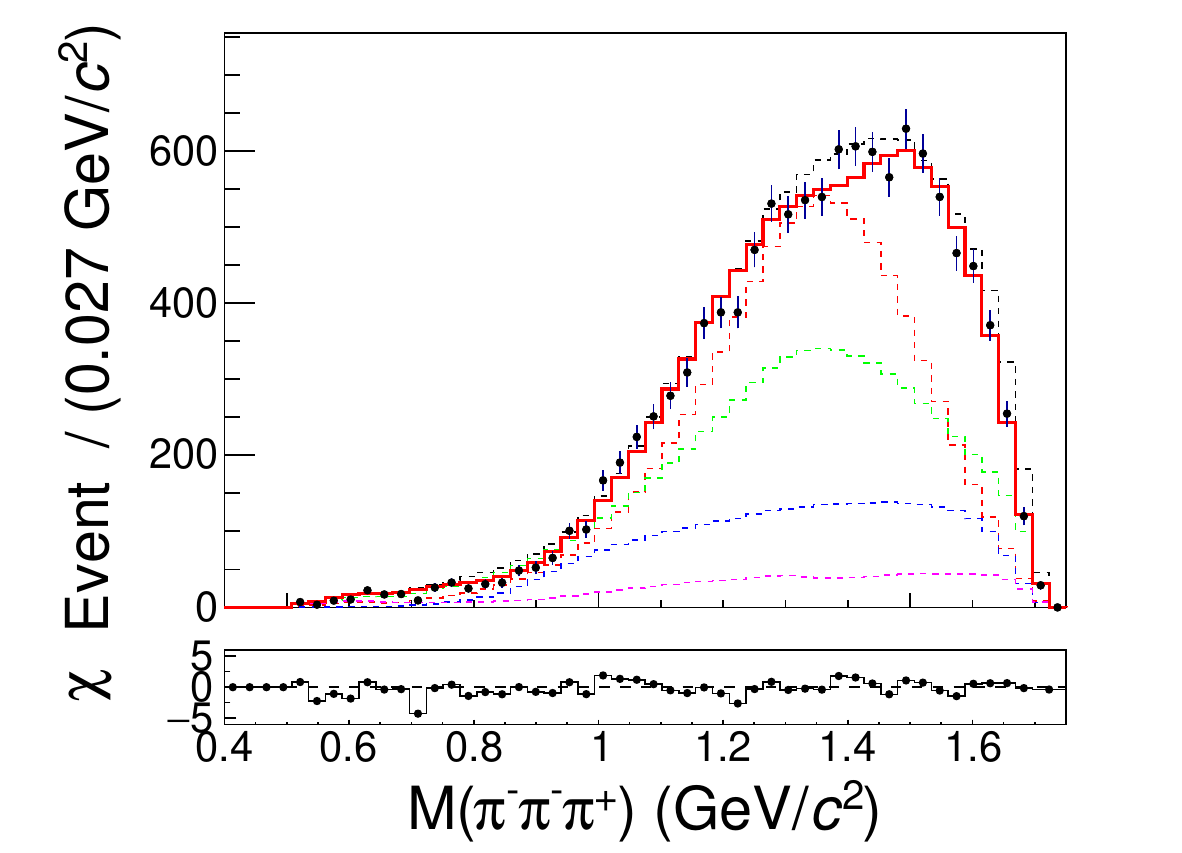}
  \end{overpic}
   \begin{overpic}[width=0.32\textwidth]{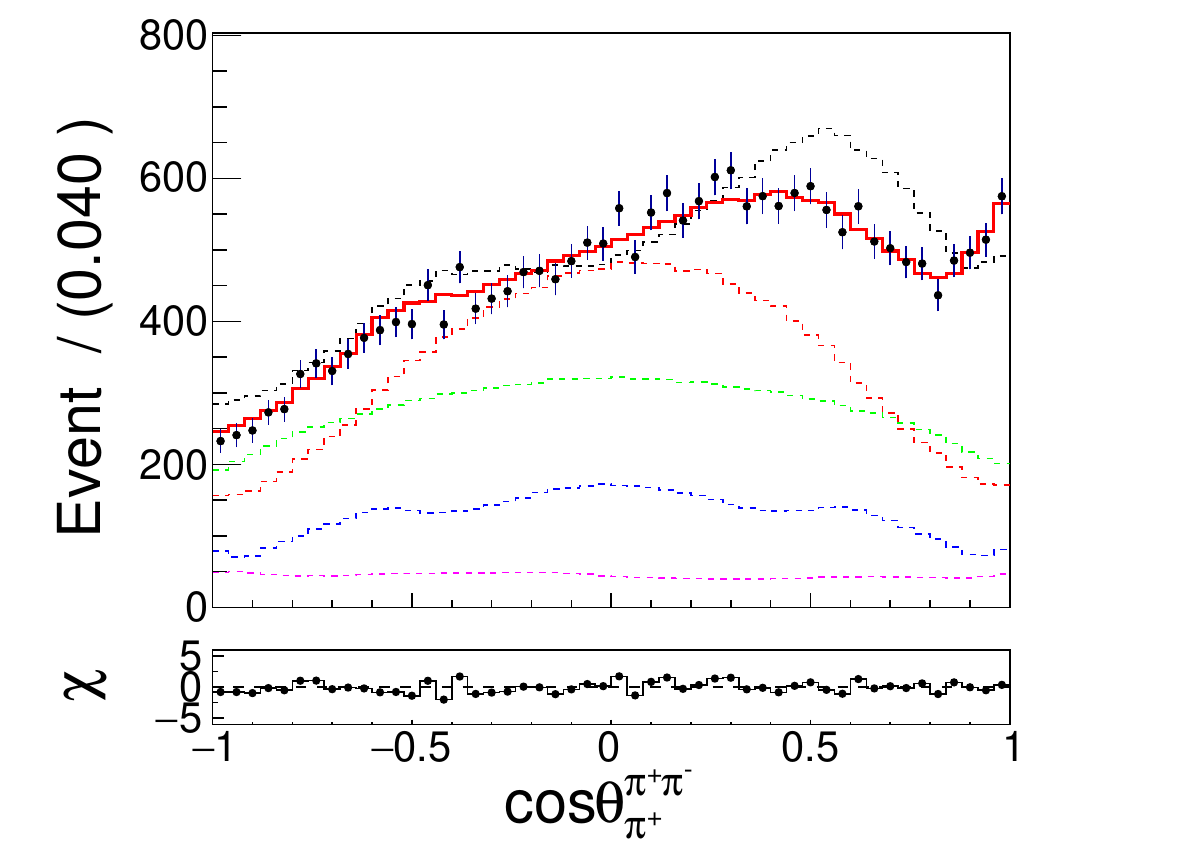}
  \end{overpic}
  \begin{overpic}[width=0.32\textwidth]{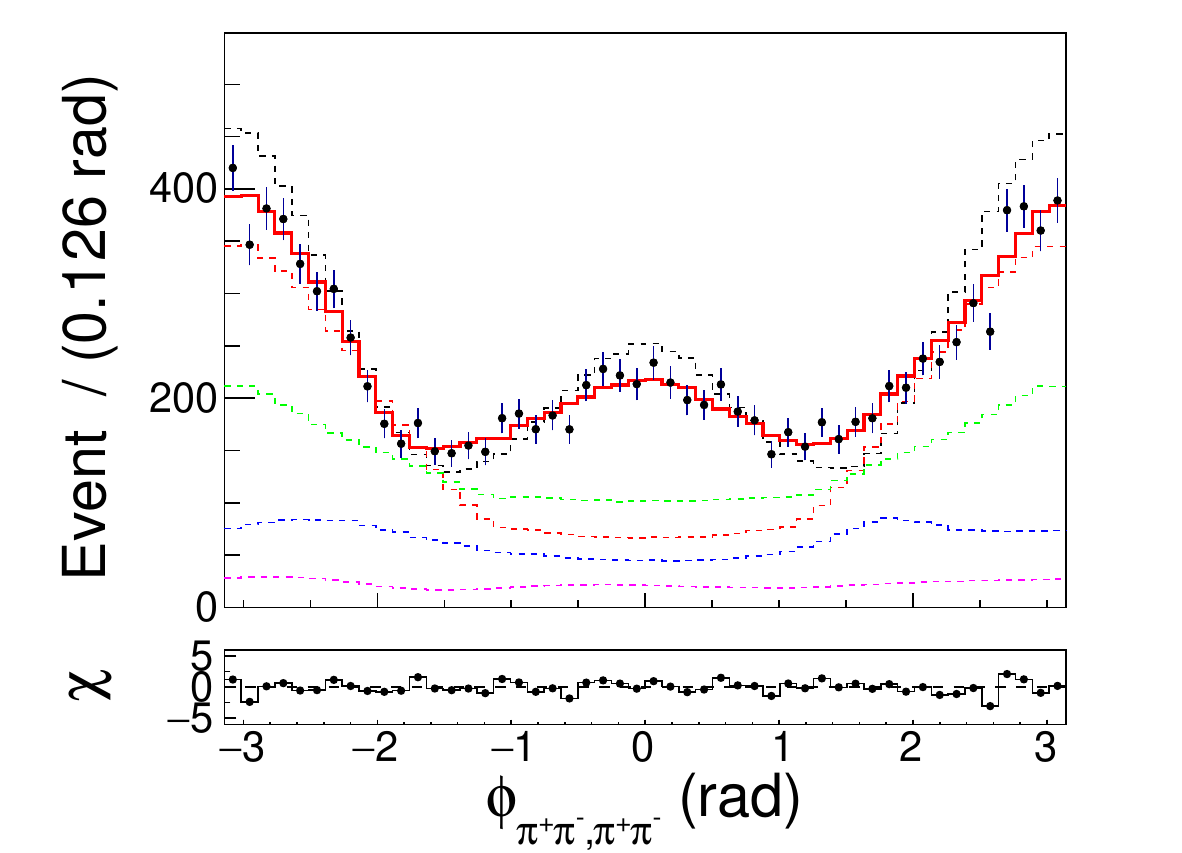} 
  \end{overpic}
\begin{overpic}[width=0.32\textwidth]{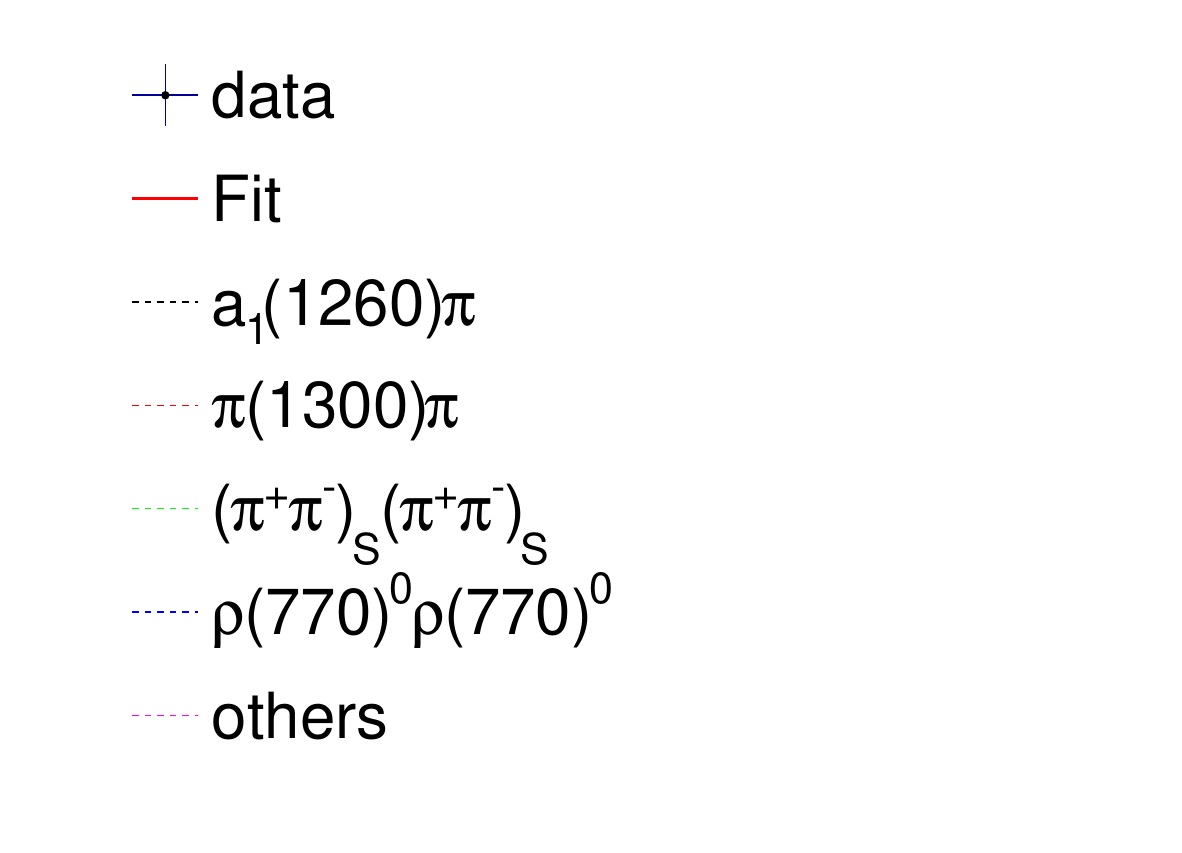}
  \end{overpic}
  \caption{The mass and angular distributions for $D^0\to \pi^+\pi^-\pi^+\pi^-$, where $\theta_{a}^{ab}$ is the helicity angle of $a$ in the $ab$ system, $\phi_{ab,cd}$ is the angle between the decay planes of $ab$ and $cd$ systems in the $D^0$ rest frame.}
  \label{fig:fit_2pip2pim}
\end{figure*}

\begin{figure*}[htbp]
  \centering
 \begin{overpic}[width=0.32\textwidth]{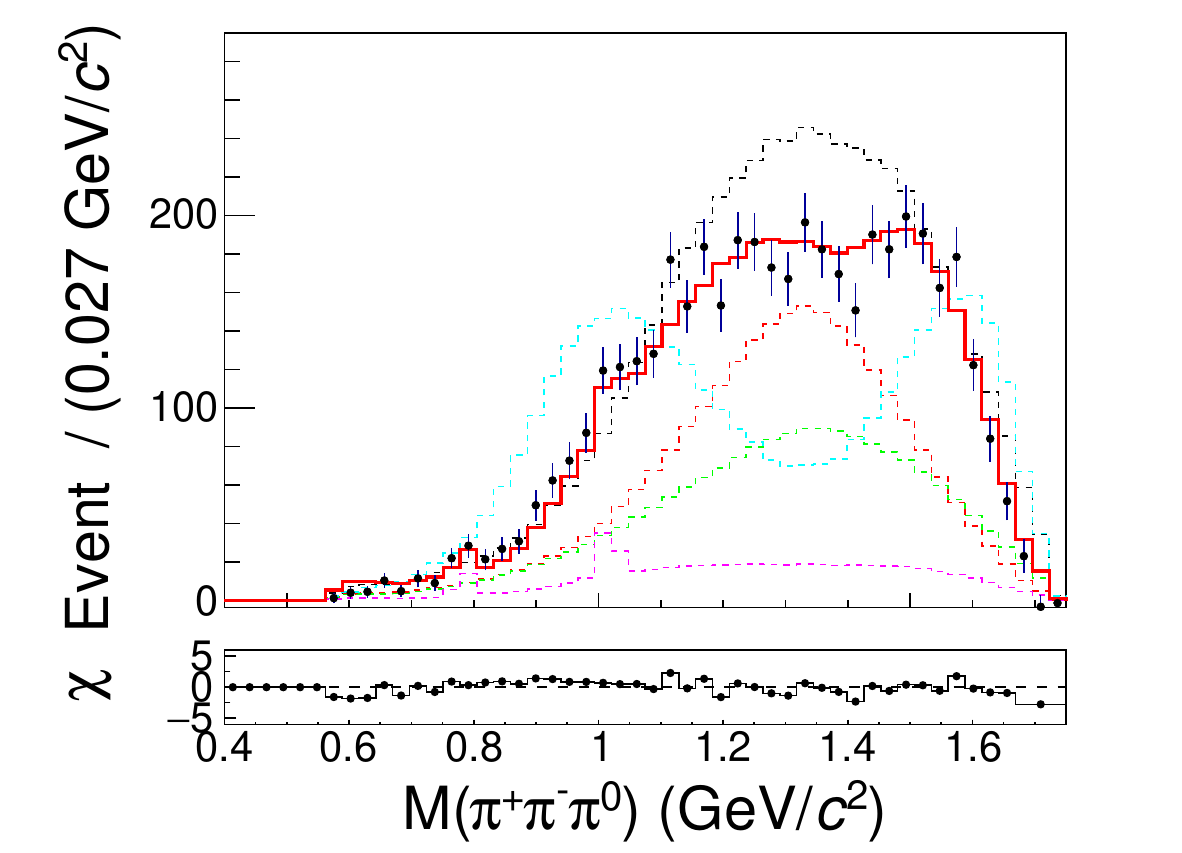}
  \end{overpic}
  \begin{overpic}[width=0.32\textwidth]{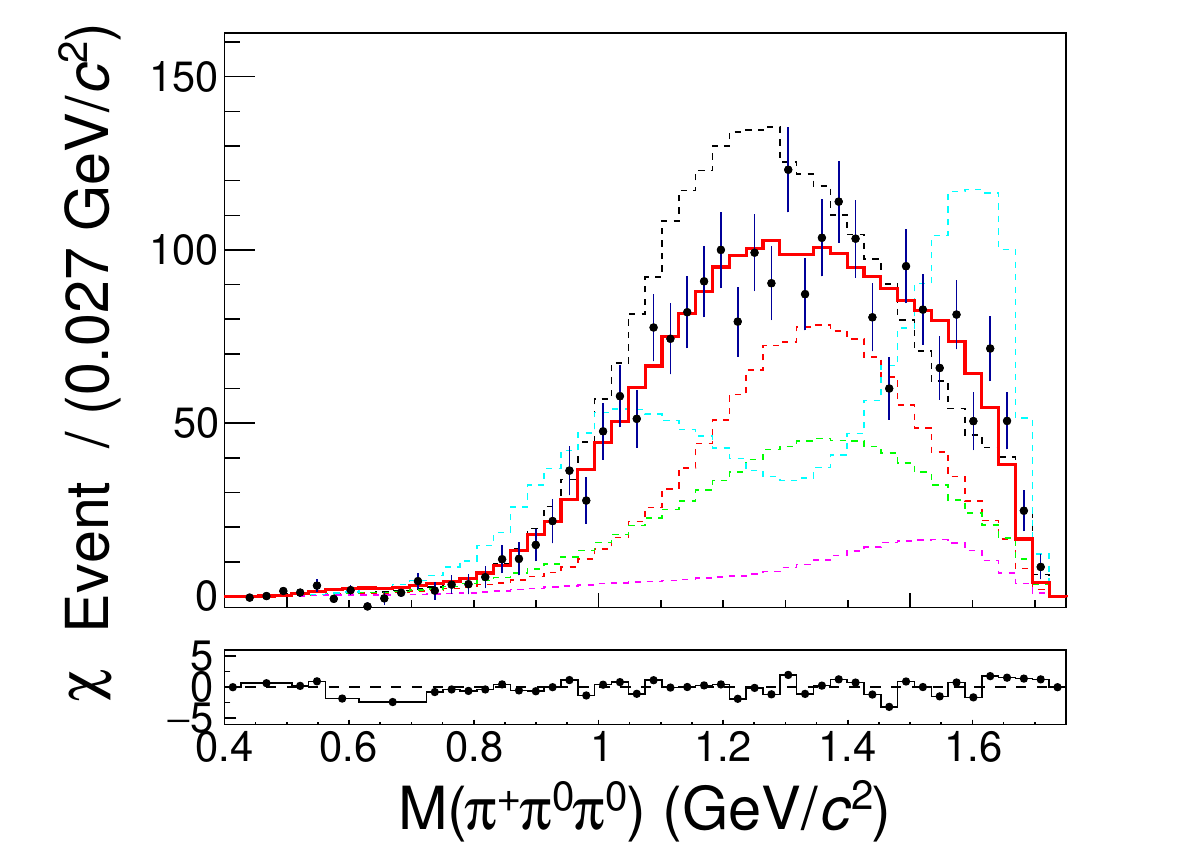}
  \end{overpic}
\begin{overpic}[width=0.32\textwidth]{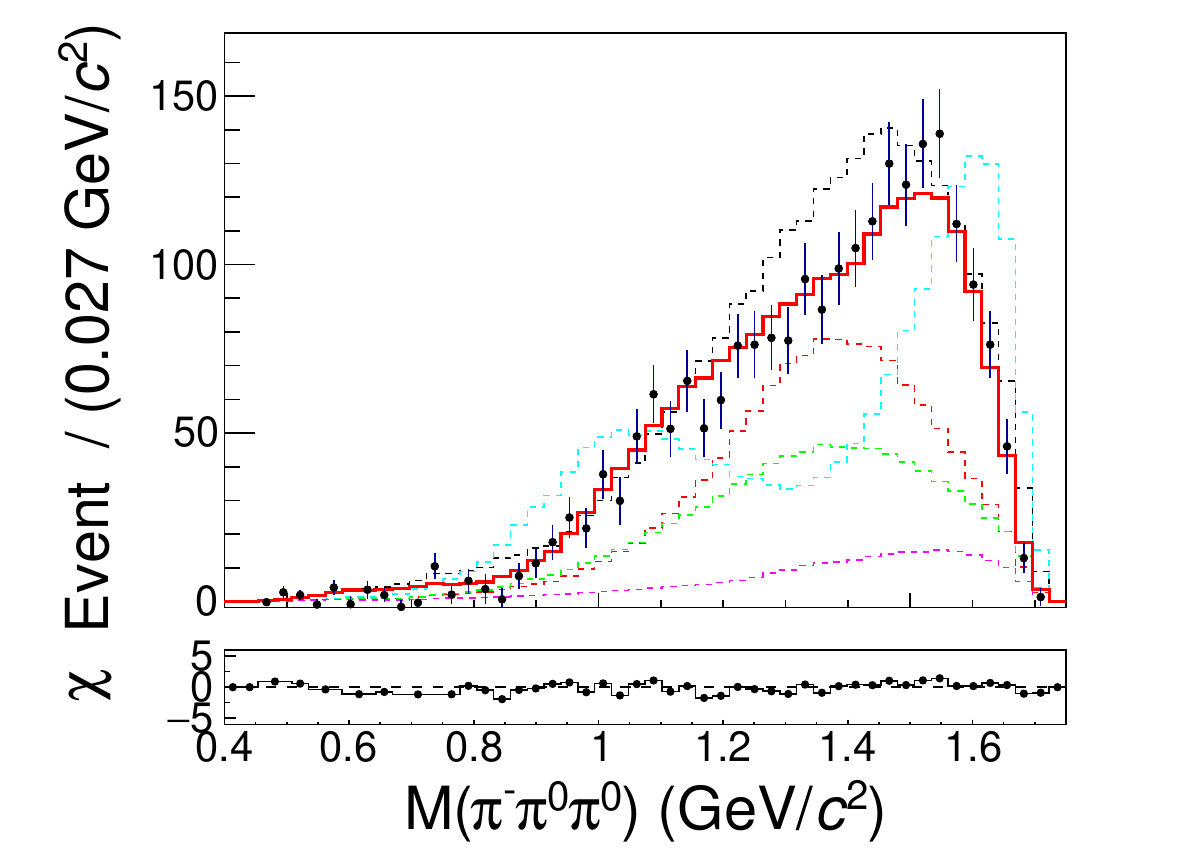}
\end{overpic}
   \begin{overpic}[width=0.32\textwidth]{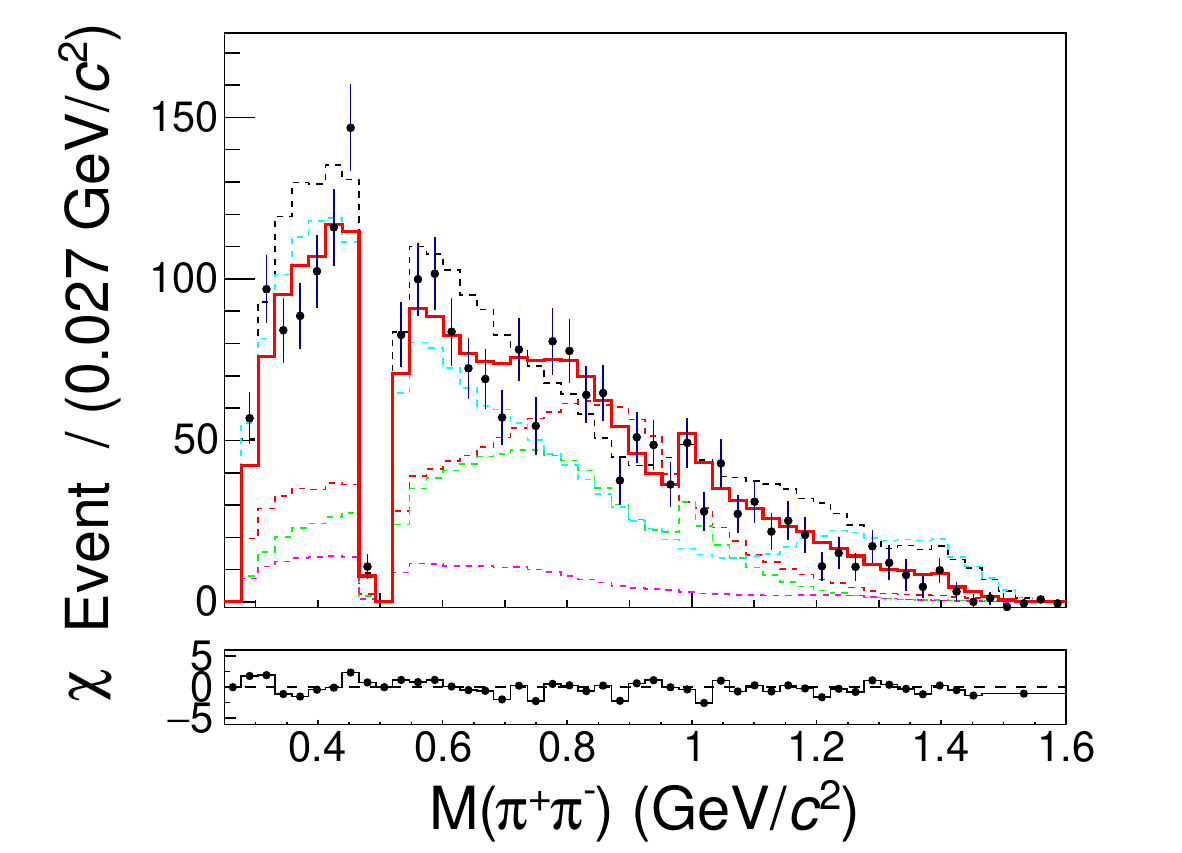}
  \end{overpic}
 \begin{overpic}[width=0.32\textwidth]{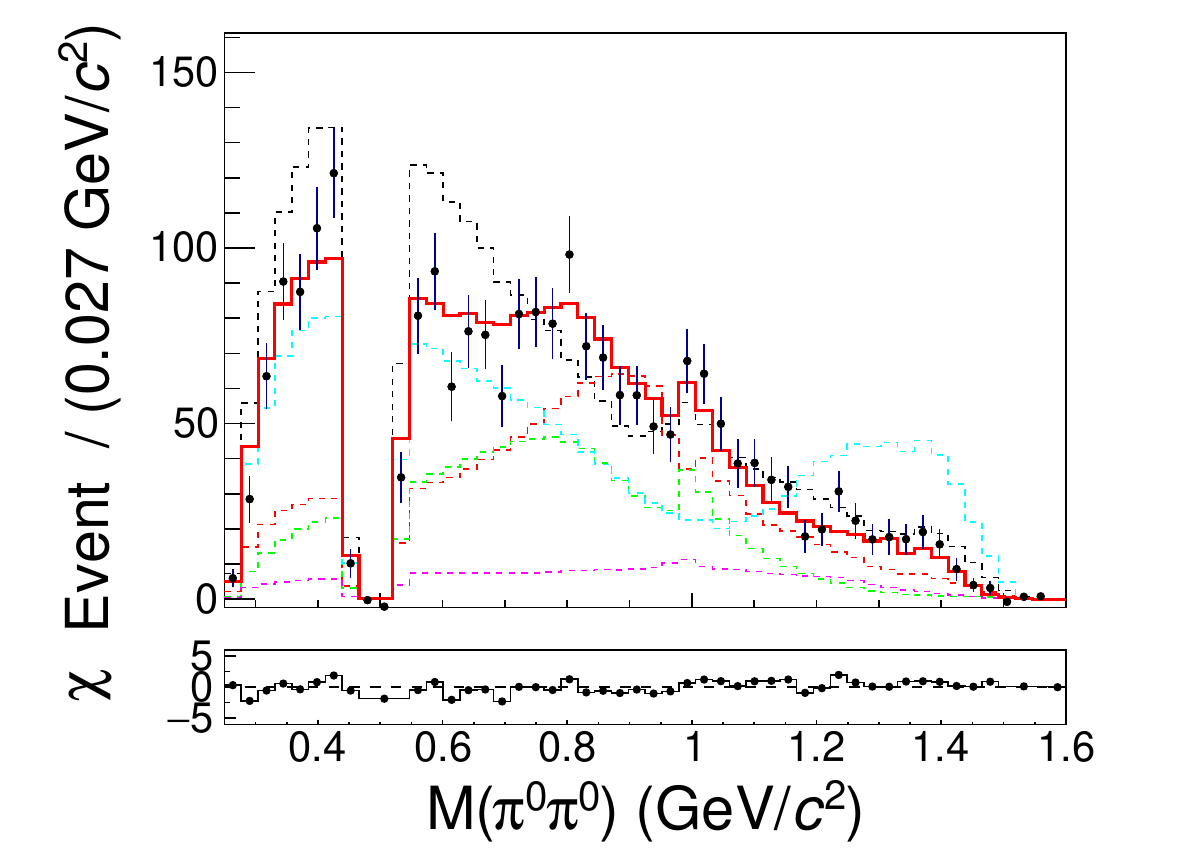}
  \end{overpic}
 \begin{overpic}[width=0.32\textwidth]{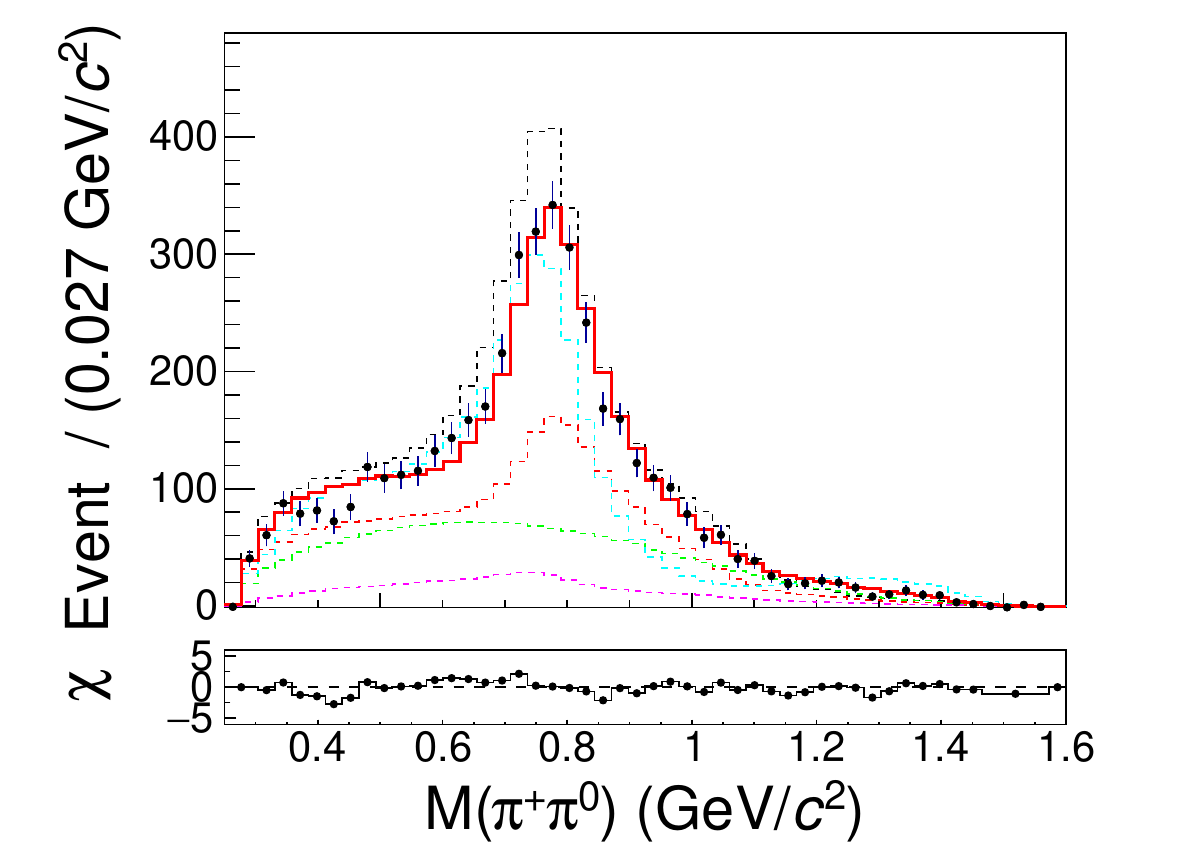}
  \end{overpic}
 \begin{overpic}[width=0.32\textwidth]{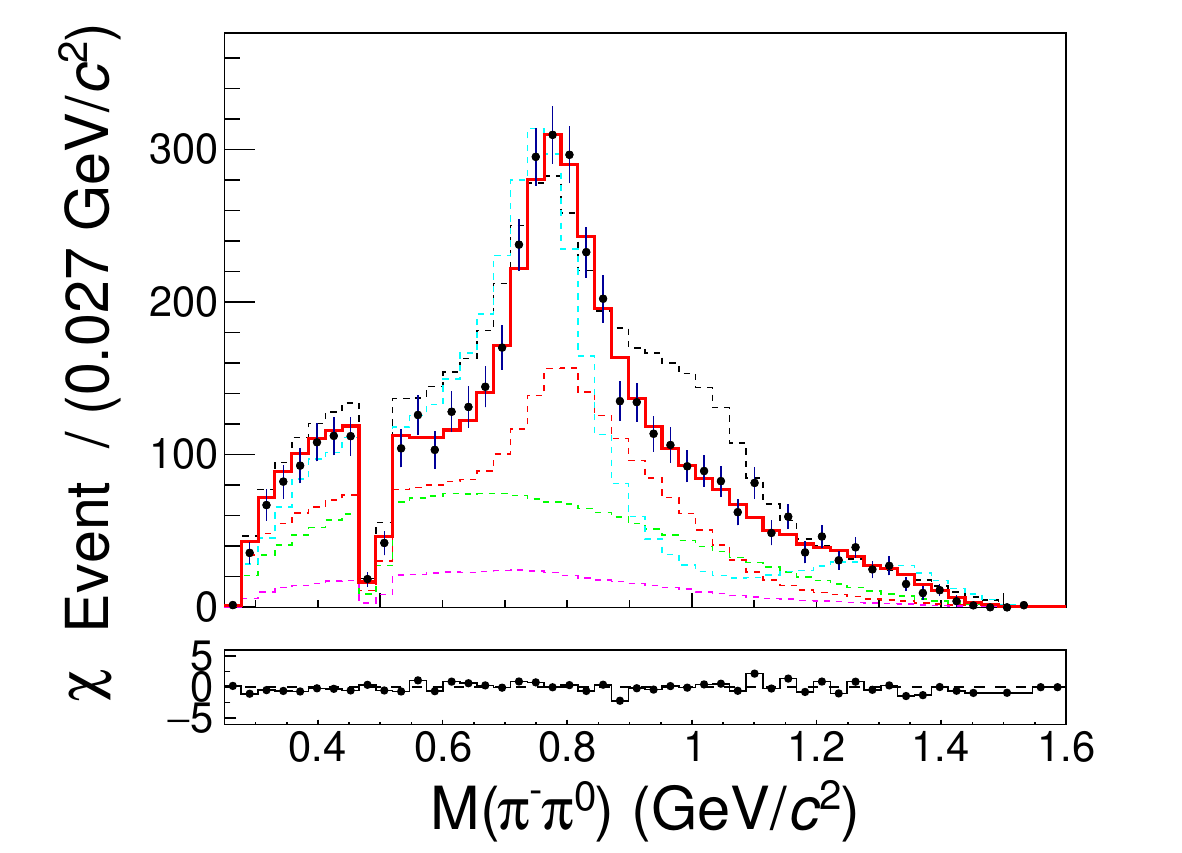}
  \end{overpic}
\begin{overpic}[width=0.32\textwidth]{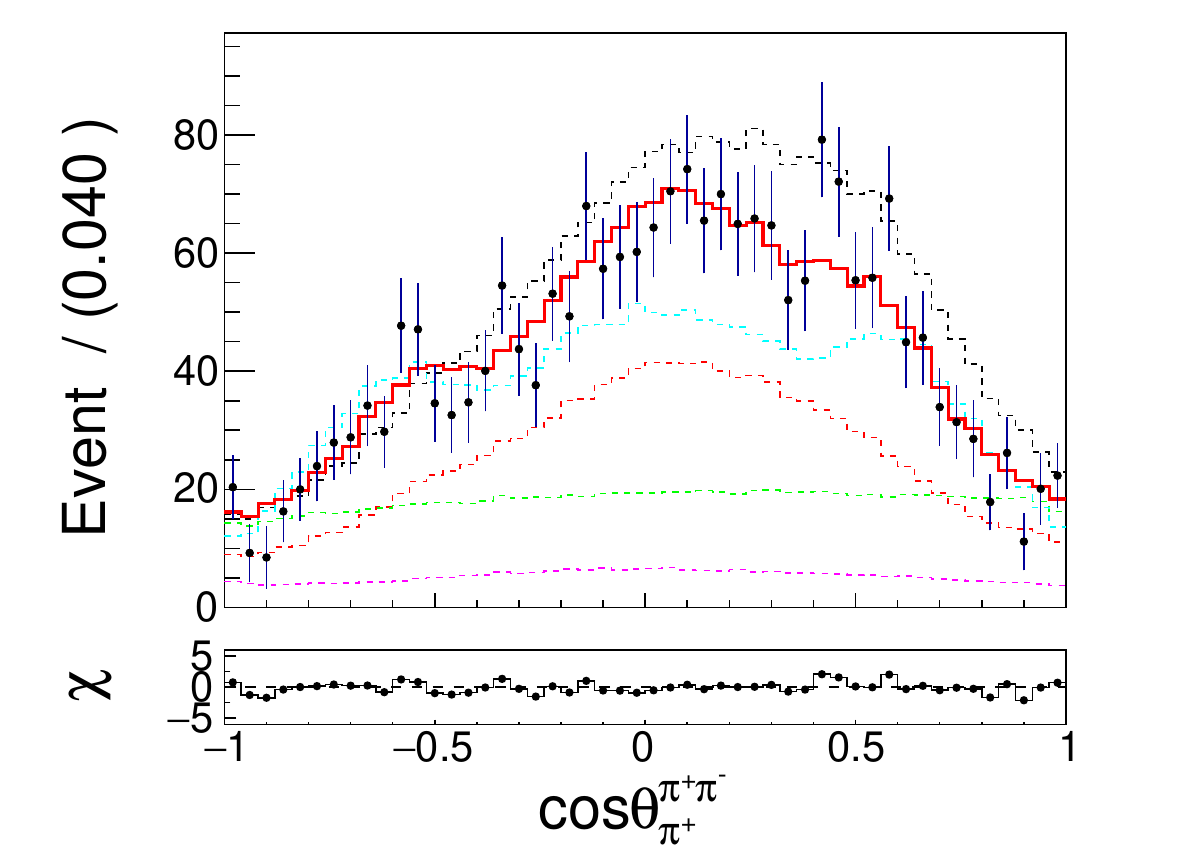}
  \end{overpic}
\begin{overpic}[width=0.32\textwidth]{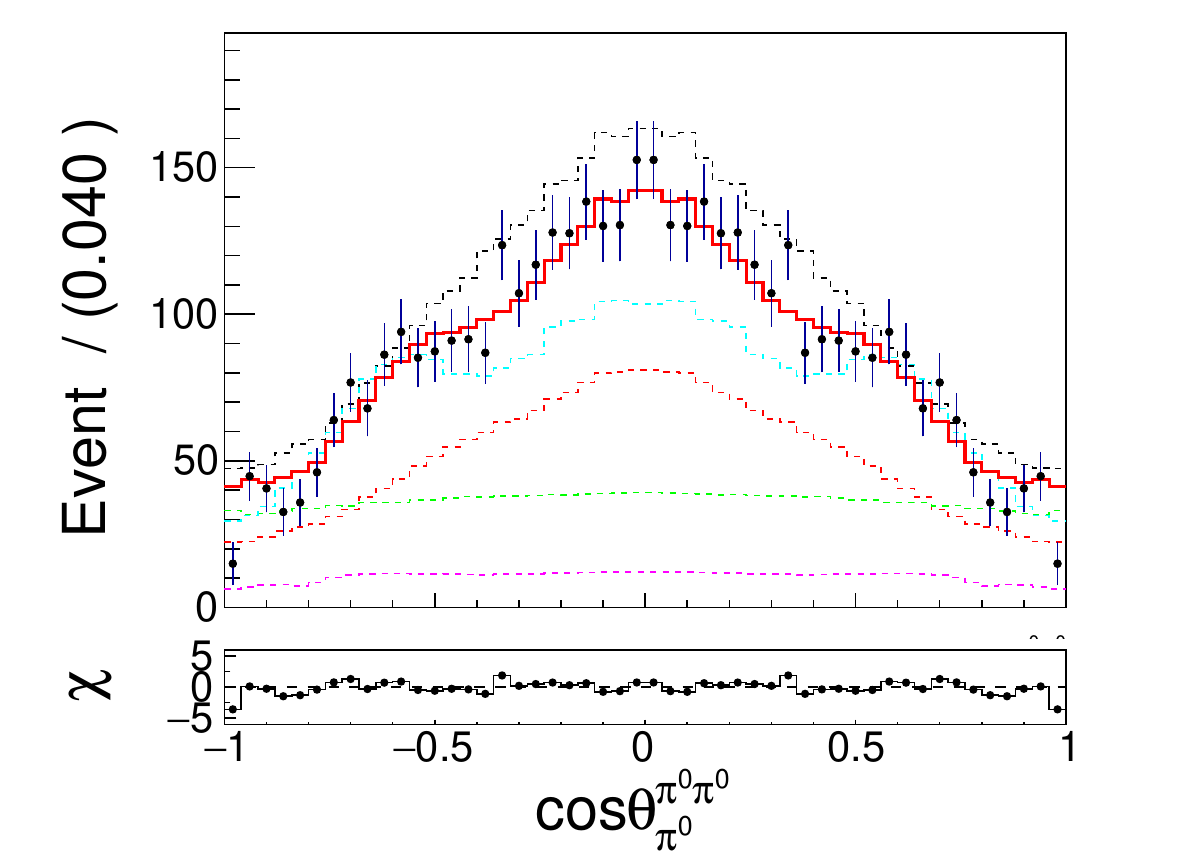}
  \end{overpic}
\begin{overpic}[width=0.32\textwidth]{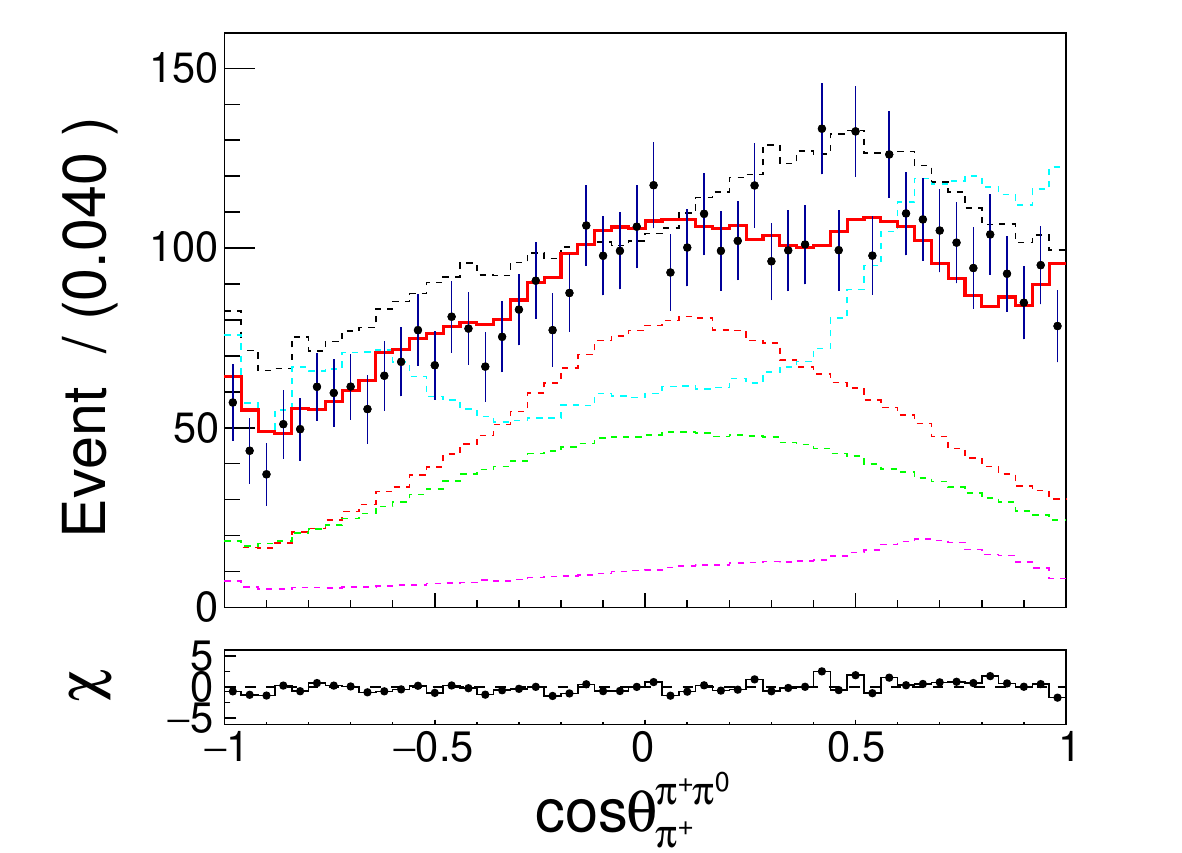}
  \end{overpic}
\begin{overpic}[width=0.32\textwidth]{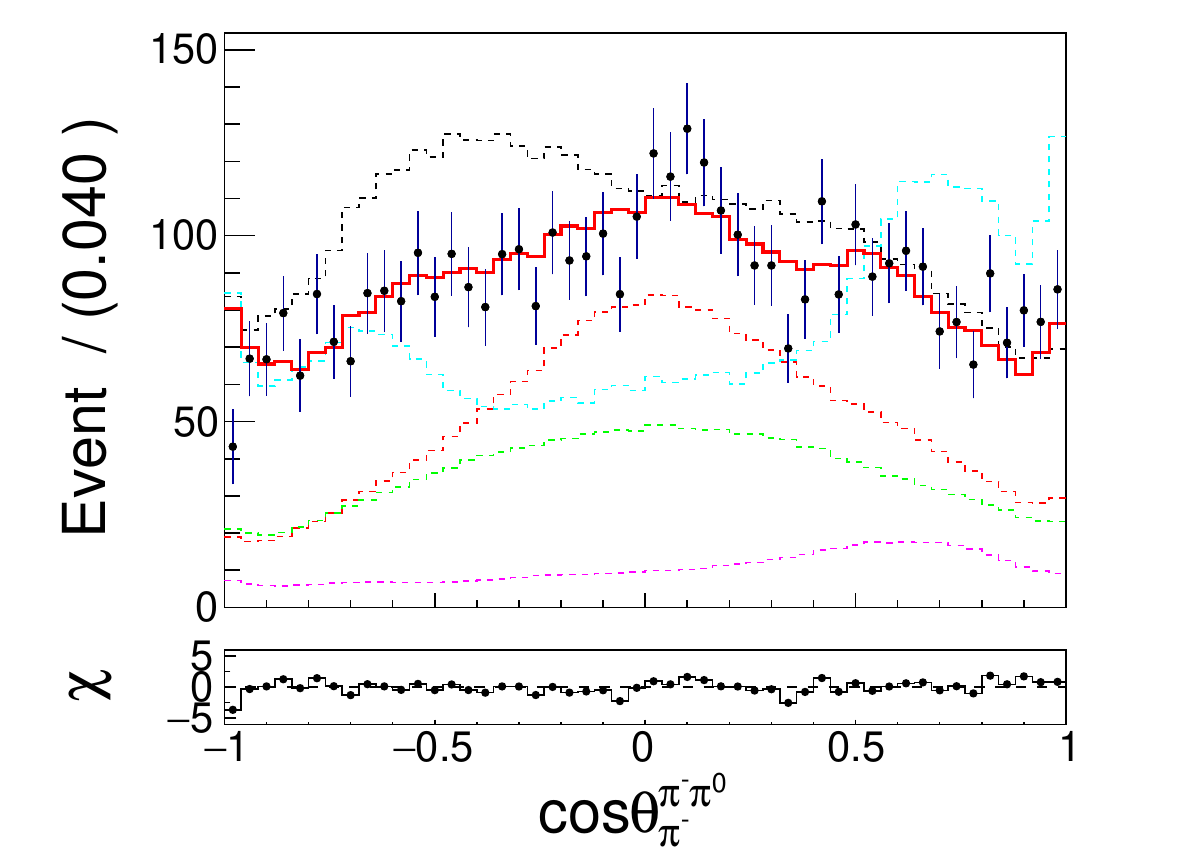}
  \end{overpic}
\begin{overpic}[width=0.32\textwidth]{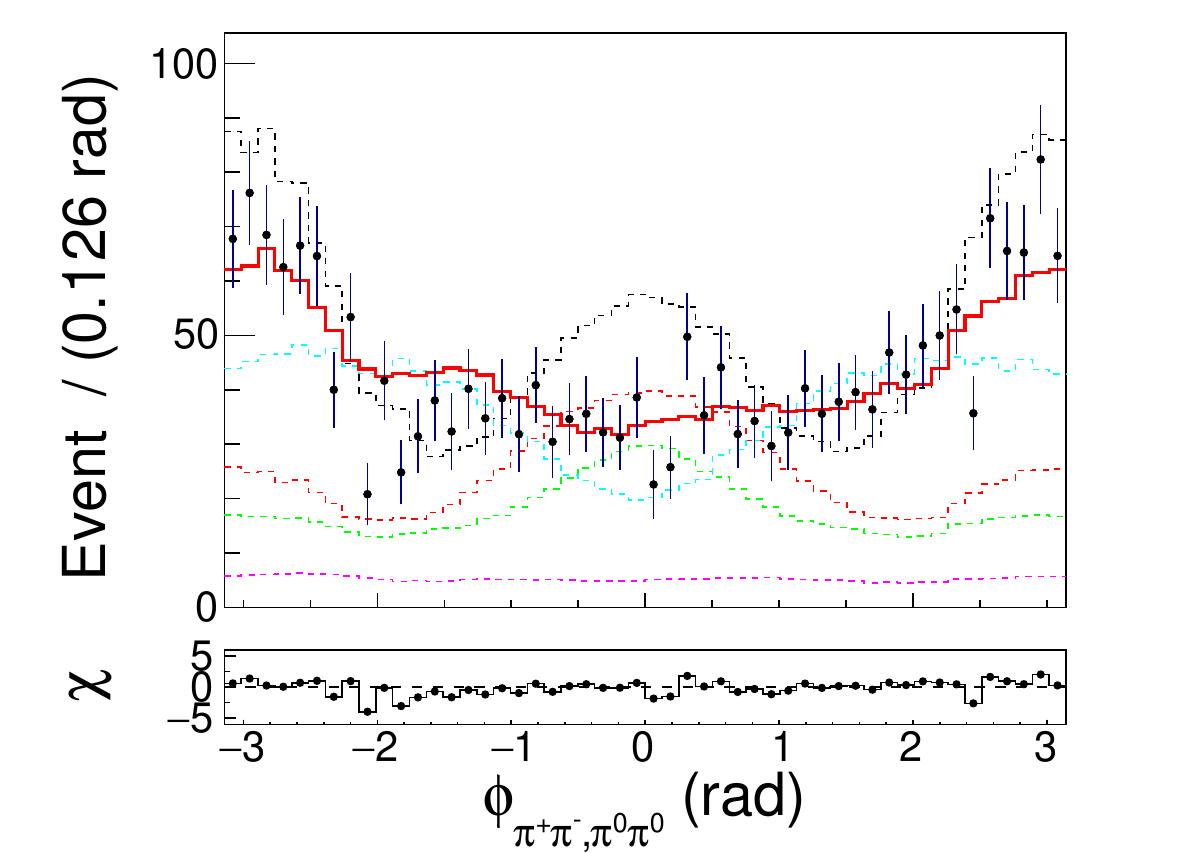}
  \end{overpic}
\begin{overpic}[width=0.32\textwidth]{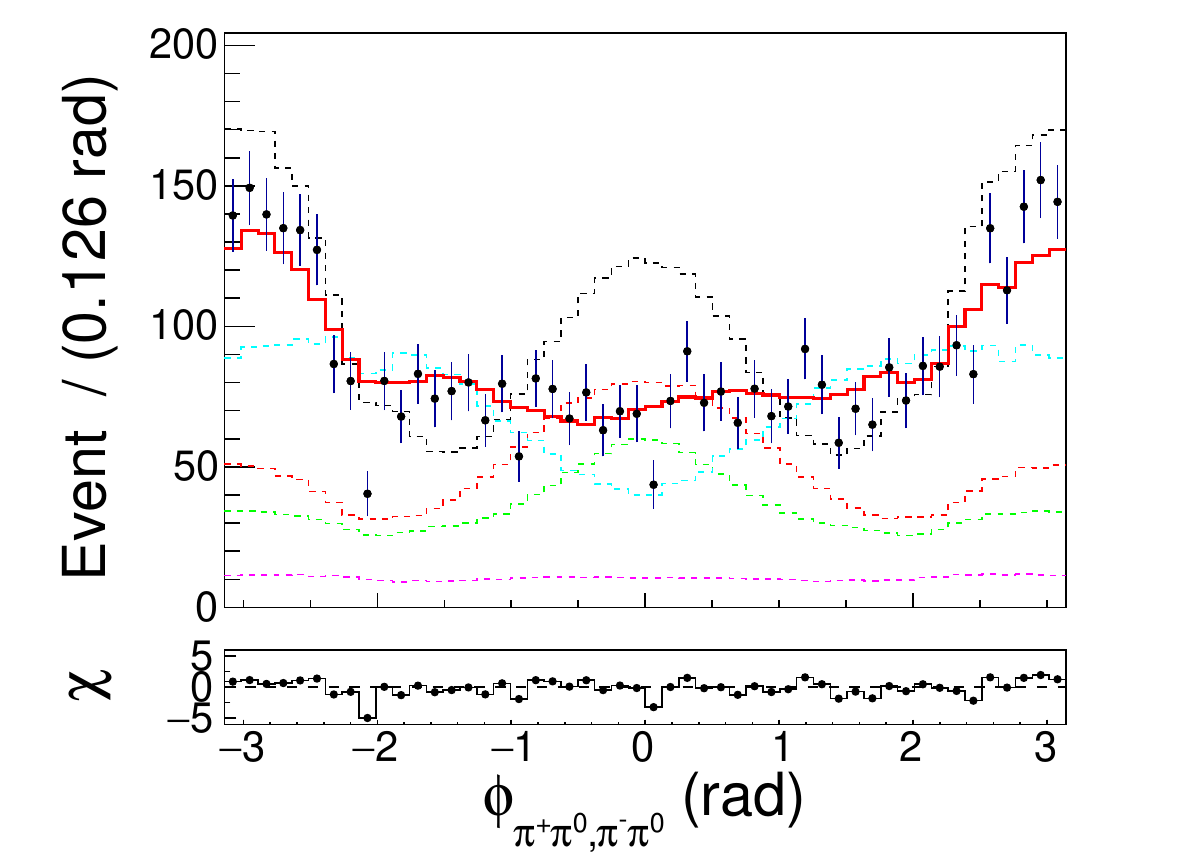}
  \end{overpic}
\begin{overpic}[width=0.32\textwidth]{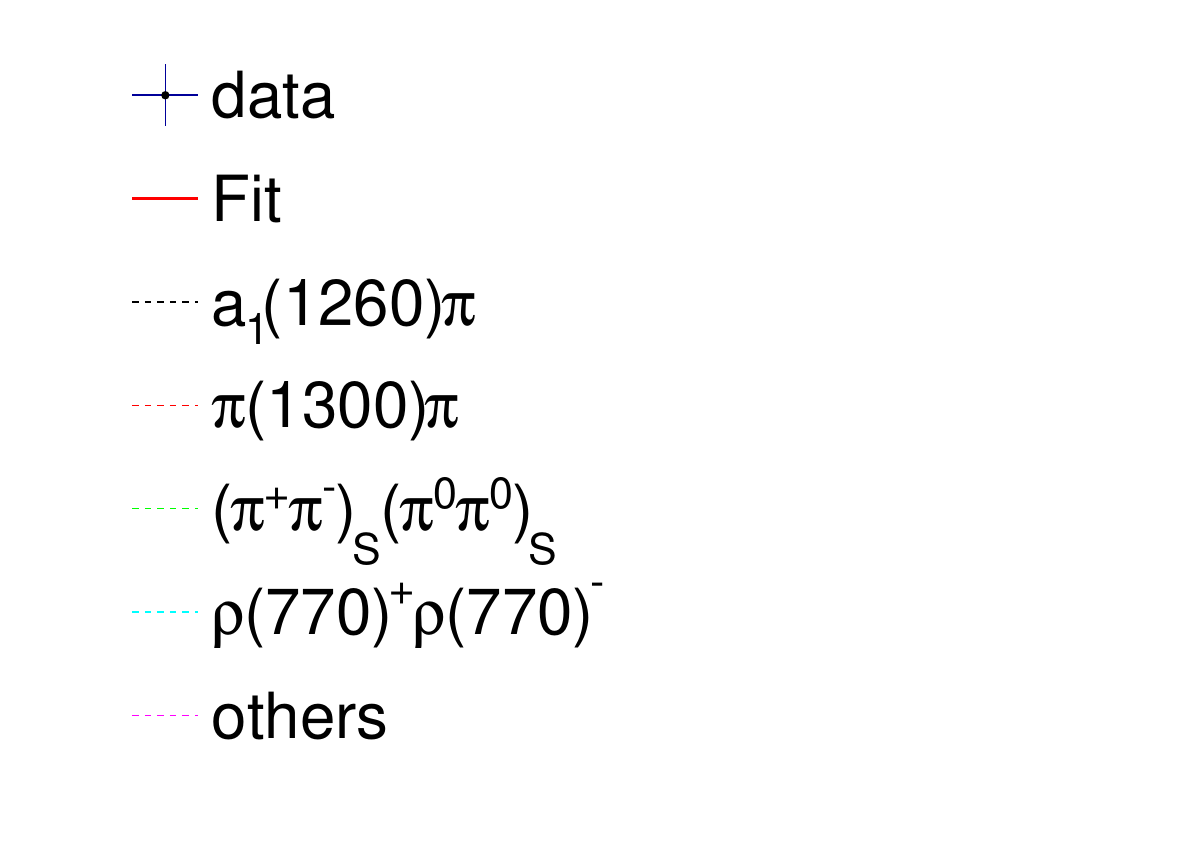}
\end{overpic}
\begin{overpic}[width=0.32\textwidth]{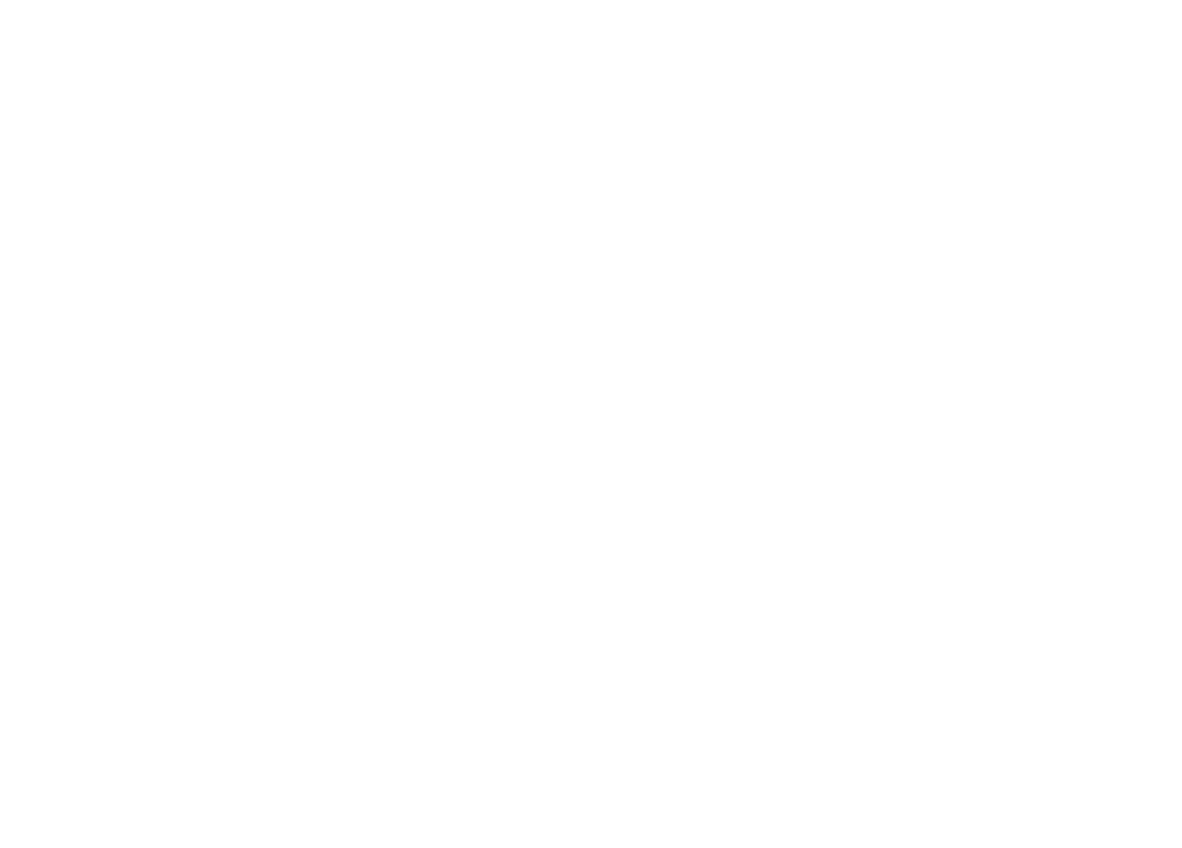}
\end{overpic}
  \caption{The mass and angular distributions in $D^0\to \pi^{+}\pi^{-}\pi^{0}\pi^{0}$, where $\theta_{a}^{ab}$ is the helicity angle of $a$ in the $ab$ system and $\phi_{ab,cd}$ is the angle between the decay planes of $ab$ and $cd$ systems in the $D^0$ rest frame.}
\label{fig:fit_pippim2pi0}
\end{figure*}

With the nominal amplitude model, the fit fraction (FF) of a specific amplitude $i$ is calculated as 
 \begin{eqnarray}
{\rm FF}_i = \frac{\int |\Lambda_iU_i(p)|^2\mathrm{d}\Phi}{\int |\sum_j \Lambda_j U_j(p)|^2\mathrm{d}\Phi}.
  \end{eqnarray}
The FF of the interference between two amplitudes $i$ and $j$ ($j\neq i$) is calculated as
 \begin{eqnarray}
{\rm FF}_{ij} = \frac{\int 2{\rm Re}[\Lambda_iU_i(p)\Lambda^{*}_jU^{*}_j(p)]\mathrm{d}\Phi}{\int |\sum_k \Lambda_k U_k(p)|^2\mathrm{d}\Phi}.
  \end{eqnarray}
The $CP$-even fraction $F_+$ is calculated as
 \begin{eqnarray}
   F_{+}^{f} = \frac{\int |A^{+}_{f}(p)|^2\mathrm{d}\Phi}{\int |A^{+}_{f}(p)|^2+|A^{-}_{f}(p)|^2 \mathrm{d}\Phi},
  \end{eqnarray}
where $A^{\pm}_{f}(p)=\frac{1}{\sqrt{2}}[A_{f}(p) ~\pm~\bar{A}_{f}(p)]$ is the amplitude of the $D^0\to f$ decay in a $CP$-even or $CP$-odd state. 
The FF results of different amplitudes are summarized in Tables~\ref{tab:fitfrac}, \ref{tab:3pires} and \ref{tab:respara}. The resonant parameters, masses, and widths of $a_1(1260)$ and $\pi(1300)$ are determined by the parameter scans as shown in Fig.~\ref{fig:respara}, which are compatible with the values from PDG.
The fit fractions of the interference terms are summarized in Table~\ref{tab:inter}. The fit results show large interferences among the dominant intermediate processes $D^{0}\to a_{1}(1260)\pi$, $D^{0}\to\pi(1300)\pi$, $D^{0}\to\rho(770)\rho(770)$, and $D^{0}\to2(\pi\pi)_{S}$. 
The results for $F_+$ obtained in this analysis are presented in Table~\ref{tab:cp+frac}, which show good agreement with other measurements. 
In all above tables, the mean values are obtained based on the output from the MINUIT fit,
while the corresponding statistical uncertainties are estimated by the bootstrap method~\cite{10.1214/aos/1176344552} based on data, since the weighted maximum likelihood fit is used, and the statistical uncertainties given by the inverse second derivative of the negative logarithmic likelihood are no longer asymptotically correct~\cite{Langenbruch:2019nwe}. In the bootstrap method, the bootstrap samples are generated by repeatedly resampling the data set with replacement for thousand times, and both sPlot and amplitude analysis are performed for these samples. The width of the distribution of estimated parameter values is used as estimator for the parameter uncertainty.
The comparisons between data and MC projection based on the nominal model for various invariant mass and angle distributions are shown in Figs.~\ref{fig:fit_2pip2pim} and \ref{fig:fit_pippim2pi0}.

\begin{table*}[htbp]
\caption{The fit parameters, FFs and significances of individual amplitudes, where the first uncertainties are statistical, the second are the experimental systematic uncertainties and the third are the model-dependent systematic uncertainties.}
\label{tab:fitfrac}
\begin{center}
\begin{lrbox}{\tablebox}
\begin{tabular}{l|c|c|c|c|c}
\hline
\hline
\multirow{2}{*}{Amplitude}			&\multirow{2}{*}{Magnitude}		&\multirow{2}{*}{Phase (rad)}		&\multicolumn{2}{c|}{FF (\%)}		&\multirow{2}{*}{Significance ($\sigma$)}\\	
\cline{4-5}	
							&							&							&$\pi^{+}\pi^{-}\pi^{+}\pi^{-}$		&$\pi^{+}\pi^{-}\pi^{0}\pi^{0}$		&			\\
\hline
$a_{1}(1260)^{+}\pi^{-}$			&100(fixed)					&0(fixed)						&82.2~$\pm$~3.3~$\pm$~2.3~$\pm$~15.8	&57.4~$\pm$~2.7~$\pm$~3.0~$\pm$~6.5		&$>$10	\\
$a_{1}(1260)^{-}\pi^{+}$			& $35.3~\pm~2.7~\pm~0.8~\pm~4.6$	& $\phantom{+}0.23~\pm~0.07~\pm~0.02~\pm~0.16$		&10.3~$\pm$~1.5~$\pm$~0.3~$\pm$~2.5	&7.2~$\pm$~1.1~$\pm$~0.2~$\pm$~2.2		&	$>$10	\\
$a_{1}(1260)^{0}\pi^{0}$			& $50.9~\pm~3.1~\pm~0.4~\pm~4.7$	& $-2.99~\pm~0.06~\pm~0.08~\pm~0.14$				&-							&32.9~$\pm$~3.2~$\pm$~1.6~$\pm$~8.3		&$>$10	\\
$a_{1}(1420)^{+}\pi^{-}$			& $19.0~\pm~3.6~\pm~1.3~\pm~3.9$	& $\phantom{+}2.70~\pm~0.18~\pm~0.05~\pm~1.09$		&0.6~$\pm$~0.2~$\pm$~0.0~$\pm$~0.2	&0.3~$\pm$~0.1~$\pm$~0.0~$\pm$~0.1		&	6.0	\\	
$a_{1}(1640)^{+}\pi^{-}$			& $20.1~\pm~3.0~\pm~2.6~\pm~5.8$	& $-2.07~\pm~0.16~\pm~0.02~\pm~0.28$				&1.7~$\pm$~0.5~$\pm$~0.4~$\pm$~0.8	&1.1~$\pm$~0.3~$\pm$~0.2~$\pm$~0.6		&	7.3	\\
$a_{1}(1640)^{-}\pi^{+}$			& $10.5~\pm~2.8~\pm~0.6~\pm~3.8$	& $-1.26~\pm~0.29~\pm~0.23~\pm~0.50$				&0.5~$\pm$~0.3~$\pm$~0.0~$\pm$~0.4	&0.3~$\pm$~0.2~$\pm$~0.0~$\pm$~0.2		&	5.2	\\
$a_{2}(1320)^{+}\pi^{-}$			& $0.23~\pm~0.07~\pm~0.03~\pm~0.05$	& $-2.92~\pm~0.30~\pm~0.14~\pm~0.23$				&0.2~$\pm$~0.1~$\pm$~0.0~$\pm$~0.1	&0.2~$\pm$~0.1~$\pm$~0.0~$\pm$~0.1		& 4.6	\\
$a_{2}(1320)^{-}\pi^{+}$			& $0.30~\pm~0.05~\pm~0.01~\pm~0.04$	& $-0.47~\pm~0.21~\pm~0.06~\pm~0.15$				&0.4~$\pm$~0.1~$\pm$~0.0~$\pm$~0.1	&0.3~$\pm$~0.1~$\pm$~0.0~$\pm$~0.1		& 6.4\\
$h_{1}(1170)^{0}\pi^{0}$			& $9.7~\pm~2.2~\pm~1.5~\pm~3.6$		& $-0.59~\pm~0.27~\pm~0.09~\pm~0.35$				&-							&1.3~$\pm$~0.6~$\pm$~0.4~$\pm$~1.0 		&	6.5	\\
$\pi(1300)^{+}\pi^{-}$			& $76.3~\pm~3.6~\pm~4.6~\pm~4.3$	& $-2.325~\pm~0.044~\pm~0.038~\pm~0.297$				&32.3~$\pm$~2.6~$\pm$~1.6~$\pm$~4.2	&15.6~$\pm$~1.4~$\pm$~1.3~$\pm$~2.2		&$>$10	\\
$\pi(1300)^{-}\pi^{+}$			& $65.1~\pm~3.4~\pm~3.3~\pm~3.9$	& $-2.631~\pm~0.045~\pm~0.083~\pm~0.208$				&23.5~$\pm$~2.3~$\pm$~0.5~$\pm$~3.9	&11.4~$\pm$~1.1~$\pm$~0.6~$\pm$~2.3		&$>$10	\\
$\pi(1300)^{0}\pi^{0}$			& $61.1~\pm~3.2~\pm~3.3~\pm~4.0$	& $\phantom{+}0.61~\pm~0.05~\pm~0.08~\pm~0.29$		&-							&23.2~$\pm$~2.8~$\pm$~1.4~$\pm$~3.1		&$>$10	\\
$\pi_{2}(1670)^{0}\pi^{0}$			& $12.2~\pm~1.5~\pm~1.5~\pm~2.1$	& $-1.11~\pm~0.14~\pm~0.13~\pm~0.36$				&-							&1.1~$\pm$~0.2~$\pm$~0.2~$\pm$~0.3		&6.9	\\	
$\rho(770)^{0}\rho(770)^{0}$		& - 							& - 							&28.0~$\pm$~1.9~$\pm$~0.6~$\pm$~3.0	&- 					& $>$10 \\
$~~[S]$						& $6.1~\pm~1.1~\pm~0.3~\pm~1.4$		& $-3.10~\pm~0.17~\pm~0.11~\pm~0.50$				&1.7~$\pm$~0.6~$\pm$~0.1~$\pm$~0.5	&-					&	6.5	\\		
$~~[P] $						& $6.17~\pm~0.36~\pm~0.13~\pm~0.58$	& $\phantom{+}1.62~\pm~0.07~\pm~0.02~\pm~0.09$		&9.8~$\pm$~1.0~$\pm$~0.4~$\pm$~0.8	&- 					&	$>$10	\\	
$~~[D] $						& $4.54~\pm~0.22~\pm~0.06~\pm~0.34$	& $-3.06~\pm~0.05~\pm~0.01~\pm~0.20$				&23.1~$\pm$~2.1~$\pm$~0.8~$\pm$~2.3	&- 					&	$>$10\\	
$\rho(770)^{0}\rho(1450)^{0}$		& - 						& - 							&2.5~$\pm$~0.9~$\pm$~0.0~$\pm$~1.2	&- 					& 8.0 \\	
$~~[P] $						& $13.9~\pm~2.5~\pm~0.7~\pm~1.5$	& $\phantom{+}0.68~\pm~0.20~\pm~0.09~\pm~0.15$		&1.0~$\pm$~0.4~$\pm$~0.1~$\pm$~0.6	&-					&	6.4	\\	
$~~[D] $						& $5.6~\pm~1.3~\pm~0.3~\pm~1.2$		& $\phantom{+}3.08~\pm~0.20~\pm~0.07~\pm~0.43$		&1.5~$\pm$~0.9~$\pm$~0.2~$\pm$~1.1	&-					&	5.0	\\
$\rho(770)^{+}\rho(770)^{-}$		& - 						& - 							&-							&90.9~$\pm$~3.9~$\pm$~3.5~$\pm$~6.6		& $>$10 \\
$~~[S] $						& $13.7~\pm~1.2~\pm~0.6~\pm~1.3$	& $\phantom{+}3.03~\pm~0.09~\pm~0.09~\pm~0.17$		&-							&13.0~$\pm$~2.0~$\pm$~1.2~$\pm$~3.2				&	$>$10	\\	
$~~[P] $						& $7.10~\pm~0.36~\pm~0.23~\pm~0.35$	& $-1.69~\pm~0.07~\pm~0.02~\pm~0.14$				&-							&19.6~$\pm$~1.3~$\pm$~1.3~$\pm$~1.3			&	$>$10	\\	
$~~[D] $						& $4.59~\pm~0.22~\pm~0.05~\pm~0.24$	& $\phantom{+}0.06~\pm~0.05~\pm~0.02~\pm~0.09$		&-							&36.0~$\pm$~3.0~$\pm$~0.6~$\pm$~2.4				&	$>$10	\\
$\rho(770)^{+}\rho(1450)^{-}[D] $	& $8.1~\pm~1.7~\pm~1.4~\pm~2.7$	 	& $-1.01~\pm~0.18~\pm~0.11~\pm~0.26$				&-							&1.7~$\pm$~0.8~$\pm$~0.6~$\pm$~1.7				&	6.3	\\
$\rho(770)^{0}(\pi\pi)_{S} $		&	-					&		-					&2.7~$\pm$~0.6~$\pm$~0.3~$\pm$~1.7	&1.0~$\pm$~0.2~$\pm$~0.1~$\pm$~0.4				&	$>$10	\\	
$~~\beta_{1}$					& $8.4~\pm~3.6~\pm~1.1~\pm~1.6$		& $-1.68~\pm~0.50~\pm~0.26~\pm~0.29$				&	-						&	-				&	-	\\
$~~f_{\pi\pi}^{\rm prod}$			& $40.7~\pm~5.0~\pm~3.7~\pm~3.9$	& $-0.50~\pm~0.14~\pm~0.09~\pm~0.14$				&	-						&	-				&	-	\\
$~~f_{KK}^{\rm prod}$				& $121~\pm~25~\pm~11~\pm~16$		& $\phantom{+}1.73~\pm~0.23~\pm~0.03~\pm~0.16$		&	-						&	-				&	-	\\
$(\pi^{+}\pi^{-})_{S}(\pi\pi)_{S} $	& 	-					&	-		 				&62.8~$\pm$~4.6~$\pm$~0.5~$\pm$~9.7	&37.4~$\pm$~3.0~$\pm$~1.8~$\pm$~4.8		&	$>$10	\\	
$~~a_{1,1}$					& $2224~\pm~35~\pm~26~\pm~33$		& $-1.044~\pm~0.019~\pm~0.008~\pm~0.074$				&		-					&	-				&	-	\\
$~~a_{1,2}$					& $7287~\pm~62~\pm~46~\pm~93$		& $\phantom{+}1.727~\pm~0.009~\pm~0.004~\pm~0.062$		&		-					&	- 				& -\\
$~~b_{2,\pi\pi}$				& $8816~\pm~120~\pm~181\pm~90$		& $-1.107~\pm~0.014~\pm~0.002~\pm~0.082$				&		-					&	-				&	-	\\	
$~~c_{[\pi\pi,\pi\pi]}$			& $2433~\pm~96~\pm~67~\pm~90$		& $\phantom{+}1.796~\pm~0.043~\pm~0.042~\pm~0.044$		&		-					&	-				&	-	\\
$~~c_{[\pi\pi,KK]}$				& $5417~\pm~477~\pm~47~\pm~462$		& $\phantom{+}2.68~\pm~0.10~\pm~0.06~\pm~0.09$		&		-					&	-				&	-	\\
$f_{2}(1270)^{0}(\pi\pi)_{S} $		&	-					&		-					&1.8~$\pm$~0.4~$\pm$~0.0~$\pm$~1.3	&1.1~$\pm$~0.2~$\pm$~0.0~$\pm$~0.7		&9.1\\	
$~~f_{\pi\pi}^{\rm prod}$			& $18.3~\pm~1.8~\pm~1.1~\pm~3.5$ 	& $-1.39~\pm~0.10~\pm~0.04~\pm~0.20$				&	-						&	-				&-\\
$~~f_{KK}^{\rm prod}$				& $56~\pm~10~\pm~14~\pm~8$			& $\phantom{+}2.29~\pm~0.20~\pm~0.05~\pm~0.39$		&	-						&	-				&-	\\
$\omega(782)\pi^{0} $				& $1.58~\pm~0.30~\pm~0.05~\pm~0.13$	& $-0.50~\pm~0.44~\pm~0.15~\pm~0.23$				&- 							&0.9~$\pm$~0.4~$\pm$~0.0~$\pm$~0.2		&	6.1	\\	
$\phi(1020)\pi^{0} $				& $0.44~\pm~0.06~\pm~0.03~\pm~0.05$	& $\phantom{+}2.51~\pm~0.41~\pm~0.10~\pm~0.22$		&- 							&1.5~$\pm$~0.4~$\pm$~0.2~$\pm$~0.2		&	7.4	\\

\hline
 \hline
\end{tabular}
\end{lrbox}
  \resizebox{1.0\textwidth}{!}{\usebox{\tablebox}}
\end{center}
\end{table*}

\begin{table*}[htbp]
\caption{The fit parameters, FFs and significance for the three-body decays of $a_{1}(1260)$, $a_1(1420)$, $a_1(1640)$, $a_2(1320)$, $h_1(1170)$, $\pi(1300)$ and $\pi_2(1670)$. The first uncertainties are the statistical uncertainties, the second are the experimental systematic uncertainties and the third are the model-dependent systematic uncertainties.}
\label{tab:3pires}
\begin{center}
\begin{lrbox}{\tablebox}
\begin{tabular}{l|c|c|c|c|c|c}
\hline
\hline
\multirow{3}{*}{Amplitude}				&\multirow{3}{*}{Magnitude}	&\multirow{3}{*}{Phase (rad)}		&\multicolumn{3}{c|}{Relative FF (\%)}											&\multirow{3}{*}{Significance ($\sigma$)}\\
\cline{4-6}
								&						&							&$\pi^{+}\pi^{-}\pi^{+}\pi^{-}$	&\multicolumn{2}{c|}{$\pi^{+}\pi^{-}\pi^{0}\pi^{0}$}			&								\\
\cline{4-6}
								&						&							&charge $=~\pm~1$			&charge $=~\pm~1$			&charge $=0$				&								\\
\hline
$a_{1}(1260)\to \rho(770)\pi[S]$		&	1(fixed)				&	0(fixed)					&79.7~$\pm$~2.2~$\pm$~1.8~$\pm$~3.1	&	81.2~$\pm$~2.0~$\pm$~1.6~$\pm$~2.8		&	78.9~$\pm$~2.2~$\pm$~1.5~$\pm$~3.5	&	$>$10	\\
$a_{1}(1260)\to \rho(770)\pi[D]$		&	$0.060~\pm~0.009~\pm~0.002~\pm~0.022$		&$\phantom{+}0.01~\pm~0.17~\pm~0.09~\pm~0.09$		&1.3~$\pm$~0.3~$\pm$~0.1~$\pm$~0.7	&	1.2~$\pm$~0.3~$\pm$~0.1~$\pm$~0.6		&	1.3~$\pm$~0.3~$\pm$~0.1~$\pm$~0.7	&	7.5	\\
$a_{1}(1260)\to  f_{2}(1270)\pi[P]$		&	$0.311~\pm~0.033~\pm~0.019~\pm~0.021$		&$-1.58~\pm~0.11~\pm~0.08~\pm~0.25$			&1.8~$\pm$~0.4~$\pm$~0.3~$\pm$~0.5	&	0.8~$\pm$~0.2~$\pm$~0.1~$\pm$~0.3		&	1.2~$\pm$~0.2~$\pm$~0.1~$\pm$~0.3	&	$>$10	\\
$a_{1}(1260)\to (\pi^{+}\pi^{-})_{S}\pi[P]$&		-				&	-						&5.4~$\pm$~0.6~$\pm$~0.5~$\pm$~0.8	&	3.1~$\pm$~0.4~$\pm$~0.4~$\pm$~0.7		&	3.9~$\pm$~0.5~$\pm$~0.5~$\pm$~0.9		&	$>$10	\\
$~~\beta_{1}$						&	$0.83~\pm~0.13~\pm~0.11~\pm~0.34$		&$-2.18~\pm~0.14~\pm~0.19~\pm~0.16$			&-						&	-					&	-					&	-\\
$~~f_{\pi\pi}^{\rm prod}$				&	$2.47~\pm~0.16~\pm~0.12~\pm~0.12$		&$\phantom{+}0.34~\pm~0.08~\pm~0.06~\pm~0.12$		&- 						&	-					&	-					&	-\\
$~~f_{KK}^{\rm prod}$					&	$6.59~\pm~0.84~\pm~0.85~\pm~1.24$		&$\phantom{+}1.76~\pm~0.12~\pm~0.08~\pm~0.13$		&- 						&	-					&	-					&	-\\
\hline
$\pi(1300)\to \rho(770)\pi$			&	1(fixed)				&	0(fixed)					&53.8~$\pm$~2.9~$\pm$~0.9~$\pm$~8.7	&	79.0~$\pm$~2.0~$\pm$~0.8~$\pm$~6.8	&	73.6~$\pm$~2.6~$\pm$~0.8~$\pm$~9.1	&	$>$10	\\	
$\pi(1300)\to (\pi^{+}\pi^{-})_{S}\pi$	&	-					&	-						&51.1~$\pm$~2.9~$\pm$~0.9~$\pm$~8.4	&	26.2~$\pm$~2.1~$\pm$~0.8~$\pm$~6.5	&	36.7~$\pm$~2.6~$\pm$~1.0~$\pm$~8.3	&	$>$10	\\
$~~\beta_{1}$						&	$5.0~\pm~0.5~\pm~0.2~\pm~0.2$	&	$-0.64~\pm~0.10~\pm~0.04~\pm~0.25$			&-						&	-					&	-					&	-\\
$~~f_{KK}^{\rm prod}$					&	$43.8~\pm~2.8~\pm~0.5~\pm~2.7$	&	$\phantom{+}0.34~\pm~0.06~\pm~0.02~\pm~0.06$		&- 						&	-					&	-					&	-	\\
\hline
$a_{1}(1420)\to  f_{0}(980)\pi[P]$		&	1(fixed)				&	0(fixed)					&100						&		100				&	-					&6.0	\\
\hline
$a_{1}(1640)\to \rho(770)\pi[S]$		&	1(fixed)				&	0(fixed)					&100						&		100				&	-					&9.1	\\
\hline
$a_{2}(1320)\to \rho(770)\pi[D]$		&	1(fixed)				&	0(fixed)					&100						&	100					&	-					&	7.3	\\
\hline
$h_{1}(1170)\to \rho(770)\pi[S]$		&	1(fixed)				&	0(fixed)					&-						&	-					&	100					&	6.5	\\
\hline
$\pi_{2}(1670)\to  f_2(1270)\pi[S]$		&	1(fixed)				&	0(fixed)					&-						&	-					&	100					&	6.9	\\

\hline
\hline
\end{tabular}
\end{lrbox}
  \resizebox{1.0\textwidth}{!}{\usebox{\tablebox}}
\end{center}
\end{table*}

\begin{table*}[htbp]
\caption{Masses and widths of $a_1(1260)$ and $\pi(1300)$ obtained via parameter scans and presented in the PDG~\cite{ParticleDataGroup:2022pth}. The first uncertainties in this fit are the statistical uncertainties, the second are the experimental systematic uncertainties and the third are the model-dependent systematic uncertainties.}
\label{tab:respara}
\begin{center}
\begin{lrbox}{\tablebox}
\begin{tabular}{c|cc|cc}
\hline
\hline
				&	\multicolumn{2}{c|}{This work}	&	\multicolumn{2}{c}{PDG}			\\
\hline
				& ${\rm Mass}~({\rm GeV}/c^{2})$	& ${\rm Width}~({\rm GeV})$	& ${\rm Mass}~({\rm GeV}/c^{2})$	& ${\rm Width}~({\rm GeV})$		\\		
\hline	
$a_1(1260)$		&	$1.193~\pm~0.005~\pm~0.003~\pm~0.023$		&$0.487~\pm~0.009~\pm~0.015~\pm~0.039$		&$1.230~\pm~0.040$	&	$0.250-0.600$				\\
$\pi(1300)$		&	$1.534~\pm~0.011~\pm~0.009~\pm~0.020$		&$0.610~\pm~0.030~\pm~0.021~\pm~0.090$		&$1.300~\pm~0.100$	&	$0.200-0.600$		\\
\hline
\hline
\end{tabular}
\end{lrbox}
  \resizebox{1.0\textwidth}{!}{\usebox{\tablebox}}
\end{center}
\end{table*}

\end{multicols}

\begin{sidewaystable*}[htbp]
\caption{The fit fractions of interference terms in $D^0\to\pi^+\pi^-\pi^+\pi^-$ (first) and $D^0\to\pi^+\pi^-\pi^0\pi^0$ (second). The uncertainties are statistical only. }
\label{tab:inter}
\begin{center}
\begin{lrbox}{\tablebox}
\begin{tabular}{l|cccccccccccccc}
\hline
\hline
FF(\%)									&1			&2			&3			&4			& 5			&6			&7			& 8 			&9 			& 10 &11 &12 &13 &14\\
\hline	                                                                                                
1. $D^{0}\to a_{1}(1260)^{+}\pi^{-}$                 &82.2 $\pm$ 3.3	&9.9 $\pm$ 0.8	&0.5 $\pm$ 0.3	&-15.3 $\pm$ 2.5&-0.4 $\pm$ 0.7	&0.0 $\pm$ 0.0	&0.0 $\pm$ 0.0	&1.7 $\pm$ 0.2	&-27.3 $\pm$ 2.1&-4.6 $\pm$ 1.9&-5.2 $\pm$ 1.5&-1.3 $\pm$ 2.3&5.2 $\pm$ 1.1&-1.4 $\pm$ 0.2\\
2. $D^{0}\to a_{1}(1260)^{-}\pi^{+}$                 &         		&10.3 $\pm$ 1.5	&-0.1 $\pm$ 0.1	&-1.4 $\pm$ 0.3	&-2.5 $\pm$ 0.8	&0.0 $\pm$ 0.0	&0.0 $\pm$ 0.0	&-10.8 $\pm$ 1.2&-0.1 $\pm$ 0.1	&-1.8 $\pm$ 0.7&-1.5 $\pm$ 0.5&-0.3 $\pm$ 0.8&3.4 $\pm$ 0.5&-0.4 $\pm$ 0.1\\
3. $D^{0}\to a_{1}(1420)^{+}\pi^{-}$                 &         		&        		&0.6 $\pm$ 0.2	&0.1 $\pm$ 0.1	&-0.0 $\pm$ 0.0	&0.0 $\pm$ 0.0	&0.0 $\pm$ 0.0	&-0.2 $\pm$ 0.0	&0.1 $\pm$ 0.1	&-0.3 $\pm$ 0.1&-0.1 $\pm$ 0.1&-0.2 $\pm$ 0.1&1.0 $\pm$ 0.2&-0.2 $\pm$ 0.0\\
4. $D^{0}\to a_{1}(1640)^{+}\pi^{-}$                 &         		&         		&         		&1.7 $\pm$ 0.5	&0.2 $\pm$ 0.1	&-0.0 $\pm$ 0.0	&-0.0 $\pm$ 0.0	&0.2 $\pm$ 0.1	&4.9 $\pm$ 0.7	&-1.1 $\pm$ 0.3&0.3 $\pm$ 0.1&0.1 $\pm$ 0.3&-1.6 $\pm$ 0.6&0.1 $\pm$ 0.0\\
5. $D^{0}\to a_{1}(1640)^{-}\pi^{+}$                 &         		&         		&         		&         		&0.5 $\pm$ 0.3	&-0.0 $\pm$ 0.0	&-0.0 $\pm$ 0.0	&2.7 $\pm$ 0.8	&0.3 $\pm$ 0.1	&-0.2 $\pm$ 0.2&0.1 $\pm$ 0.1&0.4 $\pm$ 0.3&-1.9 $\pm$ 0.5&-0.0 $\pm$ 0.0\\
6. $D^{0}\to a_{2}(1320)^{+}\pi^{-}$                 &         		&        		&         		&         		&         		&0.2 $\pm$ 0.1	&-0.1 $\pm$ 0.0	&0.0 $\pm$ 0.0	&0.0 $\pm$ 0.0	&-0.8 $\pm$ 0.3&0.2 $\pm$ 0.1&-0.0 $\pm$ 0.0&-0.0 $\pm$ 0.0&-0.0 $\pm$ 0.0\\
7. $D^{0}\to a_{2}(1320)^{-}\pi^{+}$                 &         		&        		&         		&         		&         		&         		&0.3 $\pm$ 0.1	&-0.0 $\pm$ 0.0	&-0.0 $\pm$ 0.0	&1.0 $\pm$ 0.2&-0.2 $\pm$ 0.1&0.0 $\pm$ 0.0&0.0 $\pm$ 0.0&0.0 $\pm$ 0.0\\
8. $D^{0}\to \pi(1300)^{+}\pi^{-}$                   &         		&         		&         		&         		&         		&         		&         		&32.3 $\pm$ 2.6 &9.2 $\pm$ 1.5	&-9.3 $\pm$ 0.7&-2.8 $\pm$ 0.8&4.0 $\pm$ 1.6&-49.2 $\pm$ 4.1&0.3 $\pm$ 0.3\\
9. $D^{0}\to \pi(1300)^{-}\pi^{+}$                   &         		&        		&         		&         		&         		&         		&         		&			&23.5 $\pm$ 2.3	&-6.0 $\pm$ 0.7&-2.6 $\pm$ 0.7&-2.8 $\pm$ 1.5&-39.8 $\pm$ 3.9&-0.1 $\pm$ 0.3\\
10. $D^{0}\to \rho(770)^{0}\rho(770)^{0}$            &         		&         		&         		&         		&         		&         		&         		&         		&         		&28.0 $\pm$ 1.9&0.0 $\pm$ 2.1&0.0 $\pm$ 0.0&2.0 $\pm$ 0.5&-1.2 $\pm$ 0.2\\
11. $D^{0}\to \rho(770)^{0}\rho(1450)^{0}$           &         		&         		&         		&         		&         		&         		&         		&         		&         		&         &2.5 $\pm$ 0.9&0.0 $\pm$ 0.0&-0.6 $\pm$ 0.3&-0.3 $\pm$ 0.1\\
12. $D^{0}\to \rho(770)^{0}(\pi\pi)_{S}$             &         		&         		&         		&         		&         		&         		&         		&         		&         		&         &         &2.7 $\pm$ 0.6&0.0 $\pm$ 0.0&-0.0 $\pm$ 0.0\\
13. $D^{0}\to (\pi^{+}\pi^{-})_{S}(\pi\pi)_{S}$      &         		&         		&         		&         		&         		&         		&         		&         		&         		&         &         &         &62.8 $\pm$ 4.6&-1.3 $\pm$ 0.4\\
14. $D^{0}\to f_{2}(1270)^{0}(\pi\pi)_{S}$           &         		&         		&         		&         		&         		&         		&         		&         		&        		&         &         &         &         &1.8 $\pm$ 0.4\\
\hline
\hline
\end{tabular}
\end{lrbox}
  \resizebox{0.9\textwidth}{!}{\usebox{\tablebox}}
  \vspace{18 pt}
  
  \begin{lrbox}{\tablebox}
\begin{tabular}{l|cccccccccccccccccccc}
\hline
\hline
FF(\%)								&1			&2			&3			&4			& 5			&6			&7			& 8 			&9 			& 10 &11 &12 &13 &14&15&16&17&18&19&20\\
\hline	                                                                                                
1. $D^{0}\to a_{1}(1260)^{+}\pi^{-}$           &57.5 $\pm$ 2.7	&4.6 $\pm$ 0.4	&8.5 $\pm$ 0.6	&0.7 $\pm$ 0.2	&-10.2 $\pm$ 1.7&-0.3 $\pm$ 0.3	&0.0 $\pm$ 0.0	&-0.0 $\pm$ 0.0	&-1.6 $\pm$ 0.4	&-0.0 $\pm$ 0.0	&0.8 $\pm$ 0.1	&-18.6 $\pm$ 1.8&0.6 $\pm$ 0.2&-33.5 $\pm$ 1.8&-1.4 $\pm$ 1.2&-1.8 $\pm$ 0.3&2.4 $\pm$ 0.8&-0.4 $\pm$ 0.1&0.0 $\pm$ 0.0&-0.0 $\pm$ 0.0\\
2. $D^{0}\to a_{1}(1260)^{-}\pi^{+}$           &         	&7.2 $\pm$ 1.1	&3.1 $\pm$ 0.3	&-0.0 $\pm$ 0.0	&-0.6 $\pm$ 0.1	&-1.8 $\pm$ 0.5	&-0.0 $\pm$ 0.0	&0.0 $\pm$ 0.0	&0.6 $\pm$ 0.1	&0.3 $\pm$ 0.0	&-0.0 $\pm$ 0.0	&-7.6 $\pm$ 1.0&0.1 $\pm$ 0.1&-12.1 $\pm$ 1.1&0.4 $\pm$ 0.1&0.6 $\pm$ 0.1&2.0 $\pm$ 0.4&-0.1 $\pm$ 0.0&0.0 $\pm$ 0.0&-0.0 $\pm$ 0.0\\
3. $D^{0}\to a_{1}(1260)^{0}\pi^{0}$           &         	&         	&32.9 $\pm$ 3.2	&-0.1 $\pm$ 0.0	&-1.5 $\pm$ 0.3	&-0.2 $\pm$ 0.3	&0.0 $\pm$ 0.0	&0.0 $\pm$ 0.0	&0.0 $\pm$ 0.0	&-11.4 $\pm$ 1.1&-12.0 $\pm$ 1.0&0.7 $\pm$ 0.1&0.3 $\pm$ 0.1&-34.3 $\pm$ 2.6&0.3 $\pm$ 0.8&0.0 $\pm$ 0.0&4.8 $\pm$ 1.1&-0.3 $\pm$ 0.1&-0.0 $\pm$ 0.0&-0.0 $\pm$ 0.0\\
4. $D^{0}\to a_{1}(1420)^{+}\pi^{-}$           &         	&         	&         	&0.3 $\pm$ 0.1	&0.0 $\pm$ 0.1	&-0.0 $\pm$ 0.0	&-0.0 $\pm$ 0.0	&0.0 $\pm$ 0.0	&-0.0 $\pm$ 0.0	&0.0 $\pm$ 0.0	&0.1 $\pm$ 0.0	&-0.3 $\pm$ 0.1&-0.0 $\pm$ 0.0&-0.4 $\pm$ 0.1&0.0 $\pm$ 0.0&-0.1 $\pm$ 0.0&0.7 $\pm$ 0.1&0.0 $\pm$ 0.0&-0.0 $\pm$ 0.0&0.0 $\pm$ 0.0\\
5. $D^{0}\to a_{1}(1640)^{+}\pi^{-}$           &         	&         	&         	&         	&1.1 $\pm$ 0.3	&0.1 $\pm$ 0.0	&-0.0 $\pm$ 0.0	&0.0 $\pm$ 0.0	&0.3 $\pm$ 0.1	&-0.0 $\pm$ 0.0	&-0.1 $\pm$ 0.0	&3.7 $\pm$ 0.6&0.0 $\pm$ 0.0&3.0 $\pm$ 0.5&0.2 $\pm$ 0.1&0.1 $\pm$ 0.0&-1.3 $\pm$ 0.4&0.0 $\pm$ 0.0&0.0 $\pm$ 0.0&0.0 $\pm$ 0.0\\
6. $D^{0}\to a_{1}(1640)^{-}\pi^{+}$           &         	&         	&         	&         	&         	&0.3 $\pm$ 0.2	&0.0 $\pm$ 0.0	&-0.0 $\pm$ 0.0	&-0.0 $\pm$ 0.1	&-0.1 $\pm$ 0.0	&0.0 $\pm$ 0.0	&1.7 $\pm$ 0.5&0.0 $\pm$ 0.0&1.1 $\pm$ 0.4&-0.0 $\pm$ 0.0&-0.0 $\pm$ 0.0&-1.4 $\pm$ 0.4&-0.0 $\pm$ 0.0&-0.0 $\pm$ 0.0&-0.0 $\pm$ 0.0\\
7. $D^{0}\to a_{2}(1320)^{+}\pi^{-}$           &         	&         	&         	&         	&         	&         	&0.2 $\pm$ 0.1	&-0.0 $\pm$ 0.0	&0.0 $\pm$ 0.0	&0.0 $\pm$ 0.0	&-0.0 $\pm$ 0.0	&0.0 $\pm$ 0.0&0.0 $\pm$ 0.0&-0.6 $\pm$ 0.2&0.0 $\pm$ 0.0&0.0 $\pm$ 0.0&-0.0 $\pm$ 0.0&-0.0 $\pm$ 0.0&0.0 $\pm$ 0.0&-0.0 $\pm$ 0.0\\
8. $D^{0}\to a_{2}(1320)^{-}\pi^{+}$           &         	&         	&         	&         	&         	&         	&         	&0.3 $\pm$ 0.1	&-0.0 $\pm$ 0.0	&0.0 $\pm$ 0.0	&-0.0 $\pm$ 0.0	&-0.0 $\pm$ 0.0&0.0 $\pm$ 0.0&1.0 $\pm$ 0.2&0.0 $\pm$ 0.0&-0.0 $\pm$ 0.0&-0.0 $\pm$ 0.0&-0.0 $\pm$ 0.0&0.0 $\pm$ 0.0&-0.0 $\pm$ 0.0\\
9. $D^{0}\to h_{1}(1170)^{0}\pi^{0}$           &         	&         	&         	&         	&         	&         	&         	&         	&1.3 $\pm$ 0.6	&1.5 $\pm$ 0.5	&-1.7 $\pm$ 0.4	&-0.0 $\pm$ 0.0&0.0 $\pm$ 0.0&0.0 $\pm$ 0.0&0.1 $\pm$ 0.1&0.3 $\pm$ 0.2&0.0 $\pm$ 0.0&-0.0 $\pm$ 0.0&0.0 $\pm$ 0.0&0.0 $\pm$ 0.0\\
10. $D^{0}\to \pi(1300)^{+}\pi^{-}$            &         	&         	&         	&         	&         	&         	&         	&         	&         	&15.6 $\pm$ 1.4	&4.2 $\pm$ 0.4 	&-3.7 $\pm$ 0.4&-0.2 $\pm$ 0.0&-5.3 $\pm$ 0.7&-1.4 $\pm$ 0.7&1.1 $\pm$ 0.2&-14.6 $\pm$ 1.2&0.2 $\pm$ 0.1&0.0 $\pm$ 0.0&-0.0 $\pm$ 0.0\\
11. $D^{0}\to \pi(1300)^{-}\pi^{+}$            &         	&         	&         	&         	&         	&         	&         	&         	&         	&		&11.4 $\pm$ 1.1 &-3.1 $\pm$ 0.4&-0.1 $\pm$ 0.0&-1.9 $\pm$ 0.6&0.3 $\pm$ 0.2&-1.0 $\pm$ 0.2&-11.1 $\pm$ 1.1&0.1 $\pm$ 0.1&0.0 $\pm$ 0.0&-0.0 $\pm$ 0.0\\
12. $D^{0}\to \pi(1300)^{0}\pi^{0}$            &         	&         	&         	&         	&         	&         	&         	&         	&         	&         	&         &23.2 $\pm$ 2.8&0.4 $\pm$ 0.1&-5.3 $\pm$ 1.2&-1.1 $\pm$ 0.6&0.0 $\pm$ 0.0&-21.5 $\pm$ 2.2&0.2 $\pm$ 0.2&-0.0 $\pm$ 0.0&0.0 $\pm$ 0.0\\
13. $D^{0}\to \pi_{2}(1670)^{0}\pi^{0}$        &         	&         	&         	&         	&         	&         	&         	&         	&         	&         	&         &         &1.1 $\pm$ 0.2&-0.6 $\pm$ 0.2&-0.0 $\pm$ 0.0&0.0 $\pm$ 0.0&-0.0 $\pm$ 0.0&-0.5 $\pm$ 0.2&0.0 $\pm$ 0.0&-0.0 $\pm$ 0.0\\
14. $D^{0}\to \rho(770)^{+}\rho(770)^{-}$      &         	&         	&         	&         	&         	&         	&         	&         	&         	&         	&         &         &         &90.9 $\pm$ 3.9&-5.1 $\pm$ 3.4&0.0 $\pm$ 0.0&-2.1 $\pm$ 1.0&-2.4 $\pm$ 0.3&0.0 $\pm$ 0.0&-0.0 $\pm$ 0.0\\
15. $D^{0}\to \rho(770)^{+}\rho(1450)^{-}[D]$  &         	&         	&         	&         	&         	&         	&         	&         	&         	&         	&         &         &         &         &1.7 $\pm$ 0.8&0.0 $\pm$ 0.0&-1.0 $\pm$ 0.3&0.2 $\pm$ 0.1&-0.0 $\pm$ 0.0&0.0 $\pm$ 0.0\\
16. $D^{0}\to \rho(770)^{0}(\pi\pi)_{S}$       &         	&         	&         	&         	&         	&         	&         	&         	&         	&         	&         &         &         &         &         &1.0 $\pm$ 0.2&0.0 $\pm$ 0.0&0.0 $\pm$ 0.0&0.0 $\pm$ 0.0&-0.0 $\pm$ 0.0\\
17. $D^{0}\to (\pi^{+}\pi^{-})_{S}(\pi\pi)_{S}$&         	&         	&         	&         	&         	&         	&         	&         	&         	&         	&         &         &         &         &         &         &37.4 $\pm$ 3.0&-0.0 $\pm$ 0.0&-0.0 $\pm$ 0.0&0.0 $\pm$ 0.0\\
18. $D^{0}\to f_{2}(1270)^{0}(\pi\pi)_{S}$     &         	&         	&         	&         	&         	&         	&         	&         	&         	&         	&         &         &         &         &         &         &         &1.1 $\pm$ 0.2&0.0 $\pm$ 0.0&-0.0 $\pm$ 0.0\\
19. $D^{0}\to \omega(782)\pi^{0}$              &         	&         	&         	&         	&         	&         	&         	&         	&         	&         	&         &         &         &         &         &         &         &         &0.9 $\pm$ 0.4&-0.0 $\pm$ 0.0\\
20. $D^{0}\to \phi(1020)\pi^{0}$               &         	&         	&         	&         	&         	&         	&         	&         	&         	&         	&         &         &         &         &         &         &         &         &         &1.5 $\pm$ 0.4\\
\hline
\hline
\end{tabular}
\end{lrbox}
  \resizebox{0.9\textwidth}{!}{\usebox{\tablebox}}

\end{center}
\end{sidewaystable*}

\begin{multicols}{2}

\begin{table*}[htbp]
\caption{The $CP$-even fractions obtained in this work and comparisons with the CLEO-c and prior BESIII measurements.}
\label{tab:cp+frac}
\begin{center}
\begin{lrbox}{\tablebox}
\begin{tabular}{c|c|c}
\hline
\hline
										&	$F_{+}^{\pi^{+}\pi^{-}\pi^{+}\pi^{-}}$										& $F_{+}^{\pi^{+}\pi^{-}\pi^{0}\pi^{0}}$(non-$\eta$)\\
\hline
This work (model-dependent)					&$(75.2~\pm~1.1_{\rm stat.}~\pm~1.5_{\rm syst.})\%$									&$(68.9~\pm~1.5_{\rm stat.}~\pm~2.4_{\rm syst.})\%$\\	
CLEO-c (model-dependent)						& $(72.9~\pm~0.9_{\rm stat.}~\pm~1.5_{\rm syst.}~\pm~1.0_{\rm model})\%$~\cite{dArgent:2017gzv}		&-\\	
CLEO-c (model-independent, global)				&$(73.7~\pm~2.8)\%$~\cite{Malde:2015mha}											&-\\ 	
CLEO-c (model-independent, binned)				& $(76.9~\pm~2.1_{\rm stat.}~\pm~1.0_{\rm syst.}~\pm~0.2_{K_{S}~{\rm veto}})\%$~\cite{Harnew:2017tlp}	&-\\  
BESIII (model-independent, global)				& $(73.4~\pm~1.5_{\rm stat.}~\pm~0.8_{\rm syst.})\%$~\cite{BESIII:2022wqs}					&$(68.2~\pm~7.7)\%$~\cite{BESIII:2022qrs}\\
\hline
\hline
\end{tabular}
\end{lrbox}
  \resizebox{1.0\textwidth}{!}{\usebox{\tablebox}}
\end{center}
\end{table*}

\section{SYSTEMATIC UNCERTAINTY OF THE AMPLITUDE ANALYSIS}\label{sec_sys_amp}
The systematic uncertainties of the amplitude analysis come from two aspects.
One is the experimental systematic uncertainty which includes those from background estimation, detection efficiency over PHSP and fit bias. 
Another is the model-dependent systematic uncertainty which includes those from resonance line shape, radii of Blatt-Weisskopf barrier factors, quantum correlation parameters, extra amplitudes. 
To estimate these uncertainties for each source, the fit is performed with alternative conditions, and the deviations from the nominal results are taken as the corresponding uncertainties. 
Table~\ref{tab:sys_mag} and \ref{tab:sys_pha} summary the systematic uncertainties on the magnitude and phase of the fit parameters in the unit of the statistical uncertainty, respectively. Table~\ref{tab:sys_ff} summarized the systematic uncertainties on the FFs, resonance parameters, and $CP$-even fractions in the unit of the statistical uncertainty. The total systematic uncertainties are the square roots of the quadrature sums of the individual contributions. The individual uncertainties are obtained as follows. 
\begin{itemize}
 \item Background estimation 

The sPlot technique is used to estimate the signal weight and background, the corresponding sources of systematic uncertainty are all the PDFs and the magnitudes of peaking backgrounds in BKGVI in the 2D fit on the $M_{\rm bc}^{\rm tag}$ versus $M_{\rm bc}^{\rm sig}$ distribution. To estimate the corresponding uncertainties, an alternative fit on the $M_{\rm bc}^{\rm tag}$ versus $M_{\rm bc}^{\rm sig}$ distribution is performed by varying the means and widths of two Gaussian functions by $~\pm~1\sigma$ for signal, varying the fixed parameters of the ARGUS function by $~\pm~1\sigma$ for BKGI and BKGII, changing the Student's function to the bifurcated Student's function with different $n$ and $\sigma$ on the left and right sides of the maximum value and two ARGUS functions to one ARGUS function on the ($M_{\rm bc}^{\rm sig}+M_{\rm bc}^{\rm tag}$) dimension for BKGIII, varying the means and widths of the bifurcated Gaussian functions by $~\pm~1\sigma$ for BKGIV and BKGV and varying the fixed peaking background yields in BKGVI by $~\pm~1\sigma$ according to the estimated values in Table~\ref{tab:Npeaks}. The output results are used in the sPlot technique and amplitude analysis. 
 
 \item Detection efficiency 

The systematic uncertainties due to detection efficiencies are from the $\pi^{\pm}$ tracking/PID and $\pi^0$ reconstruction. The corresponding uncertainties are obtained by weighting the PHSP signal MC sample with a factor $\epsilon_{\rm Data}/\epsilon_{\rm MC}$  when normalizing the PDF in the amplitude analysis, where $\epsilon_{\rm Data}$ and $\epsilon_{\rm MC}$ are the efficiencies of data and MC simulation, respectively.
The $\pi^{\pm}$ tracking/PID efficiencies are quoted from Ref.~\cite{BESIII:2018apz}, while the efficiency of the $\pi^0$ reconstruction is studied with the control samples of $D^0\to K^-\pi^+\pi^0$ and $D^0\to\pi^+\pi^-\pi^0$ versus three tag modes used in this analysis.

\item Fit bias 

The sPlot technique, MC integration, detection resolution, and many other potential problems from fit tools may lead to fit bias. 
To estimate this fit bias, 100 sets of samples with the same size and signal purity as the data are generated, where the signal events are generated according to the the nominal model and the background events are from the inclusive MC sample. The amplitude fit is performed on these samples, and the average deviations of the fitted parameters from their input values are taken as uncertainties due to fit bias.

 \item The parameters and models of resonances

The systematic uncertainties associated with the masses and widths of resonances in the amplitude analysis are estimated by shifting the corresponding fixed PDG~\cite{ParticleDataGroup:2022pth} values or optimized values by $~\pm~1\sigma$. The systematic uncertainties associated with the model of $\pi\pi$ S-wave are estimated by replacing the K-matrix formula with the sum of three independent resonances of $f_0(500)$, $f_0(980)$, and $f_0(1370)$, which are described with the formula in Ref.~\cite{Bugg:2006gc}, a Flatt$\rm{\acute{e}}$ parametrization, and a relative Breit-Wigner function. For the magnitude and phase of the fit parameters of $\pi\pi$ S-wave, the systematic uncertainties associated with the model of $\pi\pi$ S-wave are not considered. The overall uncertainties are the square roots of the quadrature sums of the individual contributions.

\item Radii of Blatt-Weisskopf barrier factors 

The systematic uncertainties due to radii of Blatt-Weisskopf barrier factors are obtained by varying the radii of the $D^{0}$ meson and other resonances by $~\pm~1~{\rm{GeV}^{-1}}c$ in the amplitude analysis, and the square root of the quadrature sum of the individual effects is taken as the uncertainty. The choice of radii has a significant effect on the values of the magnitudes of the fit parameters, especially for the components with high orbital angular momentum. Therefore, this part is not included in the systematic uncertainties of the magnitudes of the fit parameters.

\item Quantum correlation correction

The systematic uncertainty due to the quantum correlation correction is estimated by varying the input quantum correlation parameters by $~\pm~1\sigma$ in the amplitude analysis.

\item Extra amplitudes

The systematic uncertainties due to the extra amplitudes are estimated with the alternative fits with the models including an additional amplitude of $f_2(1270)f_2(1270)[D] $ or $1^{++}[\rho(770)^+\pi^{-} [S]]_{\rm NR}\pi^{0}$ (``NR" represents non-resonant contribution), which are of  3$\sigma$ to 5$\sigma$ significance based on the nominal model. The square root of the quadrature sum of the individual effects is taken as the uncertainty.

\end{itemize} 

\begin{table*}[htbp]
\caption{ Systematic uncertainties in the magnitude of fit parameters in units of statistical standard deviations. 1: Background estimation. 2: Detection efficiency. 3: Fit bias. 4: Resonance parameters.  5: Radii of Blatt-Weisskopf barrier factors. 6: Quantum correlation correction. 7: Extra amplitudes. In the ``Total'' column, the term before ``$\pm$'' is the total experimental systematic uncertainty (1, 2 and 3), and the term after ``$\pm$'' is total model-dependent systematic uncertainty (4, 5, 6 and 7). The systematic uncertainties from ``5''are not considered here.}
\label{tab:sys_mag}
\begin{center}
\begin{lrbox}{\tablebox}
\begin{tabular}{l|ccccccccc}
\hline
\hline
															&1	&2		&3		&4	& 5		&6		&7		& Total\\
\hline
magnitude($D^{0}\to a_{1}(1260)^{-}\pi^{+}$)                                     & 0.2   & 0.1   & 0.2  & 1.6   &  -    & 0.1   & 0.7   & 0.3 $\pm$ 1.7\\
magnitude($D^{0}\to a_{1}(1260)^{0}\pi^{0}$)                                     & 0.1   & 0.0   & 0.1  & 1.3   &  -    & 0.0   & 0.8   & 0.1 $\pm$ 1.5\\
magnitude($D^{0}\to a_{1}(1420)^{+}\pi^{-}$)                                     & 0.2   & 0.0   & 0.3   & 1.0   & -     & 0.1   & 0.5   & 0.4 $\pm$ 1.1\\
magnitude($D^{0}\to a_{1}(1640)^{+}\pi^{-}$)                                     & 0.8   & 0.2   & 0.2   & 1.4   &  -    & 0.2   & 1.3   & 0.9 $\pm$ 1.9\\
magnitude($D^{0}\to a_{1}(1640)^{-}\pi^{+}$)                                     & 0.1   & 0.1   & 0.2  & 0.6   &  -    & 0.2   & 1.2   & 0.2 $\pm$ 1.4\\
magnitude($D^{0}\to a_{2}(1320)^{+}\pi^{-}$)                                     & 0.2   & 0.1   & 0.3  & 0.7   &  -    & 0.1   & 0.2   & 0.4 $\pm$ 0.7\\
magnitude($D^{0}\to a_{2}(1320)^{-}\pi^{+}$)                                     & 0.0   & 0.0   & 0.2   & 0.6   & -     & 0.1   & 0.5   & 0.2 $\pm$ 0.8\\
magnitude($D^{0}\to h_{1}(1170)^{0}\pi^{0}$)                                     & 0.6   & 0.1   & 0.1   & 1.3   & -     & 0.0   & 1.0   & 0.7 $\pm$ 1.7\\
magnitude($D^{0}\to \pi(1300)^{+}\pi^{-}$)                                       & 1.3   & 0.0   & 0.0  & 1.1   & -     & 0.1   & 0.4   & 1.3 $\pm$ 1.2\\
magnitude($D^{0}\to \pi(1300)^{-}\pi^{+}$)                                       & 0.8   & 0.1   & 0.6   & 1.1   & -     & 0.1   & 0.1   & 1.0 $\pm$ 1.1\\
magnitude($D^{0}\to \pi(1300)^{0}\pi^{0}$)                                       & 1.0   & 0.2   & 0.1  & 1.2   &  -    & 0.2   & 0.3   & 1.0 $\pm$ 1.2\\
magnitude($D^{0}\to \pi_{2}(1670)^{0}\pi^{0}$)                                   & 0.7   & 0.1   & 0.6  & 0.3   &  -    & 0.1   & 1.4   & 1.0 $\pm$ 1.4\\
magnitude($D^{0}\to \rho(770)^{0}\rho(770)^{0}[S]$)                              & 0.0   & 0.0   & 0.3   & 1.2   & -     & 0.0   & 0.4   & 0.3 $\pm$ 1.3\\
magnitude($D^{0}\to \rho(770)^{0}\rho(770)^{0}[P]$)                              & 0.3   & 0.0   & 0.2  & 1.5   &  -    & 0.4   & 0.4   & 0.4 $\pm$ 1.6\\
magnitude($D^{0}\to \rho(770)^{0}\rho(770)^{0}[D]$)                              & 0.1   & 0.0   & 0.2  & 1.5   &  -    & 0.0   & 0.2   & 0.3 $\pm$ 1.5\\
magnitude($D^{0}\to \rho(770)^{0}\rho(1450)^{0}[P]$)                             & 0.0   & 0.2   & 0.2  & 0.5   &  -    & 0.3   & 0.2   & 0.3 $\pm$ 0.6\\
magnitude($D^{0}\to \rho(770)^{0}\rho(1450)^{0}[D]$)                             & 0.2   & 0.1   & 0.1  & 0.3   &  -    & 0.1   & 0.9   & 0.2 $\pm$ 0.9\\
magnitude($D^{0}\to \rho(770)^{+}\rho(770)^{-}[S]$)                              & 0.5   & 0.1   & 0.1  & 0.6   &  -    & 0.1   & 0.9   & 0.5 $\pm$ 1.1\\
magnitude($D^{0}\to \rho(770)^{+}\rho(770)^{-}[P]$)                              & 0.6   & 0.1   & 0.0   & 0.7   & -     & 0.0   & 0.6   & 0.6 $\pm$ 1.0\\
magnitude($D^{0}\to \rho(770)^{+}\rho(770)^{-}[D]$)                              & 0.2   & 0.1   & 0.1  & 0.9   &  -    & 0.0   & 0.6   & 0.2 $\pm$ 1.1\\
magnitude($D^{0}\to \rho(770)^{+}\rho(1450)^{-}[D]$)                             & 0.8   & 0.1   & 0.1   & 1.4   &  -    & 0.1   & 0.7   & 0.8 $\pm$ 1.6\\
magnitude($D^{0}\to \rho(770)^{0}(\pi\pi)_{S}$, $\beta_1$)                       & 0.2   & 0.0   & 0.2   & 0.2   & -     & 0.0   & 0.4   & 0.3 $\pm$ 0.4\\
magnitude($D^{0}\to \rho(770)^{0}(\pi\pi)_{S}$, $f_{\pi\pi}^{\rm prod}$)         & 0.6   & 0.0   & 0.4   & 0.4   & -     & 0.1   & 0.7   & 0.7 $\pm$ 0.8\\
magnitude($D^{0}\to \rho(770)^{0}(\pi\pi)_{S}$, $f_{KK}^{\rm prod}$)             & 0.4   & 0.0   & 0.0   & 0.2   & -     & 0.1   & 0.6   & 0.4 $\pm$ 0.7\\
magnitude($D^{0}\to (\pi^{+}\pi^{-})_{S}(\pi\pi)_{S}$, $a_{1,1}$)                & 0.7   & 0.1   & 0.0  & 0.5   &  -    & 0.1   & 0.8   & 0.7 $\pm$ 0.9\\
magnitude($D^{0}\to (\pi^{+}\pi^{-})_{S}(\pi\pi)_{S}$, $a_{1,2}$)                & 0.6   & 0.2   & 0.3  & 1.2   &  -    & 0.2   & 0.9   & 0.7 $\pm$ 1.5\\
magnitude($D^{0}\to (\pi^{+}\pi^{-})_{S}(\pi\pi)_{S}$, $b_{2,\pi\pi}$)           & 1.5   & 0.0   & 0.0  & 0.3   &  -    & 0.1   & 0.7   & 1.5 $\pm$ 0.7\\
magnitude($D^{0}\to (\pi^{+}\pi^{-})_{S}(\pi\pi)_{S}$, $c_{[\pi\pi,\pi\pi]}$)    & 0.6   & 0.1   & 0.2   & 0.4   &  -    & 0.1   & 0.8   & 0.7 $\pm$ 0.9\\
magnitude($D^{0}\to (\pi^{+}\pi^{-})_{S}(\pi\pi)_{S}$, $c_{[\pi\pi,KK]}$)        & 0.1   & 0.0   & 0.1  & 0.6   & -     & 0.2   & 0.8   & 0.1 $\pm$ 1.0\\
magnitude($D^{0}\to f_{2}(1270)^{0}(\pi\pi)_{S}$, $f_{\pi\pi}^{\rm prod}$)       & 0.5   & 0.0   & 0.3  & 0.4   &  -    & 0.1   & 1.9   & 0.6 $\pm$ 1.9\\
magnitude($D^{0}\to f_{2}(1270)^{0}(\pi\pi)_{S}$, $f_{KK}^{\rm prod}$)           & 1.2   & 0.1   & 0.7  & 0.3   &  -    & 0.1   & 0.7   & 1.4 $\pm$ 0.8\\
magnitude($D^{0}\to \omega(782)\pi^{0}$)                                         & 0.1   & 0.0   & 0.1  & 0.2   &   -   & 0.0   & 0.4   & 0.2 $\pm$ 0.4\\
magnitude($D^{0}\to \phi(1020)\pi^{0}$)                                          & 0.2   & 0.1   & 0.4  & 0.6   &  -    & 0.1   & 0.5   & 0.5 $\pm$ 0.8\\
magnitude($a_{1}(1260)\to \rho(770)\pi[D]$)                                      & 0.1   & 0.0   & 0.1   & 2.4   &  -    & 0.0   & 0.2   & 0.2 $\pm$ 2.4\\
magnitude($a_{1}(1260)\to f_{2}(1270)\pi[P]$)                                    & 0.6   & 0.0   & 0.1   & 0.5   &   -   & 0.1   & 0.3   & 0.6 $\pm$ 0.6\\
magnitude($a_{1}(1260)\to (\pi^{+}\pi^{-})_{S}\pi[P]$, $\beta_1$)                & 0.9   & 0.0   & 0.0   & 0.6   &  -    & 0.0   & 2.5   & 0.9 $\pm$ 2.6\\
magnitude($a_{1}(1260)\to (\pi^{+}\pi^{-})_{S}\pi[P]$, $f_{\pi\pi}^{\rm prod}$)  & 0.7   & 0.0   & 0.0  & 0.7   &   -   & 0.1   & 0.2   & 0.7 $\pm$ 0.7\\
magnitude($a_{1}(1260)\to (\pi^{+}\pi^{-})_{S}\pi[P]$, $f_{KK}^{\rm prod}$)      & 1.0   & 0.0   & 0.2  & 0.8   &   -   & 0.1   & 1.3   & 1.0 $\pm$ 1.5\\
magnitude($\pi(1300)\to (\pi^{+}\pi^{-})_{S}\pi$, $\beta_1$)                     & 0.2   & 0.1   & 0.3   & 0.4   &  -    & 0.0   & 0.1   & 0.3 $\pm$ 0.5\\
magnitude($\pi(1300)\to (\pi^{+}\pi^{-})_{S}\pi$, $f_{KK}^{\rm prod}$)           & 0.2   & 0.0   & 0.1  & 0.8   &   -   & 0.1   & 0.5   & 0.2 $\pm$ 1.0\\
\hline
\hline
\end{tabular}
\end{lrbox}
\resizebox{0.56\textheight}{!}{\usebox{\tablebox}}
\end{center}
\end{table*}

\begin{table*}[htbp]
\caption{ Systematic uncertainties in the phase of fit parameters in units of statistical standard deviations. 1: Background estimation. 2: Detection efficiency. 3: Fit bias. 4: Resonance parameters.  5: Radii of Blatt-Weisskopf barrier factors. 6: Quantum correlation correction. 7: Extra amplitudes. In the ``Total'' column, the term before ``$\pm$'' is the total experimental systematic uncertainty (1, 2 and 3), and the term after ``$\pm$'' is total model-dependent systematic uncertainty (4, 5, 6 and 7). }
\label{tab:sys_pha}
\begin{center}
\begin{lrbox}{\tablebox}
\begin{tabular}{l|ccccccccc}
\hline
\hline
															&1		&2		&3		&4		& 5		&6		&7		& Total\\
\hline
phase($D^{0}\to a_{1}(1260)^{-}\pi^{+}$)                                         & 0.2   & 0.0   & 0.0   & 2.2   & 0.2   & 0.1   & 0.2   & 0.2 $\pm$ 2.2\\
phase($D^{0}\to a_{1}(1260)^{0}\pi^{0}$)                                         & 1.4   & 0.1   & 0.1  & 0.7   & 1.8   & 0.0   & 1.3   & 1.4 $\pm$ 2.3\\
phase($D^{0}\to a_{1}(1420)^{+}\pi^{-}$)                                         & 0.0   & 0.0   & 0.3  & 6.0   & 0.6   & 0.0   & 0.6   & 0.3 $\pm$ 6.1\\
phase($D^{0}\to a_{1}(1640)^{+}\pi^{-}$)                                         & 0.0   & 0.0   & 0.1   & 1.6   & 0.6   & 0.0   & 0.2   & 0.1 $\pm$ 1.7\\
phase($D^{0}\to a_{1}(1640)^{-}\pi^{+}$)                                         & 0.8   & 0.0   & 0.2   & 1.1   & 1.2   & 0.2   & 0.6   & 0.8 $\pm$ 1.7\\
phase($D^{0}\to a_{2}(1320)^{+}\pi^{-}$)                                         & 0.5   & 0.1   & 0.1  & 0.6   & 0.2   & 0.0   & 0.5   & 0.5 $\pm$ 0.8\\
phase($D^{0}\to a_{2}(1320)^{-}\pi^{+}$)                                         & 0.3   & 0.0   & 0.0  & 0.5   & 0.3   & 0.1   & 0.4   & 0.3 $\pm$ 0.7\\
phase($D^{0}\to h_{1}(1170)^{0}\pi^{0}$)                                         & 0.3   & 0.1   & 0.2  & 0.4   & 0.5   & 0.1   & 1.1   & 0.3 $\pm$ 1.3\\
phase($D^{0}\to \pi(1300)^{+}\pi^{-}$)                                           & 0.9   & 0.0   & 0.0  & 6.2   & 2.4   & 0.1   & 1.3   & 0.9 $\pm$ 6.8\\
phase($D^{0}\to \pi(1300)^{-}\pi^{+}$)                                           & 1.8   & 0.1   & 0.1   & 3.6   & 2.5   & 0.0   & 1.4   & 1.9 $\pm$ 4.6\\
phase($D^{0}\to \pi(1300)^{0}\pi^{0}$)                                           & 1.6   & 0.1   & 0.0  & 5.1   & 2.2   & 0.1   & 1.6   & 1.6 $\pm$ 5.8\\
phase($D^{0}\to \pi_{2}(1670)^{0}\pi^{0}$)                                       & 0.9   & 0.0   & 0.0   & 1.7   & 0.6   & 0.1   & 1.8   & 0.9 $\pm$ 2.6\\
phase($D^{0}\to \rho(770)^{0}\rho(770)^{0}[S]$)                                  & 0.6   & 0.0   & 0.1  & 2.9   & 0.6   & 0.2   & 0.4   & 0.6 $\pm$ 2.9\\
phase($D^{0}\to \rho(770)^{0}\rho(770)^{0}[P]$)                                  & 0.1   & 0.2   & 0.1  & 1.2   & 0.0   & 0.2   & 0.5   & 0.2 $\pm$ 1.3\\
phase($D^{0}\to \rho(770)^{0}\rho(770)^{0}[D]$)                                  & 0.2   & 0.0   & 0.1   & 3.7   & 0.7   & 0.1   & 1.0   & 0.2 $\pm$ 3.9\\
phase($D^{0}\to \rho(770)^{0}\rho(1450)^{0}[P]$)                                 & 0.4   & 0.2   & 0.1   & 0.7   & 0.1   & 0.1   & 0.1   & 0.4 $\pm$ 0.8\\
phase($D^{0}\to \rho(770)^{0}\rho(1450)^{0}[D]$)                                 & 0.4   & 0.0   & 0.0  & 1.7   & 1.0   & 0.2   & 0.6   & 0.4 $\pm$ 2.1\\
phase($D^{0}\to \rho(770)^{+}\rho(770)^{-}[S]$)                                  & 1.0   & 0.0   & 0.1   & 0.3   & 0.8   & 0.1   & 1.7   & 1.0 $\pm$ 1.9\\
phase($D^{0}\to \rho(770)^{+}\rho(770)^{-}[P]$)                                  & 0.1   & 0.1   & 0.3  & 1.8   & 0.8   & 0.2   & 0.4   & 0.3 $\pm$ 2.1\\
phase($D^{0}\to \rho(770)^{+}\rho(770)^{-}[D]$)                                  & 0.4   & 0.0   & 0.1  & 1.7   & 0.1   & 0.1   & 0.4   & 0.4 $\pm$ 1.8\\
phase($D^{0}\to \rho(770)^{+}\rho(1450)^{-}[D]$)                                 & 0.6   & 0.1   & 0.0   & 1.0   & 0.9   & 0.0   & 0.5   & 0.6 $\pm$ 1.4\\
phase($D^{0}\to \rho(770)^{0}(\pi\pi)_{S}$, $\beta_1$)                           & 0.5   & 0.1   & 0.2  & 0.1   & 0.5   & 0.0   & 0.3   & 0.5 $\pm$ 0.6\\
phase($D^{0}\to \rho(770)^{0}(\pi\pi)_{S}$, $f_{\pi\pi}^{\rm prod}$)             & 0.6   & 0.0   & 0.2  & 0.3   & 0.3   & 0.0   & 0.9   & 0.6 $\pm$ 1.0\\
phase($D^{0}\to \rho(770)^{0}(\pi\pi)_{S}$, $f_{KK}^{\rm prod}$)                 & 0.1   & 0.0   & 0.0  & 0.2   & 0.5   & 0.0   & 0.3   & 0.1 $\pm$ 0.7\\
phase($D^{0}\to (\pi^{+}\pi^{-})_{S}(\pi\pi)_{S}$, $a_{1,1}$)                    & 0.4   & 0.0   & 0.1   & 0.4   & 3.8   & 0.0   & 0.5   & 0.4 $\pm$ 3.9\\
phase($D^{0}\to (\pi^{+}\pi^{-})_{S}(\pi\pi)_{S}$, $a_{1,2}$)                    & 0.4   & 0.0   & 0.0  & 1.0   & 6.6   & 0.1   & 1.5   & 0.4 $\pm$ 6.9\\
phase($D^{0}\to (\pi^{+}\pi^{-})_{S}(\pi\pi)_{S}$, $b_{2,\pi\pi}$)               & 0.1   & 0.0   & 0.0  & 0.4   & 5.6   & 0.1   & 1.5   & 0.2 $\pm$ 5.8\\
phase($D^{0}\to (\pi^{+}\pi^{-})_{S}(\pi\pi)_{S}$, $c_{[\pi\pi,\pi\pi]}$)        & 1.0   & 0.0   & 0.1  & 0.3   & 0.7   & 0.0   & 0.7   & 1.0 $\pm$ 1.0\\
phase($D^{0}\to (\pi^{+}\pi^{-})_{S}(\pi\pi)_{S}$, $c_{[\pi\pi,KK]}$)            & 0.4   & 0.0   & 0.3  & 0.5   & 0.5   & 0.1   & 0.6   & 0.6 $\pm$ 0.9\\
phase($D^{0}\to f_{2}(1270)^{0}(\pi\pi)_{S}$, $f_{\pi\pi}^{\rm prod}$)           & 0.4   & 0.1   & 0.1  & 0.4   & 1.4   & 0.0   & 1.3   & 0.4 $\pm$ 2.0\\
phase($D^{0}\to f_{2}(1270)^{0}(\pi\pi)_{S}$, $f_{KK}^{\rm prod}$)               & 0.2   & 0.0   & 0.2   & 0.3   & 1.6   & 0.1   & 1.1   & 0.2 $\pm$ 2.0\\
phase($D^{0}\to \omega(782)\pi^{0}$)                                             & 0.3   & 0.0   & 0.1   & 0.5   & 0.3   & 0.0   & 0.0   & 0.3 $\pm$ 0.5\\
phase($D^{0}\to \phi(1020)\pi^{0}$)                                              & 0.1   & 0.0   & 0.2   & 0.3   & 0.4   & 0.0   & 0.1   & 0.2 $\pm$ 0.5\\
phase($a_{1}(1260)\to \rho(770)\pi[D]$)                                          & 0.5   & 0.1   & 0.1  & 0.4   & 0.3   & 0.1   & 0.0   & 0.5 $\pm$ 0.5\\
phase($a_{1}(1260)\to f_{2}(1270)\pi[P]$)                                        & 0.7   & 0.0   & 0.1   & 2.2   & 0.2   & 0.0   & 0.7   & 0.7 $\pm$ 2.3\\
phase($a_{1}(1260)\to (\pi^{+}\pi^{-})_{S}\pi[P]$, $\beta_1$)                    & 1.3   & 0.2   & 0.3  & 0.6   & 0.9   & 0.2   & 0.3   & 1.4 $\pm$ 1.1\\
phase($a_{1}(1260)\to (\pi^{+}\pi^{-})_{S}\pi[P]$, $f_{\pi\pi}^{\rm prod}$)      & 0.7   & 0.1   & 0.1   & 0.5   & 0.5   & 0.1   & 1.2   & 0.7 $\pm$ 1.5\\
phase($a_{1}(1260)\to (\pi^{+}\pi^{-})_{S}\pi[P]$, $f_{KK}^{\rm prod}$)          & 0.6   & 0.0   & 0.0  & 0.3   & 0.4   & 0.1   & 0.9   & 0.6 $\pm$ 1.0\\
phase($\pi(1300)\to (\pi^{+}\pi^{-})_{S}\pi$, $\beta_1$)                         & 0.3   & 0.1   & 0.2  & 0.3   & 2.4   & 0.1   & 0.8   & 0.4 $\pm$ 2.5\\
phase($\pi(1300)\to (\pi^{+}\pi^{-})_{S}\pi$, $f_{KK}^{\rm prod}$)               & 0.2   & 0.0   & 0.2   & 0.5   & 0.3   & 0.1   & 0.8   & 0.3 $\pm$ 1.0\\
\hline
\hline
\end{tabular}
\end{lrbox}
\resizebox{0.53\textheight}{!}{\usebox{\tablebox}}
\end{center}
\end{table*}

\begin{table*}[htbp]
\caption{ Systematic uncertainties in the FFs, resonance parameters and $CP$-even fractions in units of statistical standard deviations. 1: Background estimation. 2: Detection efficiency. 3: Fit bias. 4: Resonance parameters.  5: Radii of Blatt-Weisskopf barrier factors. 6: Quantum correlation correction. 7: Extra amplitudes.  For the items with ``/", the quantities before and after  ``/" are for $D^0\to\pi^{+}\pi^{-}\pi^{+}\pi^{-}$ and $D^0\to\pi^{+}\pi^{-}\pi^{0}\pi^{0}$, respectively. In the ``Total'' column, the term before ``$\pm$'' is the total experimental systematic uncertainty (1, 2 and 3), and the term after ``$\pm$'' is total model-dependent systematic uncertainty (4, 5, 6 and 7). }
\label{tab:sys_ff}
\begin{center}
\begin{lrbox}{\tablebox}
\begin{tabular}{l|ccccccccc}
\hline
\hline
										&1		&2		&3		&4		& 5		&6		&7		& Total\\
\hline	                                                                                                
FF($D^{0}\to  a_{1}(1260)^{+}\pi^{-}$)			& 0.6/1.1 	& 0.0/0.1 	& 0.3/0.1 	& 3.8/0.5 	& 3.0/2.1 	& 0.0/0.1 	& 0.2/1.1 	& 0.7~$\pm$~4.8/1.1~$\pm$~2.4\\
FF($D^{0}\to  a_{1}(1260)^{-}\pi^{+}$)			& 0.1/0.1 	& 0.1/0.1 	& 0.2/0.2 	& 0.7/1.5 	& 1.4/1.2 	& 0.1/0.1 	& 0.7/0.7 	& 0.2~$\pm$~1.7/0.2~$\pm$~2.0\\
FF($D^{0}\to  a_{1}(1260)^{0}\pi^{0}$)			& -/0.4 	& -/0.1 	& -/0.2 	& -/1.8 	& -/1.7 	& -/0.0 	& -/0.9 	& -/0.5~$\pm$~2.6\\
FF($D^{0}\to  a_{1}(1420)^{+}\pi^{-}$)			& 0.1/0.2 	& 0.0/0.0 	& 0.1/0.0 	& 0.4/0.7 	& 0.4/0.5 	& 0.0/0.1 	& 0.6/0.5 	& 0.1~$\pm$~0.8/0.2~$\pm$~1.0\\
FF($D^{0}\to  a_{1}(1640)^{+}\pi^{-}$)			& 0.7/0.8 	& 0.2/0.2 	& 0.2/0.1 	& 0.6/0.9 	& 1.1/1.1 	& 0.2/0.2 	& 1.2/1.3 	& 0.8~$\pm$~1.7/0.8~$\pm$~1.9\\
FF($D^{0}\to  a_{1}(1640)^{-}\pi^{+}$)			& 0.1/0.1 	& 0.1/0.1 	& 0.0/0.0 	& 0.4/0.2 	& 0.6/0.6 	& 0.2/0.2 	& 0.9/1.0 	&0.1~$\pm$~1.2/0.1~$\pm$~1.2 \\
FF($D^{0}\to  a_{2}(1320)^{+}\pi^{-}$)			& 0.1/0.2 	& 0.2/0.2 	& 0.3/0.3 	& 0.5/0.4 	& 0.2/0.1 	& 0.1/0.1 	& 0.3/0.2 	& 0.4~$\pm$~0.6/0.4~$\pm$~0.5 \\
FF($D^{0}\to  a_{2}(1320)^{-}\pi^{+}$)			& 0.1/0.0 	& 0.0/0.0 	& 0.1/0.1 	& 0.7/0.3 	& 0.5/0.6 	& 0.1/0.1 	& 0.4/0.5 	& 0.1~$\pm$~1.0/0.1~$\pm$~0.8 \\
FF($D^{0}\to  h_{1}(1170)^{0}\pi^{0}$)			&  -/0.7 	& -/0.1 	& -/0.2 	& -/1.1 	& -/0.7 	& -/0.0 	& -/1.0 	& -/0.7~$\pm$~1.6 \\
FF($D^{0}\to  \pi(1300)^{+}\pi^{-}$)				& 0.6/0.9 	& 0.1/0.1 	& 0.0/0.0 	& 0.9/1.0 	& 1.1/1.2 	& 0.2/0.1 	& 0.6/0.2 	& 0.6~$\pm$~1.6/0.9~$\pm$~1.6 \\
FF($D^{0}\to  \pi(1300)^{-}\pi^{+}$)				& 0.1/0.4 	& 0.2/0.2 	& 0.0/0.1 	& 1.1/1.4 	& 1.2/1.5 	& 0.1/0.1 	& 0.5/0.1 	& 0.2~$\pm$~1.7/0.5~$\pm$~2.1 \\
FF($D^{0}\to  \pi(1300)^{0}\pi^{0}$)				& -/0.5 	& -/0.2 	& -/0.0 	& -/0.9 	& -/0.6 	& -/0.0 	& -/0.2 	& -/0.5~$\pm$~1.1 \\
FF($D^{0}\to  \pi_{2}(1670)^{0}\pi^{0}$)			& -/0.8 	& -/0.1 	& -/0.1 	& -/0.3 	& -/0.4 	& -/0.1 	& -/1.4 	& -/0.8~$\pm$~1.5 \\	
FF($D^{0}\to  \rho(770)^{0}\rho(770)^{0}$)			& 0.3/- 	& 0.0/- 	& 0.0/- 	& 1.5/- 	& 0.2/- 	& 0.1/- 	& 0.6/- 	& 0.3~$\pm$~1.6/- \\
FF($D^{0}\to  \rho(770)^{0}\rho(770)^{0}[S] $)		& 0.1/- 	& 0.0/- 	& 0.1/- 	& 0.7/- 	& 0.3/- 	& 0.1/- 	& 0.3/- 	& 0.1~$\pm$~0.8/- \\
FF($D^{0}\to  \rho(770)^{0}\rho(770)^{0}[P] $)		& 0.0/- 	& 0.0/- 	& 0.4/- 	& 0.5/- 	& 0.5/- 	& 0.3/- 	& 0.1/- 	& 0.4~$\pm$~0.8/- \\
FF($D^{0}\to  \rho(770)^{0}\rho(770)^{0}[D] $)		& 0.3/- 	& 0.0/- 	& 0.2/- 	& 0.9/- 	& 0.3/- 	& 0.1/- 	& 0.5/- 	& 0.4~$\pm$~1.1/- \\
FF($D^{0}\to  \rho(770)^{0}\rho(1450)^{0}$)		& 0.0/- 	& 0.0/- 	& 0.0/- 	& 0.7/- 	& 0.7/- 	& 0.1/- 	& 0.8/- 	& 0.0~$\pm$~1.3/- \\
FF($D^{0}\to  \rho(770)^{0}\rho(1450)^{0}[P] $)		& 0.1/- 	& 0.2/- 	& 0.2/- 	& 1.3/- 	& 0.6/- 	& 0.3/- 	& 0.1/- 	& 0.3~$\pm$~1.5/- \\
FF($D^{0}\to  \rho(770)^{0}\rho(1450)^{0}[D] $)		& 0.1/- 	& 0.1/- 	& 0.1/- 	& 0.2/- 	& 0.9/- 	& 0.1/- 	& 0.8/- 	& 0.2~$\pm$~1.2/- \\
FF($D^{0}\to  \rho(770)^{+}\rho(770)^{-}$)			& -/0.9 	& -/0.0 	& -/0.3 	& -/1.3 	& -/0.5 	& -/0.1 	& -/1.0 	& -/0.9~$\pm$~1.7 \\
FF($D^{0}\to  \rho(770)^{+}\rho(770)^{-}[S] $)		& -/0.6 	& -/0.1 	& -/0.1 	& -/1.1 	& -/0.6 	& -/0.0 	& -/1.0 	& -/0.6~$\pm$~1.6 \\
FF($D^{0}\to  \rho(770)^{+}\rho(770)^{-}[P] $)		& -/1.0 	& -/0.2 	& -/0.1 	& -/0.5 	& -/0.4 	& -/0.1 	& -/0.8 	& -/1.0~$\pm$~1.0 \\
FF($D^{0}\to  \rho(770)^{+}\rho(770)^{-}[D] $)		& -/0.1 	& -/0.1 	& -/0.2 	& -/0.4 	& -/0.3 	& -/0.1 	& -/0.6 	& -/0.2~$\pm$~0.8 \\
FF($D^{0}\to  \rho(770)^{+}\rho(1450)^{-}[D] $)		& -/0.7 	& -/0.1 	& -/0.4 	& -/1.9 	& -/0.8 	& -/0.1 	& -/0.7 	& -/0.8~$\pm$~2.2 \\
FF($D^{0}\to  \rho(770)^{0}(\pi\pi)_{S} $)			& 0.5/0.3 	& 0.0/0.0 	& 0.2/0.2 	& 2.7/1.8 	& 0.1/0.1 	& 0.0/0.0 	& 1.0/0.7 	& 0.5~$\pm$~2.9/0.4~$\pm$~1.9 \\
FF($D^{0}\to  (\pi^{+}\pi^{-})_{S}(\pi\pi)_{S}$)	& 0.1/0.6 	& 0.0/0.0 	& 0.1/0.2 	& 2.0/1.3 	& 0.6/0.8 	& 0.1/0.0 	& 0.3/0.5 	& 0.1~$\pm$~2.1/0.6~$\pm$~1.6 \\
FF($D^{0}\to  f_{2}(1270)^{0}(\pi\pi)_{S} $)		& 0.0/0.1 	& 0.0/0.0 	& 0.1/0.1 	& 2.6/2.9 	& 1.1/1.0 	& 0.0/0.0 	& 1.4/1.6 	& 0.1~$\pm$~3.2/0.1~$\pm$~3.5 \\
FF($D^{0}\to  \omega(782)\pi^{0} $)				& -/0.1 	& -/0.0 	& -/0.1 	& -/0.3 	& -/0.1 	& -/0.0 	& -/0.4 	& -/0.1~$\pm$~0.5 \\
FF($D^{0}\to  \phi(1020)\pi^{0} $)				& -/0.2 	& -/0.1 	& -/0.4 	& -/0.3 	& -/0.2 	& -/0.1 	& -/0.5 	& -/0.5~$\pm$~0.6 \\
FF($a_{1}(1260)^{\pm}\to \rho(770)\pi[S]$)			& 0.8/0.8 	& 0.1/0.1 	& 0.1/0.0 	& 1.2/1.2 	& 0.4/0.4 	& 0.1/0.1 	& 0.6/0.6 	& 0.8~$\pm$~1.4/0.8~$\pm$~1.4 \\
FF($a_{1}(1260)^{\pm}\to \rho(770)\pi[D]$)			& 0.1/0.1 	& 0.0/0.0 	& 0.2/0.2 	& 2.2/2.1 	& 0.5/0.4 	& 0.0/0.0 	& 0.1/0.1 	& 0.2~$\pm$~2.3/0.2~$\pm$~2.1 \\
FF($a_{1}(1260)^{\pm}\to  f_{2}(1270)\pi[P]$)		& 0.7/0.7 	& 0.0/0.0 	& 0.1/0.1 	& 0.4/0.3 	& 1.2/1.2 	& 0.1/0.1 	& 0.3/0.3 	& 0.7~$\pm$~1.3/0.7~$\pm$~1.3 \\
FF($a_{1}(1260)^{\pm}\to (\pi^{+}\pi^{-})_{S}\pi[P]$)	& 0.8/0.9 	& 0.1/0.0 	& 0.2/0.2 	& 1.0/1.6 	& 0.3/0.1 	& 0.2/0.1 	& 0.9/0.8 	& 0.8~$\pm$~1.4/0.9~$\pm$~1.8 \\
FF($a_{1}(1260)^{0}\to \rho(770)\pi[S]$)			& -/0.7 	& -/0.1 	& -/0.0 	& -/1.4 	& -/0.3 	& -/0.1 	& -/0.6 	& -/0.7~$\pm$~1.6 \\
FF($a_{1}(1260)^{0}\to \rho(770)\pi[D]$)			& -/0.1 	& -/0.0 	& -/0.2 	& -/2.2 	& -/0.4 	& -/0.0 	& -/0.1 	& -/0.2~$\pm$~2.2 \\
FF($a_{1}(1260)^{0}\to  f_{2}(1270)\pi[P]$)		& -/0.7 	& -/0.0 	& -/0.1 	& -/0.4 	& -/1.4 	& -/0.1 	& -/0.3 	& -/0.7~$\pm$~1.5 \\
FF($a_{1}(1260)^{0}\to (\pi^{+}\pi^{-})_{S}\pi[P]$)	& -/0.8 	& -/0.0 	& -/0.3 	& -/1.6 	& -/0.1 	& -/0.1 	& -/0.8 	& -/0.9~$\pm$~1.8 \\
FF($\pi(1300)^{\pm}\to \rho(770)\pi$)				& 0.0/0.2 	& 0.1/0.1 	& 0.3/0.3 	& 2.0/0.8 	& 2.2/3.3 	& 0.1/0.1 	& 0.4/0.3 	& 0.3~$\pm$~3.0/0.4~$\pm$~3.4 \\
FF($\pi(1300)^{\pm}\to (\pi^{+}\pi^{-})_{S}\pi$)	& 0.1/0.1 	& 0.1/0.1 	& 0.3/0.4 	& 1.8/1.2 	& 2.3/2.9 	& 0.1/0.1 	& 0.3/0.2 	& 0.3~$\pm$~2.9/0.4~$\pm$~3.1 \\
FF($\pi(1300)^{0}\to \rho(770)\pi$)				& -/0.2 	& -/0.1 	& -/0.2 	& -/1.0 	& -/3.3 	& -/0.1 	& -/0.4 	& -/0.3~$\pm$~3.5 \\
FF($\pi(1300)^{0}\to (\pi^{+}\pi^{-})_{S}\pi$)		& -/0.0 	& -/0.1 	& -/0.4 	& -/1.9 	& -/2.6 	& -/0.1 	& -/0.1 	& -/0.4~$\pm$~3.2 \\
\hline                                                                                                  
$M_{a_{1}(1260)}$							& 0.5 	& 0.1 	& 0.0 	& 4.0 	& 1.2 	& 0.1 	& 1.6 	&  0.5~$\pm$~4.5\\
$\Gamma_{a_{1}(1260)}$						& 1.7 	& 0.2 	& 0.0 	& 3.2 	& 2.3 	& 0.4 	& 1.7 	&  1.7~$\pm$~4.3\\
$M_{\pi(1300)}$								& 0.8 	& 0.1 	& 0.0 	& 1.1 	& 1.1 	& 0.0 	& 1.0 	&  0.8~$\pm$~1.8\\
$\Gamma_{\pi(1300)}$							& 0.7 	& 0.1 	& 0.0 	& 2.9 	& 0.4 	& 0.2 	& 0.5 	&  0.7~$\pm$~3.0\\
$F^{+}$									& 0.7/1.0 	& 0.0/0.2 	& 0.1/0.1 	& 0.7/0.8 	& 0.3/0.8 	& 0.1/0.1 	& 1.0/0.6 	&  0.7~$\pm$~1.3/1.0~$\pm$~1.3\\
\hline
\hline
\end{tabular}
\end{lrbox}
  \resizebox{0.6\textheight}{!}{\usebox{\tablebox}}
\end{center}
\end{table*}

\section{MEASUREMENT OF BRANCHING FRACTIONS}
The absolute branching fractions of $D^0\to \pi^+\pi^-\pi^+\pi^-$ and $D^0\to\pi^+\pi^-\pi^0\pi^0$(non-$\eta$) are measured with the DT method. The numbers of ST events ($N_{g}^{\rm ST}$) for the tag mode $g$ and DT events ($N_{fg}^{\rm DT}$) for the self-conjugated signal model $f$ with the tag mode $g$ are given by
\end{multicols}
\vspace{-0.28cm}
\begin{eqnarray}
 &N_{g}^{\rm ST} =&2N_{D^{0}\bar{D}^{0}}\mathcal{B}_{g}\epsilon_{g}^{ST}(1+y_D^2)(1+r_{g}^{2} -2r_{g}R_{g}y_D\mathrm{cos}\delta_{g}),\label{eq_ST} \\
 &N_{fg}^{\rm DT}=&2N_{D^{0}\bar{D}^{0}}\mathcal{B}_{f}\mathcal{B}_{g}\epsilon_{fg}^{DT}(1+y_D^2)[1+r_{g}^{2} -2r_{g}R_{g}\mathrm{cos}\delta_{g}(2F_{+}^{f}-1)]. \label{eq_DT}
  \end{eqnarray}
\vspace{-0.28cm}
\begin{multicols}{2}
\noindent Here, $N_{D^{0}\bar{D}^{0}}$ is the total number of $D^{0}\bar{D}^{0}$ pairs in data, $\mathcal{B}_{f}$ and $\mathcal{B}_{g}$ are the branching fractions of the signal mode $f$ and tag mode $g$, respectively, $\epsilon_{f}^{\rm ST}$ and $\epsilon_{fg}^{\rm ST}$ are the corresponding ST and DT efficiencies,
and $y_D$ is the $D^0$-$\bar{D}^0$ mixing parameter. By combining Eqs.~\eqref{eq_ST} and \eqref{eq_DT} and ignoring the term $2r_{g}R_{g}y_D\mathrm{cos}\delta_{g}$ in Eq.~\eqref{eq_ST}, the branching fraction of $D^0\to f$ is given by
 \begin{eqnarray}
 \mathcal{B}_{f} = \frac{\sum_{g}N_{fg}^{\rm DT}}{\sum_{g}N_{g}^{\rm ST}(\epsilon_{fg}^{\rm DT}/\epsilon_{g}^{\rm ST})\left[1-\frac{2r_{g}R_{g}\mathrm{cos}\delta_{g}}{1+r_{g}^{2}}(2F_{+}^{f}-1)\right]}. \label{eq_BF}
  \end{eqnarray}

As described in Sec.~\ref{signalyield}, the ST and DT yields are extracted by performing an unbinned maximum likelihood fit on the $M_{\rm bc}^{\rm tag}$ distribution and the 2D distribution of $M_{\rm bc}^{\rm tag}$ versus $M_{\rm bc}^{\rm sig}$, respectively. The corresponding ST and DT efficiencies are obtained with the similar fit processes on the signal MC sample as described in Sec.~\ref{MCsimulation}. The corresponding ST and DT yields as well as the ST and DT efficiencies are summarized in Table~\ref{tab:Neff}. According to Eq.~\eqref{eq_BF}, the values in Table~\ref{tab:Neff} and the $CP$-even fractions obtained in the amplitude analysis, the branching fractions of $D^0\to \pi^+\pi^-\pi^+\pi^-$ and $D^0\to \pi^+\pi^-\pi^0\pi^0$(non-$\eta$) are calculated and summarized in Table~\ref{tab:Br}. According to the FFs obtained in the amplitude analysis, the branching fractions of the intermediate processes are also determined. The obtained results are summarized in Table~\ref{tab:subbfs}.

\begin{table*}[htbp]
\caption{Branching fractions of $D^0\to \pi^+\pi^-\pi^+\pi^-$ and $D^0\to\pi^+\pi^-\pi^0\pi^0$(non-$\eta$) from this work compared to PDG~\cite{ParticleDataGroup:2022pth} values, where the first uncertainties are statistical and the second systematic.}
\label{tab:Br}
\begin{center}
\begin{tabular}{c|cc}
\hline
\hline     	
                                     &   This work                                      &  PDG\\   \hline
$ D^0 \to \pi^+\pi^-\pi^+\pi^-$   &   $(0.688~\pm~0.010~\pm~0.010)\%$   &    $(0.756~\pm~0.020)\%$   \\
$ D^0 \to \pi^+\pi^-\pi^0\pi^0$(non-$\eta$)  &   $(0.951~\pm~0.025~\pm~0.021)\%$   &     $(1.005~\pm~0.090)\%$  \\  
 \hline
\hline
\end{tabular}
\end{center}
\end{table*}

\begin{table*}[htbp]
\caption{ Branching fractions of the intermediate processes with FFs$>$1\% in $D^0\to \pi^+\pi^-\pi^+\pi^-$ and $D^0\to \pi^+\pi^-\pi^0\pi^0$. The first uncertainties are the statistical uncertainties, the second are the systematic uncertainties from the measurement of total branching fractions and the third are the systematic uncertainties of the FFs from amplitude analysis.}
\label{tab:subbfs}
\begin{center}
\begin{tabular}{l|c|c}
\hline
\hline
\multirow{2}{*}{Component}			&\multicolumn{2}{c}{Branching fraction (\%)}\\
\cline{2-3}
								&$\pi^+\pi^-\pi^+\pi^-$  		&$\pi^+\pi^-\pi^0\pi^0$ 	\\
\hline	
$D^0\to a_{1}(1260)^{+}\pi^-$			&$0.566~\pm~0.024~\pm~0.008~\pm~0.110$				&$0.546~\pm~0.027~\pm~0.011~\pm~0.069$\\
$D^0\to a_{1}(1260)^{-}\pi^+$			&$0.071~\pm~0.010~\pm~0.001~\pm~0.017$				&$0.068~\pm~0.011~\pm~0.001~\pm~0.021$	\\
$D^0\to a_{1}(1260)^{0}\pi^0$			&-								&$0.313~\pm~0.031~\pm~0.007~\pm~0.082$	\\
$D^0\to a_{1}(1640)^{+}\pi^-$			&$0.012~\pm~0.003~\pm~0.000~\pm~0.006$				&$0.010~\pm~0.003~\pm~0.000~\pm~0.007$	\\
$D^0\to h_{1}(1170)^{0}\pi^0$			&-								&$0.012~\pm~0.006~\pm~0.000~\pm~0.010$	\\
$D^0\to \pi(1300)^{+}\pi^-$			&$0.222~\pm~0.018~\pm~0.003~\pm~0.031$				&$0.148~\pm~0.014~\pm~0.003~\pm~0.025$	\\
$D^0\to \pi(1300)^{-}\pi^+$			&$0.162~\pm~0.016~\pm~0.002~\pm~0.028$				&$0.108~\pm~0.011~\pm~0.002~\pm~0.021$	\\
$D^0\to \pi(1300)^{0}\pi^0$			&-								&$0.221~\pm~0.027~\pm~0.005~\pm~0.033$	\\
$D^0\to \pi_{2}(1670)^{0}\pi^0$			&-								&$0.010~\pm~0.002~\pm~0.000~\pm~0.004$	\\	
$D^0\to \rho(770)^{0}\rho(770)^{0}$		&$0.193~\pm~0.013~\pm~0.003~\pm~0.022$				&-	\\
$D^0\to \rho(770)^{0}\rho(770)^{0}[S] $	&$0.012~\pm~0.004~\pm~0.000~\pm~0.003$				&-	\\
$D^0\to \rho(770)^{0}\rho(770)^{0}[P] $	&$0.067~\pm~0.007~\pm~0.001~\pm~0.006$				&-	\\
$D^0\to \rho(770)^{0}\rho(770)^{0}[D] $	&$0.159~\pm~0.015~\pm~0.002~\pm~0.017$				&-	\\
$D^0\to \rho(770)^{0}\rho(1450)^{0}$		&$0.017~\pm~0.006~\pm~0.000~\pm~0.008$				&-	\\
$D^0\to \rho(770)^{0}\rho(1450)^{0}[P] $	&$0.007~\pm~0.003~\pm~0.000~\pm~0.003$				&-	\\
$D^0\to \rho(770)^{0}\rho(1450)^{0}[D] $	&$0.010~\pm~0.006~\pm~0.000~\pm~0.008$				&-	\\
$D^0\to \rho(770)^{+}\rho(770)^{-}$		&-								&$0.864~\pm~0.040~\pm~0.018~\pm~0.075$	\\
$D^0\to \rho(770)^{+}\rho(770)^{-}[S] $	&-								&$0.124~\pm~0.019~\pm~0.003~\pm~0.033$	\\
$D^0\to \rho(770)^{+}\rho(770)^{-}[P] $	&-								&$0.186~\pm~0.013~\pm~0.004~\pm~0.019$	\\
$D^0\to \rho(770)^{+}\rho(770)^{-}[D] $	&-								&$0.342~\pm~0.029~\pm~0.007~\pm~0.024$	\\
$D^0\to \rho(770)^{+}\rho(1450)^{-}[D] $	&-								&$0.016~\pm~0.008~\pm~0.000~\pm~0.016$	\\
$D^0\to \rho(770)^{0}(\pi\pi)_{S} $		&$0.019~\pm~0.004~\pm~0.000~\pm~0.012$				&$0.010~\pm~0.002~\pm~0.000~\pm~0.004$	\\
$D^0\to (\pi^+\pi^-)_{S}(\pi\pi)_{S} $	&$0.432~\pm~0.032~\pm~0.006~\pm~0.066$				&$0.356~\pm~0.029~\pm~0.007~\pm~0.049$	\\
$D^0\to f_{2}(1270)^{0}(\pi\pi)_{S} $		&$0.012~\pm~0.003~\pm~0.000~\pm~0.008$				&$0.010~\pm~0.002~\pm~0.000~\pm~0.008$	\\
$D^0\to \omega(782)\pi^0 $				&-								&$0.009~\pm~0.004~\pm~0.000~\pm~0.002$	\\
$D^0\to \phi(1020)\pi^0 $				&-								&$0.014~\pm~0.004~\pm~0.000~\pm~0.003$	\\
\hline
\hline
\end{tabular}
\end{center}
\end{table*}

Benefiting from the DT method, most of the systematic uncertainties associated with the ST selection are cancelled. 
The relative systematic uncertainties in the branching fraction measurement are summarized in Table~\ref{tab:brsys}, where the total uncertainties are the square roots of the quadrature sums of the individual ones. Details of the systematic uncertainties in the branching fraction measurements are described below.

\begin{itemize}
\item Tracking efficiency 

The tracking efficiency of $\pi^{\pm}$ has been studied in Ref.~\cite{BESIII:2018apz}. The systematic uncertainties are assigned by re-weighting the data-MC difference in $\pi^{\pm}$ tracking efficiencies according to the momentum distribution of $\pi^{\pm}$ in the signal MC sample. They are assigned to be 0.4\% and 0.2\% for $D^0\to\pi^{+}\pi^{-}\pi^{+}\pi^{-}$ and $D^0\to\pi^{+}\pi^{-}\pi^{0}\pi^{0}$, respectively.

\item PID efficiency

The PID efficiency of $\pi^{\pm}$ has been studied in Ref.~\cite{BESIII:2018apz}. The systematic uncertainties are assigned by re-weighting the data-MC difference in $\pi^{\pm}$ PID efficiencies according to the momentum distribution of $\pi^{\pm}$ in the signal MC sample. They are assigned to be 1.2\% and 0.6\% for $D^0\to\pi^{+}\pi^{-}\pi^{+}\pi^{-}$ and $D^0\to\pi^{+}\pi^{-}\pi^{0}\pi^{0}$, respectively.

\item $\pi^0$ reconstruction efficiency

The $\pi^0$ reconstruction efficiency is studied by the control samples $\bar{D}^{0}\to K^+\pi^-$, $K^+\pi^-\pi^0$, $K^+\pi^-\pi^+\pi^-$ versus $D^{0}\to K^-\pi^+\pi^0$, $\pi^+\pi^-\pi^0$. The corresponding systematic uncertainty for $D^0\to \pi^+\pi^-\pi^0\pi^0$ is assigned to be 1.3\%, by re-weighting the data-MC difference in $\pi^0$ reconstruction efficiencies according to the momentum distribution of $\pi^0$ in the signal MC sample.

\item $\Delta E$ requirement in signal side 

To study the systematic uncertainty from the $\Delta E$ requirement of the signal side, the $\Delta E$ of the signal MC sample is smeared by a Gaussian function, while the mean and width are obtained by performing a fit to the $\Delta E$ distribution of data with the signal MC shape convolved with a Gaussian function. The resultant changes of the efficiencies with respect to the nominal values, 0.1\% for both $D^0\to\pi^{+}\pi^{-}\pi^{+}\pi^{-}$ and $D^0\to\pi^{+}\pi^{-}\pi^{0}\pi^{0}$, are taken as the systematic uncertainties.

\item ST fit 

    The systematic uncertainty of the ST fit is studied by changing the signal shape and background shape. The signal shape is changed by replacing the Gaussian resolution with a Crystal Ball function~\cite{Skwarnicki:1986xj}. The background shape is changed with a floating cut-off parameter for the ARGUS function. The square root of the quadrature sum of the relative change of ST yields, which gives 0.5\%, is taken as the systematic uncertainty.

\item DT fit

As described in Sec.~\ref{sec_sys_amp}, the systematic uncertainties of the DT fit are obtained by varying the signal and background shapes, and the yields of peaking backgrounds in BKGVI in the fit. The square roots of the quadrature sums of the relative changes of DT yields in the individual changes are 0.6\% and 1.4\% for $D^0\to \pi^{+}\pi^{-}\pi^{+}\pi^{-}$ and $D^0\to\pi^{+}\pi^{-}\pi^{0}\pi^{0}$, respectively, which are taken as the systematic uncertainties.

\item Amplitude model

The DT efficiencies are obtained by varying the parameters of the amplitude model within their uncertainties. The resultant standard deviations of  DT efficiencies, which are 0.3\% and 0.6\% for $D^0\to \pi^{+}\pi^{-}\pi^{+}\pi^{-}$ and $D^0\to\pi^{+}\pi^{-}\pi^{0}\pi^{0}$, respectively, are taken as the systematic uncertainties. 

\item Quantum correlation corrections 

The uncertainties associated with the input parameters of quantum correlation ($r$, $R$, $\delta$, and $F_{+}$) are 0.3\% and 0.4\% for $D^0\to \pi^{+}\pi^{-}\pi^{+}\pi^{-}$ and $D^0\to\pi^{+}\pi^{-}\pi^{0}\pi^{0}$, respectively, which are assigned according to uncertainty propagation.

\item MC statistics

The systematic uncertainties related with the limited statistics of the signal MC samples are 0.1\% and 0.2\% for $D^0\to \pi^{+}\pi^{-}\pi^{+}\pi^{-}$ and $D^0\to\pi^{+}\pi^{-}\pi^{0}\pi^{0}$, respectively.
\end{itemize}

\begin{table*}[htbp]
\caption{Relative systematic uncertainties (\%) of the measured branching fractions.}
\label{tab:brsys}
\begin{center}
\begin{tabular}{c|c|c}
\hline
\hline
Source					&$\mathcal{B}(D^0\to\pi^{+}\pi^{-}\pi^{+}\pi^{-})$  	&$\mathcal{B}(D^0\to\pi^{+}\pi^{-}\pi^{0}\pi^{0})$\\
\hline
$\pi^{\pm}$ tracking			&	0.4					&	0.2\\
$\pi^{\pm}$ PID			      	&	1.2					&	0.6\\
$\pi^0$ reconstruction		&	-					&	1.3\\
$\Delta E$ cut in signal side	&	0.1					&	0.1	\\
ST fit				                	&	0.5					&	0.5\\
DT fit				               	&	0.6					&	1.4\\
Amplitude model			&	0.3					&	0.6\\
Quantum correlation correction	&	0.3					&	0.4\\
MC statistics				&	0.1					&	0.2\\
\hline
Total						&	1.5					&	2.2\\
\hline
\hline
\end{tabular}
\end{center}
\end{table*}

\section{SUMMARY}
Using 2.93~$\rm fb^{-1}$ of $e^+e^-$ collision data taken at $\sqrt{s}=3.773~\rm{GeV}$ with the BESIII detector, a joint amplitude analysis of $D^{0}\to\pi^{+}\pi^{-}\pi^{+}\pi^{-}$ and $D^{0}\to\pi^{+}\pi^{-}\pi^{0}\pi^{0}$(non-$\eta$) is performed. Large interferences between the dominant amplitudes of  $D^{0}\to a_{1}(1260)\pi$, $D^0\to\pi(1300)\pi$, $D^0\to\rho(770)\rho(770)$ and $D^0\to2(\pi\pi)_{S}$ are observed. Based on the amplitude model, the model dependent $CP$-even fractions of $D^{0}\to\pi^{+}\pi^{-}\pi^{+}\pi^{-}$ and $D^{0}\to\pi^{+}\pi^{-}\pi^{0}\pi^{0}$(non-$\eta$) are determined to be $(75.2~\pm~1.1_{\rm stat.}~\pm~1.5_{\rm syst.})\%$ and $(68.9~\pm~1.5_{\rm stat.}~\pm~2.4_{\rm syst.})\%$, respectively, which are consistent with the previous measurements carried out by the CLEO~\cite{dArgent:2017gzv,Malde:2015mha,Harnew:2017tlp} and BESIII~\cite{BESIII:2022wqs,BESIII:2022qrs} collaboration. The branching fractions of $D^{0}\to\pi^{+}\pi^{-}\pi^{+}\pi^{-}$ and $D^{0}\to\pi^{+}\pi^{-}\pi^{0}\pi^{0}$(non-$\eta$) are measured to be $(0.688~\pm~0.010_{\rm stat.}~\pm~0.010_{\rm syst.})\%$ and $(0.951~\pm~0.025_{\rm stat.}~\pm~0.021_{\rm syst.})\%$, respectively, where the former one is $3\sigma$ lower than the PDG~\cite{ParticleDataGroup:2022pth} value. This $3\sigma$ deviation may be due to the differences in the amplitude models which affect the global reconstruction efficiency. These results provide essential information to the search for $CP$ violation~\cite{LHCb:2016qbq} and measurements of the binned strong phase parameter~\cite{Harnew:2017tlp}, which are important inputs in the $\gamma$ measurement via the $B^{-}\to D K^{-}$ decay.

\vspace{6mm}
{\it The BESIII Collaboration thanks the staff of BEPCII, the IHEP computing center and the supercomputing center of the University of Science and Technology of China (USTC) for their strong support.}
\vspace{6mm}
\end{multicols}
\clearpage

\bibliographystyle{cpc}

\bibliography{reference}

\begin{thebibliography}{10}

\bibitem{Cabibbo:1963yz}
Cabibbo, N.,
\newblock Phys. Rev. Lett. {\bf 10} (1963) 531.

\bibitem{Kobayashi:1973fv}
Kobayashi, M. and Maskawa, T.,
\newblock Prog. Theor. Phys. {\bf 49} (1973) 652.

\bibitem{Gronau:1991dp}
Gronau, M. and Wyler, D.,
\newblock Phys. Lett. B {\bf 265} (1991) 172.

\bibitem{Atwood:1996ci}
Atwood, D., Dunietz, I., and Soni, A.,
\newblock Phys. Rev. Lett. {\bf 78} (1997) 3257.

\bibitem{Giri:2003ty}
Giri, A., Grossman, Y., Soffer, A., and Zupan, J.,
\newblock Phys. Rev. D {\bf 68} (2003) 054018.

\bibitem{Harnew:2017tlp}
Harnew, S., Naik, P., Prouve, C., Rademacker, J., and Asner, D.,
\newblock JHEP {\bf 01} (2018) 144.

\bibitem{LHCb:2019yan}
Aaij, R. et~al.,
\newblock JHEP {\bf 08} (2019) 041.

\bibitem{Belle:2004bbr}
Poluektov, A. et~al.,
\newblock Phys. Rev. D {\bf 70} (2004) 072003.

\bibitem{LHCb:2016qbq}
Aaij, R. et~al.,
\newblock Phys. Lett. B {\bf 769} (2017) 345.

\bibitem{Cheng:2003bn}
Cheng, H.-Y.,
\newblock Phys. Rev. D {\bf 67} (2003) 094007.

\bibitem{Cheng:2010ry}
Cheng, H.-Y. and Chiang, C.-W.,
\newblock Phys. Rev. D {\bf 81} (2010) 074021.

\bibitem{FOCUS:2007ern}
Link, J.~M. et~al.,
\newblock Phys. Rev. D {\bf 75} (2007) 052003.

\bibitem{dArgent:2017gzv}
d'Argent, P. and Skidmore, N. et~al.,
\newblock JHEP {\bf 05} (2017) 143.

\bibitem{MARK-III:1985hbd}
Baltrusaitis, R.~M. et~al.,
\newblock Phys. Rev. Lett. {\bf 56} (1986) 2140.

\bibitem{Li:2021iwf}
Li, H.-B. and Lyu, X.-R.,
\newblock Natl. Sci. Rev. {\bf 8} (2021) nwab181.

\bibitem{BESIII:2009fln}
Ablikim, M. et~al.,
\newblock Nucl. Instrum. Meth. A {\bf 614} (2010) 345.

\bibitem{Yu:2016cof}
Yu, C. et~al.,
\newblock {BEPCII Performance and Beam Dynamics Studies on Luminosity},
\newblock in {\em {7th International Particle Accelerator Conference}}, page
  TUYA01, 2016.

\bibitem{BESIII:2020nme}
Ablikim, M. et~al.,
\newblock Chin. Phys. C {\bf 44} (2020) 040001.

\bibitem{GEANT4:2002zbu}
Agostinelli, S. et~al.,
\newblock Nucl. Instrum. Meth. A {\bf 506} (2003) 250.

\bibitem{Huang:2022wuo}
Huang, K.-X. et~al.,
\newblock Nucl. Sci. Tech. {\bf 33} (2022) 142.

\bibitem{Jadach:2000ir}
Jadach, S., Ward, B. F.~L., and Was, Z.,
\newblock Phys. Rev. D {\bf 63} (2001) 113009.

\bibitem{Jadach:1999vf}
Jadach, S., Ward, B. F.~L., and Was, Z.,
\newblock Comput. Phys. Commun. {\bf 130} (2000) 260.

\bibitem{Lange:2001uf}
Lange, D.~J.,
\newblock Nucl. Instrum. Meth. A {\bf 462} (2001) 152.

\bibitem{Ping:2008zz}
Ping, R.-G.,
\newblock Chin. Phys. C {\bf 32} (2008) 599.

\bibitem{ParticleDataGroup:2022pth}
Workman, R.~L. et~al.,
\newblock PTEP {\bf 2022} (2022) 083C01.

\bibitem{Chen:2000tv}
Chen, J.~C., Huang, G.~S., Qi, X.~R., Zhang, D.~H., and Zhu, Y.~S.,
\newblock Phys. Rev. D {\bf 62} (2000) 034003.

\bibitem{Yang:2014vra}
Yang, R.-L., Ping, R.-G., and Chen, H.,
\newblock Chin. Phys. Lett. {\bf 31} (2014) 061301.

\bibitem{Richter-Was:1992hxq}
Richter-Was, E.,
\newblock Phys. Lett. B {\bf 303} (1993) 163.

\bibitem{BESIII:2014rtm}
Ablikim, M. et~al.,
\newblock Phys. Lett. B {\bf 734} (2014) 227.

\bibitem{ARGUS:1990hfq}
Albrecht, H. et~al.,
\newblock Phys. Lett. B {\bf 241} (1990) 278.

\bibitem{HFLAV:2019otj}
Amhis, Y.~S. et~al.,
\newblock Eur. Phys. J. C {\bf 81} (2021) 226.

\bibitem{BESIII:2021eud}
Ablikim, M. et~al.,
\newblock JHEP {\bf 05} (2021) 164.

\bibitem{Berger:2010zza}
Berger, N., Liu, B., and Wang, J.,
\newblock J. Phys. Conf. Ser. {\bf 219} (2010) 042031.

\bibitem{VonHippel:1972fg}
Von~Hippel, F. and Quigg, C.,
\newblock Phys. Rev. D {\bf 5} (1972) 624.

\bibitem{Mandelstam:1962ols}
Mandelstam, S., Paton, J.~E., Peierls, R.~F., and Sarker, A.~Q.,
\newblock Annals Phys. {\bf 18} (1962) 198.

\bibitem{Herndon:1973yn}
Herndon, D., Soding, P., and Cashmore, R.~J.,
\newblock Phys. Rev. D {\bf 11} (1975) 3165.

\bibitem{Brehm:1977yr}
Brehm, J.~J.,
\newblock Annals Phys. {\bf 108} (1977) 454.

\bibitem{Rarita:1941mf}
Rarita, W. and Schwinger, J.,
\newblock Phys. Rev. {\bf 60} (1941) 61.

\bibitem{Zemach:1965ycj}
Zemach, C.,
\newblock Phys. Rev. {\bf 140} (1965) B97.

\bibitem{Chung:1997jn}
Chung, S.~U.,
\newblock Phys. Rev. D {\bf 57} (1998) 431.

\bibitem{Zou:2002ar}
Zou, B.~S. and Bugg, D.~V.,
\newblock Eur. Phys. J. A {\bf 16} (2003) 537.

\bibitem{Gounaris:1968mw}
Gounaris, G.~J. and Sakurai, J.~J.,
\newblock Phys. Rev. Lett. {\bf 21} (1968) 244.

\bibitem{BES:2004twe}
Ablikim, M. et~al.,
\newblock Phys. Lett. B {\bf 607} (2005) 243.

\bibitem{Anisovich:2002ij}
Anisovich, V.~V. and Sarantsev, A.~V.,
\newblock Eur. Phys. J. A {\bf 16} (2003) 229.

\bibitem{BaBar:2008inr}
Aubert, B. et~al.,
\newblock Phys. Rev. D {\bf 78} (2008) 034023.

\bibitem{COMPASS:2020yhb}
Alexeev, G.~D. et~al.,
\newblock Phys. Rev. Lett. {\bf 127} (2021) 082501.

\bibitem{COMPASS:2018uzl}
Aghasyan, M. et~al.,
\newblock Phys. Rev. D {\bf 98} (2018) 092003.

\bibitem{Pivk:2004ty}
Pivk, M. and Le~Diberder, F.~R.,
\newblock Nucl. Instrum. Meth. A {\bf 555} (2005) 356.

\bibitem{James:1975dr}
James, F. and Roos, M.,
\newblock Comput. Phys. Commun. {\bf 10} (1975) 343.

\bibitem{10.1214/aos/1176344552}
Efron, B.,
\newblock The Annals of Statistics {\bf 7} (1979) 1 .

\bibitem{Langenbruch:2019nwe}
Langenbruch, C.,
\newblock Eur. Phys. J. C {\bf 82} (2022) 393.

\bibitem{Malde:2015mha}
Malde, S. et~al.,
\newblock Phys. Lett. B {\bf 747} (2015) 9.

\bibitem{BESIII:2022wqs}
Ablikim, M. et~al.,
\newblock Phys. Rev. D {\bf 106} (2022) 092004.

\bibitem{BESIII:2022qrs}
Ablikim, M. et~al.,
\newblock Phys. Rev. D {\bf 106} (2022) 092005.

\bibitem{BESIII:2018apz}
Ablikim, M. et~al.,
\newblock Phys. Rev. D {\bf 97} (2018) 072004.

\bibitem{Bugg:2006gc}
Bugg, D.~V.,
\newblock J. Phys. G {\bf 34} (2007) 151.

\bibitem{Skwarnicki:1986xj}
Skwarnicki, T.,
\newblock {\em {A study of the radiative CASCADE transitions between the
  Upsilon-Prime and Upsilon resonances}},
\newblock PhD thesis, Cracow, INP, 1986.

\end{thebibliography}

\end{CJK*}
\end{document}